\documentclass[journal]{IEEEtran}

\usepackage{soul} %

\usepackage[pdftex]{graphicx}
\usepackage{subcaption}
\graphicspath{{./figures/}}
\DeclareGraphicsExtensions{.pdf,.jpg,.png}

\usepackage{amsmath}
\usepackage{siunitx}
\DeclareSIUnit{\dBm}{\deci\belmilliwatt} %
\DeclareSIUnit{\dBi}{dBi} %
\DeclareSIUnit{\dBsm}{dBsm} \DeclareSIUnit{\belmilliwatt}{Bm} \DeclareSIUnit{\dBc}{dBc}

\usepackage{url}
\usepackage[hidelinks]{hyperref}
\usepackage{amssymb}
\usepackage{tablefootnote}
\usepackage{orcidlink}

\usepackage{array}
\newcolumntype{H}{>{\setbox0=\hbox\bgroup}c<{\egroup}@{}}

\usepackage{booktabs}
\usepackage[dvipsnames, svgnames, table]{xcolor} %
\usepackage{colortbl}
\usepackage{nicematrix}
\usepackage[hidelinks]{hyperref} %

\usepackage[capitalize]{cleveref}
\usepackage{multirow}
\usepackage{makecell}

\usepackage{pgfplots}
\usepgfplotslibrary{groupplots,dateplot}
\usetikzlibrary{patterns, shapes.arrows}
\usetikzlibrary{spy}
\pgfplotsset{compat=newest}
\usetikzlibrary{external}
\usetikzlibrary{calc}
\usetikzlibrary{shadings}
\usetikzlibrary{shadows.blur}
\usetikzlibrary{decorations.pathreplacing}

\usepackage{pgfplotstable}
\pgfplotsset{compat=1.18}
\usepgfplotslibrary{groupplots} 

\definecolor{mplblue}{RGB}{31,119,180}
\definecolor{mplorange}{RGB}{255,127,14}
\definecolor{mplgreen}{RGB}{44,160,44}
\definecolor{mplred}{RGB}{214,39,40}
\definecolor{mplpurple}{RGB}{148,103,189}
\definecolor{mplbrown}{RGB}{140,86,75}
\definecolor{mplpink}{RGB}{227,119,194}
\definecolor{mplgray}{RGB}{127,127,127}
\definecolor{mployellow}{RGB}{188,189,34}
\definecolor{mplcyan}{RGB}{23,190,207}

\pgfplotsset{
  cycle list={
    {mplblue,solid},
    {mplorange,solid},
    {mplgreen,solid},
    {mplred,solid},
    {mplpurple,solid},
    {mplbrown,solid},
    {mplpink,solid},
    {mplgray,solid},
    {mployellow,solid},
    {mplcyan,solid},
  },
  every axis plot/.append style={very thick}
}

\usepackage[xindy, nonumberlist]{glossaries}

\newacronym{2d}{2D}{two-dimensional}
\newacronym{3d}{3D}{three-dimensional}
\newacronym{3gpp}{3GPP}{3rd generation partnership project}
\newacronym{3msk}{3MSK}{3-level minimum-shift-keying}
\newacronym{4g}{4G}{fourth-generation}
\newacronym{5g}{5G}{fifth-generation}
\newacronym{6dof}{6DoF}{six degrees of freedom}
\newacronym{6g}{6G}{sixth-generation}
\newacronym{6tisch}{6TiSCH}{IPv6 over the TSCH mode of IEEE 802.15.4e}
\newacronym{abd}{ABD}{ambient-backscattering device}
\newacronym{abp}{ABP}{authentication by personalisation}
\newacronym{ac}{AC}{alternating current}
\newacronym{aci}{ACI}{adjacent channel interference}
\newacronym{aclr}{ACLR}{adjacent channel leakage ratio}
\newacronym{acpr}{ACPR}{adjacent channel power ratio}
\newacronym{adc}{ADC}{analog-to-digital converter}
\newacronym{adr}{ADR}{adaptive data rate}
\newacronym{aec}{AEC}{aluminum electrolytic capacitor}
\newacronym{aes}{AES}{Advanced Encryption Standard}
\newacronym{afa}{AFA}{adaptive frequency agility}
\newacronym{afe}{AFE}{analog front end}
\newacronym{ag}{AG}{array gain}
\newacronym{agc}{AGC}{automatic gain controller}
\newacronym{agps}{A-GPS}{Assisted Global Positioning System} %
\newacronym{ai}{AI}{artificial intelligence}
\newacronym{aimd}{AIMD}{Additive-Increase/Multiplicative-Decrease}
\newacronym{aiot}{A-IoT}{Ambient IoT}
\newacronym{alk}{ALK}{Alkaline}
\newacronym{am}{AM}{amplitude modulation}
\newacronym{amam}{AM/AM}{amplitude modulation to amplitude modulation}
\newacronym{ambc}{AmBC}{ambient backscatter communication}
\newacronym{amc}{AMC}{automatic modulation classification}
\newacronym{amp}{AMP}{approximate message passing}
\newacronym{ampm}{AM/PM}{amplitude modulation to phase modulation}
\newacronym{aoa}{AOA}{angle-of-arrival}
\newacronym{aod}{AOD}{angle-of-departure}
\newacronym{ap}{AP}{access point}
\newacronym{api}{API}{application program interface}
\newacronym{apt}{APT}{acoustic power transfer}
\newacronym{apu}{APU}{access point unit}
\newacronym{ar}{AR}{augmented reality}
\newacronym{arima}{ARIMA}{Auto-Regressive Integrated Moving Average}
\newacronym{arp}{ARP}{Antenna Reference Point}
\newacronym{asic}{ASIC}{application specific integrated circuit}
\newacronym{ask}{ASK}{amplitude-shift keying}
\newacronym{at}{AT}{ATtention}
\newacronym{auv}{AUV}{autonomous underwater vehicle}
\newacronym{awg}{AWG}{arbitrary waveform generator}
\newacronym{awgn}{AWGN}{additive white Gaussian noise}
\newacronym{baw}{BAW}{bulk acoustic wave}
\newacronym{bb}{BB}{base-band}
\newacronym{bc}{BC}{backscatter communication}
\newacronym{bcjr}{BCJR}{Bahl-Cocke-Jelinek-Raviv}
\newacronym{bd}{BD}{backscatter device}
\newacronym{be}{BE}{Belgium}
\newacronym{ber}{BER}{bit error rate}
\newacronym{bf}{BF}{beamforming}
\newacronym{bga}{BGA}{ball grid array}
\newacronym{bibc}{BiBC}{bistatic backscatter communication}
\newacronym{bjt}{BJT}{bipolar junction transistor}
\newacronym{bldc}{BLDC}{brushless DC}
\newacronym{ble}{BLE}{Bluetooth Low Energy}
\newacronym{bler}{BLER}{block error rate}
\newacronym{bod}{BOD}{Brown-Out Detection}
\newacronym{bom}{BOM}{bill of materials}
\newacronym{bpsk}{BPSK}{binary phase-shift keying}
\newacronym{brp}{BRP}{beam refinement process}
\newacronym{bs}{BS}{base station}
\newacronym{btwt}{bTWT}{broadcast TWT}
\newacronym{bu}{BU}{booster unit}
\newacronym{bw}{BW}{bandwidth}
\newacronym{ca}{CA}{carrier emitter}
\newacronym{cad}{CAD}{channel activity detection}
\newacronym{cars}{CARS}{calibration reference signal}
\newacronym{cbm}{CBM}{condition based maintenance}
\newacronym{cbra}{CBRA}{contention-based random access}
\newacronym{cc}{CC}{constant current}
\newacronym{ccdf}{CCDF}{complementary cumulative distribution function}
\newacronym{ccnn}{CCNN}{circular convolutional neural network}
\newacronym{ccs}{CCS}{correlative channel sounder}
\newacronym{cdf}{CDF}{cumulative distribution function}
\newacronym{cdma}{CDMA}{code division-multiple access}
\newacronym{cdrx}{CDRX}{connected mode DRX}
\newacronym{ce}{CE}{coverage enhancement}
\newacronym{cea}{CEA}{Controlled Environment Agriculture}
\newacronym{ced}{CED}{cumulative energy density}
\newacronym{ceofdm}{CE-OFDM}{constant-envelope OFDM}
\newacronym{cept}{CEPT}{European Conference of Postal and Telecommunications Administrations}
\newacronym{cf}{CF}{cell-free}
\newacronym{cfmmimo}{CF-mMIMO}{cell-free massive MIMO}
\newacronym{cfo}{CFO}{carrier frequency offset}
\newacronym{cfra}{CFRA}{contention-free random access}
\newacronym{cir}{CIR}{channel impulse response}
\newacronym{cla}{CLA}{closed-loop approach}
\newacronym{clk}{CLK}{clock}
\newacronym{cmos}{CMOS}{complementary metal oxide semiconductor}
\newacronym{cn}{CN}{Core network}
\newacronym{cnn}{CNN}{convolutional neural network}
\newacronym{co}{CO}{combinatorial optimization}
\newacronym{cordic}{CORDIC}{coordinate rotation digital computer}
\newacronym{cost}{COST}{commercial off-the-shelf}
\newacronym{cots}{COTS}{commercial off-the-shelf}
\newacronym{cp}{CP}{cyclic prefix}
\newacronym{cpe}{CPE}{common phase error}
\newacronym{cpfsk}{CPFSK}{continuous phase frequency shift keying}
\newacronym{cpm}{CPM}{continuous phase modulation}
\newacronym{cpo}{CPO}{carrier phase offset}
\newacronym{cpt}{CPT}{capacitive power transfer}
\newacronym{cpu}{CPU}{central-processing unit}
\newacronym{cpw}{CPW}{coplanar waveguide}
\newacronym{cqi}{CQI}{channel quality indicator}
\newacronym{cr}{CR}{coding rate}
\newacronym{crc}{CRC}{cyclic redundancy check}
\newacronym{crlb}{CRLB}{Cram\'er-Rao lower bound}
\newacronym{crs}{CRS}{cell reference signal}
\newacronym{crtwt}{C-RTWT}{coordinated restricted TWT}
\newacronym{cs}{CS}{compressed sensing}
\newacronym{cse}{CSE}{channel state estimation}
\newacronym{csi}{CSI}{channel state information}
\newacronym{csp}{CSP}{contact service point}
\newacronym{css}{CSS}{chirp spread spectrum}
\newacronym{cu}{CU}{central unit}
\newacronym{cv}{CV}{constant voltage}
\newacronym{cw}{CW}{continuous wave}
\newacronym{d2d}{D2D}{device-to-device}
\newacronym{d2r}{D2R}{device-to-reader}
\newacronym{dab}{DAB}{Digital Audio Broadcasting}
\newacronym{dac}{DAC}{digital-to-analog converter}
\newacronym{daq}{DAQ}{data acquisition system}
\newacronym{das}{DAS}{distributed antenna systems}
\newacronym{db}{DB}{duty-cycled barebone}
\newacronym{dbpsk}{DBPSK}{Differential Binary Phase Shift Keying}
\newacronym{dc}{DC}{direct current}
\newacronym{dcc}{DCC}{dynamic cooperation clustering}
\newacronym{ddc}{DDC}{digital down conversion}
\newacronym{dds}{DDS}{direct digital synthesis}
\newacronym{de}{DE}{drain efficiency}
\newacronym{de-ewma}{DE-EWMA}{Dynamic Error-Exponentially Weighted Moving Average}
\newacronym{dect}{DECT}{Digital Cordless Telecommunications}
\newacronym{dft}{DFT}{discrete Fourier transform}
\newacronym{dftsofdm}{DFT-s-OFDM}{discrete Fourier Transform spread OFDM}
\newacronym{dftsofdmfdss}{DFT-s-OFDM-FDSS}{DFT-s-OFDM with frequency-domain spectral shaping}
\newacronym{dftsofdmfdssse}{DFT-s-OFDM-FDSS-SE}{DFT-s-OFDM with FDSS and spectral extension}
\newacronym{dl}{DL}{downlink}
\newacronym{dlc}{DLC}{Distributed Laser Charging}
\newacronym{dli}{DLI}{direct link interference}
\newacronym{dlpf}{DLPF}{Dual Differential Low Pass Filter}
\newacronym{dlt}{DLT}{Distributed Ledger Technology}
\newacronym{dm}{DM}{diffuse multipath}
\newacronym{dma}{DMA}{Direct Memory Access}
\newacronym{dmac}{DMAC}{Direct Memory Access Controller}
\newacronym{dmc}{DMC}{diffuse multipath component}
\newacronym{dmimo}{D-MIMO}{distributed MIMO}
\newacronym{dnn}{DNN}{deep neural network}
\newacronym{doa}{DOA}{direction-of-arrival}
\newacronym{dp}{DP}{differencial privacy}
\newacronym{dpd}{DPD}{digital pre-distortion}
\newacronym{dpdk}{DPDK}{Data Plane Development Kit}
\newacronym{dram}{DRAM}{dynamic random-access memory}
\newacronym{drcs}{$\Delta$RCS}{differential-radar cross section}
\newacronym{drx}{DRX}{Discontinuous Reception Mode}
\newacronym{dsb}{DSB}{double-sideband}
\newacronym{dsl}{DSL}{digital subscriber line}
\newacronym{dsp}{DSP}{digital signal processing}
\newacronym{dss}{DSS}{Dataset Storage Standard}
\newacronym{dsss}{DSSS}{direct-sequence spread spectrum}
\newacronym{duc}{DUC}{digital up-converter}
\newacronym{dvb}{DVB}{Digital Video Broadcasting}
\newacronym{e2e}{E2E}{end-to-end}
\newacronym{easa}{EASA}{European Union Aviation Safety Agency}
\newacronym{ebg}{EBG}{electromagnetic bandgap}
\newacronym{ebw}{EBW}{excess bandwidth}
\newacronym{ec}{EC}{European Commission}
\newacronym{ecc}{ECC}{elliptic curve cryptography}
\newacronym{ecdf}{eCDF}{empirical cumulative distribution function}
\newacronym{ecdlp}{ECDLP}{elliptic curve discrete logarithm problem}
\newacronym{ecsp}{ECSP}{edge computing service point}
\newacronym{edlc}{EDLC}{electrostatic double-layer capacitors}
\newacronym{edrx}{eDRX}{Extended Discontinuous Reception Mode}
\newacronym{ee}{EE}{energy efficiency}
\newacronym{egprs}{EGPRS}{Enhanced Data Rates for GSM Evolution}
\newacronym{eh}{EH}{energy harvesting}
\newacronym{eipm}{EIPM}{Energy Intelligence Platform Module}
\newacronym{eirp}{EIRP}{equivalent isotropically radiated power}
\newacronym{em}{EM}{electromagnetic}
\newacronym{embb}{eMBB}{enhanced Mobile Broadband}
\newacronym{en}{EN}{energy neutral}
\newacronym{end}{END}{energy-neutral device}
\newacronym{enob}{ENOB}{effective number of bits}
\newacronym{eol}{EoL}{end of life}
\newacronym{ep}{EP}{energy profiler}
\newacronym{epd}{EPD}{electronic paper display}
\newacronym{epu}{EPU}{edge processing unit}
\newacronym{er}{ER}{Energy Receiver}
\newacronym{erc}{ERC}{European Radiocommunications Committee}
\newacronym{erp}{ERP}{effective radiation power}
\newacronym[plural=ESCs,firstplural=electronic speed controllers (ESCs)]{esc}{ESC}{electronic speed control}
\newacronym{esd}{ESD}{electrostatic discharge}
\newacronym{esl}{ESL}{electronic shelf label}
\newacronym{esr}{ESR}{equivalent series resistance}
\newacronym{et}{ET}{Energy Transmitter}
\newacronym{etsi}{ETSI}{European Telecommunications Standards Institute}
\newacronym{evd}{EVD}{eigenvalue decomposition}
\newacronym{evm}{EVM}{Error Vector Magnitude}
\newacronym{ewlb}{eWLB}{Embedded Wafer Level Ball Grid Array}
\newacronym{ewma}{EWMA}{Exponentially Weighted Moving Average}
\newacronym{fa}{FA}{federation anchor}
\newacronym{fair}{FAIR}{Findability, Accessibility, Interoperability, and Reuse of digital assets}
\newacronym{fc}{FC}{fusion center}
\newacronym{fcc}{FCC}{Federal Communications Commission}
\newacronym{fcf}{FCF}{frequency correlation function}
\newacronym{fd}{FD}{front-haul distance}
\newacronym{fdd}{FDD}{frequency-division duplexing}
\newacronym{fde}{FDE}{frequency domain equalizer}
\newacronym{fdm}{FDM}{frequency-division multiplexing}
\newacronym{fdma}{FDMA}{frequency division multiple access}
\newacronym{fdss}{FDSS}{frequency-domain spectral shaping}
\newacronym{fec}{FEC}{forward-error correction}
\newacronym{fem}{FEM}{finite element analysis}
\newacronym{fembb}{feMBB}{further enhanced mobile broadband}
\newacronym{fft}{FFT}{fast Fourier transform}
\newacronym{fh}{FH}{fronthaul}
\newacronym{fhss}{FHSS}{frequency hopping spread spectrum}
\newacronym{fifo}{FIFO}{First In, First Out}
\newacronym{fim}{FIM}{Fisher information matrix}
\newacronym{fits}{FITS}{Flexible Image Transport System}
\newacronym{fl}{FL}{federated learning}
\newacronym{fm}{FM}{frequency modulation}
\newacronym{fom}{FoM}{figure of merit}
\newacronym{fov}{FOV}{field of view}
\newacronym{fpga}{FPGA}{field-programmable gate array}
\newacronym{fqt}{FQT}{Fully Quantized Training}
\newacronym{fr1}{FR1}{frequency range 1}
\newacronym{fr2}{FR2}{frequency range 2}
\newacronym{fram}{FRAM}{Ferroelectric Random Access Memory}
\newacronym{fsk}{FSK}{frequency shift keying}
\newacronym{fspl}{FSPL}{free space path loss}
\newacronym{fss}{FSS}{frequency selective surface}
\newacronym{gan}{GaN}{gallium nitride}
\newacronym{gb}{GB}{grant-based}
\newacronym{gdpr}{GDPR}{general data protection regulation}
\newacronym{gf}{GF}{grant-free}
\newacronym{gmsk}{GMSK}{Gaussian minimum-shift keying}
\newacronym{gnb}{gNB}{Next Generation Node B}
\newacronym{gni}{GNI}{gross national income}
\newacronym{gnn}{GNN}{graph neural network}
\newacronym{gnss}{GNSS}{global navigation satellite system}
\newacronym{gpclk}{GPCLK}{general purpose clock}
\newacronym{gpio}{GPIO}{General-Purpose Input/Output}
\newacronym{gpl}{GPL}{GNU General Public License}
\newacronym{gprs}{GPRS}{General Packet Radio Services}
\newacronym{gps}{GPS}{Global Positioning System}
\newacronym{gpu}{GPU}{graphical processing unit}
\newacronym{gr}{GR}{GNU~Radio}
\newacronym{grc}{GRC}{GNU Radio Companion}
\newacronym{gscm}{GSCM}{geometry‐based stochastic model}
\newacronym{gsm}{GSM}{Global System for Mobile Communications}  %
\newacronym{gspm}{GSpM}{Generalized spatial modulation}
\newacronym{gwp}{GWP}{Global Warming Potential}
\newacronym{har}{HAR}{human activity recognition}
\newacronym{harq}{HARQ}{hybrid automatic repeat request}
\newacronym{hat}{HAT}{hardware attached on top}
\newacronym{hcs}{HCS}{human-centric services}
\newacronym{hdf5}{HDF5}{Hierarchical Data Format version 5}
\newacronym{hdr}{HDR}{high data rate}
\newacronym{hf}{HF}{high frequency}
\newacronym{hfss}{HFSS}{High Frequency Simulator Software}
\newacronym{hmd}{HMD}{head-mounted display}
\newacronym{hpbm}{HPBM}{half power beam width}
\newacronym{hpbw}{HPBW}{half-power beamwidth}
\newacronym{HyMPRo}{HyMPRo}{Hybrid Multi-Path Routing algorithm}
\newacronym{i.i.d.}{i.i.d.}{independent and identically distributed}
\newacronym{i2c}{I2C}{Inter-Integrated Circuit}
\newacronym{iaq}{IAQ}{Indoor Air Quality}
\newacronym{ib}{IB}{in-band}
\newacronym{ibbc}{IBBC}{inter-band beam configuration}
\newacronym{ibo}{IBO}{input back-off}
\newacronym{ic}{IC}{integrated circuit}
\newacronym{iccs}{ICCS}{Ilmsens correlative channel sounder}
\newacronym{ici}{ICI}{intercarrier interference}
\newacronym{icnirp}{ICNIRP}{International Commission on Non-Ionizing Radiation Protection}
\newacronym{id}{ID}{information decoding}
\newacronym{idft}{IDFT}{inverse discrete Fourier transform}
\newacronym{idxm}{IDXM}{index modulation}
\newacronym{ieee}{IEEE}{Institute of Electrical and Electronics Engineers}
\newacronym{if}{IF}{intermediate-frequency}
\newacronym{ifft}{IFFT}{inverse fast-Fourier-transform}
\newacronym{iid}{i.i.d.}{independently and identically distributed}
\newacronym{iis}{IIS}{integrated information system}
\newacronym{im}{IM}{intermodulation}
\newacronym{imd}{IMD}{intermodulation distortion}
\newacronym{imu}{IMU}{inertial measurement unit}
\newacronym{inh}{InH}{indoor hotspot office}
\newacronym{io}{IO}{input/output}
\newacronym{ioe}{IoE}{Internet of Everything}
\newacronym{iot}{IoT}{Internet of Things}
\newacronym{ipt}{IPT}{inductive power transfer}
\newacronym{ipy}{IPY}{Interventions per Year}
\newacronym{iq}{IQ}{in-phase and quadrature}
\newacronym{iqi}{IQI}{IQ imbalance}
\newacronym{ir}{IR}{infrared}
\newacronym{isac}{ISAC}{integrated sensing and communication}
\newacronym{isi}{ISI}{intersymbol interference}
\newacronym{ism}{ISM}{industrial, scientific and medical}
\newacronym{isp}{ISP}{internet service provider}
\newacronym{itu}{ITU}{International Telecommunication Union}
\newacronym{itu-r}{ITU-R}{ITU – Radiocommunication Sector}
\newacronym{jesd}{JESD}{Joint Electron Devices Engineering Council}
\newacronym{jfet}{JFET}{junction field effect transistor}
\newacronym{keh}{KEH}{kinetic energy harvester}
\newacronym{kpi}{KPI}{key performance indicator}
\newacronym{ktofdm}{KT-DFT-s-OFDM}{known-tail-DFT-s-OFDM}
\newacronym{kvi}{KVI}{key value indicator}
\newacronym{larva}{LARVA}{LARge Virtual Array}
\newacronym{lbt}{LBT}{listen-before-talk}
\newacronym{lca}{LCA}{life cycle assessment}
\newacronym{lch}{LCH}{logical channel}
\newacronym{lco}{LCO}{lithium cobalt oxide}
\newacronym{ldo}{LDO}{Low-dropout voltage regulator}
\newacronym{ldpc}{LDPC}{low-density parity-check}
\newacronym{ldr}{LDR}{low data rate}
\newacronym{led}{LED}{Light Emitting Diode}
\newacronym{less}{LESS}{Low Energy Scheduler Solution}
\newacronym{lev}{LEV}{light electric vehicle}
\newacronym{lfp}{LFP}{lithium iron phosphate}
\newacronym{lib}{LIB}{Lithium-Ion Battery}
\newacronym{lic}{LIC}{lithium-ion capacitor}
\newacronym{lid}{LID}{Lithium Iron Disulfide}
\newacronym{lidar}{LiDAR}{light detection and ranging}
\newacronym{liion}{Li-ion}{lithium-ion}
\newacronym{lipo}{LiPo}{lithium polymer}
\newacronym{lis}{LIS}{large intelligent surface}
\newacronym{llh}{LLH}{log-likelihood}
\newacronym{llr}{LLR}{log-likelihood ratio}
\newacronym{lls}{LLS}{link-level simulation}
\newacronym{lmd}{LMD}{Lithium Manganese Dioxide}
\newacronym{lmmse}{LMMSE}{least minimum mean square error}
\newacronym{lmo}{LMO}{lithium ion manganese oxide}
\newacronym{lna}{LNA}{low-noise amplifier}
\newacronym{lo}{LO}{local oscillator}
\newacronym{lora}{LoRa}{long range}
\newacronym{lorawan}{LoRaWAN}{long-range wide-area network}
\newacronym{los}{LoS}{line-of-sight}
\newacronym{lp}{LP}{linear programming}
\newacronym{lpf}{LPF}{low-pass filter}
\newacronym{lpt}{LPT}{laser power transfer}
\newacronym{lpwa}{LPWA}{Low Power Wide Area}
\newacronym{lpwan}{LPWAN}{low-power wide-area network}
\newacronym{lpwans}{LPWANs}{Low-Power Wide-Area Networks}
\newacronym{lqi}{LQI}{link quality indicator}
\newacronym{lrelu}{LReLU}{leaky rectified linear unit}
\newacronym{lrt}{LRT}{likelihood-ratio test}
\newacronym{ls}{LS}{least squares}
\newacronym{lsa}{LSA}{large synthetic array}
\newacronym{lsf}{LSF}{large-scale fading}
\newacronym{lsfc}{LSFC}{large-scale fading component}
\newacronym{lstm}{LSTM}{Long Short-Term Memory}
\newacronym{ltc}{LTC}{lithium thionyl chloride}
\newacronym{lte}{LTE}{Long Term Evolution}
\newacronym{ltem}{LTE-M}{Long-Term Evolution Machine Type Communication}
\newacronym{lti}{LTI}{linear time-invariant}
\newacronym{lto}{LTO}{lithium titanate}
\newacronym{lusta}{LUSTA 5G }{Logistique mUltimodale Sécuritaire Téléopérée \& Autonome 5G}
\newacronym{m2m}{M2M}{machine to machine}
\newacronym{ma}{MA}{Movable Antennas}
\newacronym{mac}{MAC}{Medium Access Control}
\newacronym{mate}{MATE}{millimeter-wave MIMO testbed}
\newacronym{mc}{MC}{Monte Carlo}
\newacronym{mcl}{MCL}{Maximum Coupling Loss}
\newacronym{mcs}{MCS}{modulation and coding scheme}
\newacronym{mcu}{MCU}{microcontroller unit}
\newacronym{mec}{MEC}{multi-access edge computing}
\newacronym{mems}{MEMS}{micro-electromechanical systems}
\newacronym{mf}{MF}{matched filter}
\newacronym{milp}{MILP}{Mixed Integer Linear Programming}
\newacronym{mimo}{MIMO}{multiple-input multiple-output}
\newacronym{miso}{MISO}{multiple-input single-output}
\newacronym{ml}{ML}{machine learning}
\newacronym{mlp}{MLP}{multilayer perceptron}
\newacronym{mmic}{MMIC}{monolithic microwave integrated circuit}
\newacronym{mmimo}{mMIMO}{massive MIMO}
\newacronym{mmse}{MMSE}{minimum mean square error}
\newacronym{mmtc}{mMTC}{massive machine-typed communication}
\newacronym{mmwave}{mmWave}{millimeter wave}
\newacronym{mn}{MN}{matching network}
\newacronym{mobc}{MoBC}{monostatic backscatter communication}
\newacronym{mosfet}{MOSFET}{metal-oxide semiconductor field effect transistor}
\newacronym{mpc}{MPC}{multipath component}
\newacronym{mpp}{MPP}{magnetic power profile}
\newacronym{mppt}{MPPT}{maximum power point tracking}
\newacronym{mqtt}{MQTT}{Message Queuing Telemetry Transport}
\newacronym{mr}{MR}{maximum ratio}
\newacronym{mrc}{MRC}{maximum ratio combining}
\newacronym{mrc_em}{MRC}{maximum ratio combining}
\newacronym{mrc_EM}{MRC}{Magnetic Resonance Coupling}
\newacronym{mrt}{MRT}{maximum ratio transmission}
\newacronym{mse}{MSE}{mean square error}
\newacronym{msk}{MSK}{Minimum-Shift Keying}
\newacronym{mtc}{MTC}{Machine-Type Communication}
\newacronym{multi-rat}{Multi-RAT}{multiple radio access technology}
\newacronym{multirat}{Multi-RAT}{Multiple Radio Access Technology}
\newacronym{music}{MUSIC}{MUltiple SIgnal Classification}
\newacronym{nas}{NAS}{Neural Architecture Search}
\newacronym{navauwall}{NAVAUWALL}{AUtomated NAVigation in WALLonia}
\newacronym{nb}{NB}{narrowband}
\newacronym{nbiot}{NB-IoT}{narrowband IoT}
\newacronym{nca}{NCA}{nickel cobalt aluminum}
\newacronym{netcdf}{NetCDF}{Network Common Data Form}
\newacronym{nf}{NF}{noise figure}
\newacronym{nfc}{NFC}{near-field communication}
\newacronym{nfv}{NFV}{network function virtualization}
\newacronym{ngmn}{NGMN}{Next Generation Mobile Networks }
\newacronym{ni}{NI}{National Instruments}
\newacronym{nicd}{NiCd}{nikkel cadmium}
\newacronym{nimh}{NiMH}{nikkel metal hydride}
\newacronym{nlos}{NLoS}{non-line-of-sight}
\newacronym{nmc}{NMC}{nickel manganese cobalt}
\newacronym{nmos}{nMOS}{n-channel metal-oxide semiconductor}
\newacronym{nn}{NN}{neural network}
\newacronym{nnls}{NNLS}{non-negative least squares}
\newacronym{noma}{NOMA}{non-orthogonal multiple access}
\newacronym{np}{NP}{Neyman-Pearson}
\newacronym{npbch}{NPBCH}{Narrowband Physical Broadcast Channel}
\newacronym{npss}{NPSS}{Narrow Band Primay Synchronization Signal}
\newacronym{nr}{NR}{New Radio}
\newacronym{nrs}{NRS}{Narrow Band Reference Signal}
\newacronym{nsss}{NSSS}{Narrowband Secondary Synchronization Signal}
\newacronym{ntp}{NTP}{network time protocol}
\newacronym{oai}{OAI}{OpenAirInterface} %
\newacronym{obw}{OBW}{occupied bandwidth}
\newacronym{odi}{O-DI}{On-Device Intelligence}
\newacronym{oef}{OEF}{Organisation Environmental Footprint}
\newacronym{ofa}{OFA}{Once-for-All}
\newacronym{ofdm}{OFDM}{orthogonal frequency-division multiplexing}
\newacronym{ofdma}{OFDMA}{orthogonal frequency-division multiple access}
\newacronym{ofdmim}{OFDM-IM}{OFDM with index modulation}
\newacronym{olos}{OLoS}{obstructed-line-of-sight}
\newacronym{oma}{OMA}{orthogonal multiple access}
\newacronym{oob}{OOB}{out-of-band}
\newacronym{ook}{OOK}{on-off keying}
\newacronym{opbo}{OPBO}{output power backoff}
\newacronym{oran}{O-RAN}{open radio-access network}
\newacronym{os}{OS}{operating system}
\newacronym{ota}{OTA}{over-the-air}
\newacronym{otaa}{OTAA}{over-the-air authentication}
\newacronym{otaml}{OTA-TinyML}{Over-the-air TinyML}
\newacronym{ova}{OVA}{One-Versus-All}
\newacronym{p1}{P1}{Phase 1}
\newacronym{p2}{P2}{Phase 2}
\newacronym{p2p}{P2P}{point-to-point}
\newacronym{pa}{PA}{power amplifier}
\newacronym{pae}{PAE}{power-added efficiency}
\newacronym{pam}{PAM}{pulse amplitude modulation}
\newacronym{pana}{PanA}{Panel A}
\newacronym{panb}{PanB}{Panel B}
\newacronym{papr}{PAPR}{peak-to-average power ratio}
\newacronym{pb}{PB}{power beacon}
\newacronym{pc}{PC}{pilot count}
\newacronym{pcb}{PCB}{printed circuit board}
\newacronym{pcg}{PCG}{power consumption gain}
\newacronym{pcie}{PCIe}{Peripheral Component Interconnect Express}
\newacronym{pcsi}{PCSI}{perfect channel state information}
\newacronym{pd}{PD}{powered device}
\newacronym{pdcch}{PDCCH}{physical downlink control channel}
\newacronym{pdcp}{PDCP}{packet data convergence protocol}
\newacronym{pdf}{PDF}{probability density function}
\newacronym{pdp}{PDP}{power delay profile}
\newacronym{pdsch}{PDSCH}{physical downlink shared channel}
\newacronym{pdu}{PDU}{Protocol Data Unit}
\newacronym{pe}{PE}{processing element}
\newacronym{peb}{PEB}{positioning error bound}
\newacronym{peet}{PET}{Polyethylene terephthalate}
\newacronym{pef}{PEF}{Product Environmental Footprint}
\newacronym{per}{PER}{packet error rate}
\newacronym{pet}{PET}{privacy enhancing technology}
\newacronym{pg}{PG}{path gain}
\newacronym{pgd}{PGD}{proximal gradient descent}
\newacronym{phy}{PHY}{physical}
\newacronym{pin}{PIN}{positive-intrinsic-negative}
\newacronym{pki}{PKI}{public key infrastucture}
\newacronym{pl}{PL}{path loss}
\newacronym{pla}{PLA}{physically large array}
\newacronym{pla2}{PLA}{polylactic acid}
\newacronym{pll}{PLL}{phase-locked loop}
\newacronym{pls}{PLS}{physical layer security}
\newacronym[plural=PMs,firstplural=person months (PMs)]{pm}{PM}{person month}
\newacronym{pmf}{PMF}{polymer microwave fiber}
\newacronym{pmic}{PMIC}{power management integrated circuit}
\newacronym{pmu}{PMU}{power management unit}
\newacronym{pn}{PN}{pseudo-noise}
\newacronym{po}{PO}{phase offset}
\newacronym{poc}{PoC}{proof of concept}
\newacronym{poe}{PoE}{power-over-Ethernet}
\newacronym{pps}{1PPS}{pulse per second}
\newacronym{pr}{PR}{phase reversal}
\newacronym{prb}{PRB}{Physical Resource Block}
\newacronym{prbs}{PRBs}{Physical Resource Blocks}
\newacronym{prs}{PRS}{Peripheral Reflex System}
\newacronym{ps}{PS}{Processing System}
\newacronym{psd}{PSD}{power spectral density}
\newacronym{pse}{PSE}{power sourcing equipment}
\newacronym{psk}{PSK}{phase shift keying}
\newacronym{psm}{PSM}{power saving mode}
\newacronym{pss}{PSS}{primary synchronisation signal}
\newacronym{ptp}{PTP}{precision-time protocol}
\newacronym{ptrs}{PTRS}{Phase-Tracking Reference Signals}
\newacronym{ptw}{PTW}{paging time window}
\newacronym{pv}{PV}{photovoltaic}
\newacronym{pw}{PW}{planar wavefront}
\newacronym{pwm}{PWM}{pulse width modulation}
\newacronym{qam}{QAM}{quadrature amplitude modulation}
\newacronym{qos}{QoS}{quality-of-service}
\newacronym{qpsk}{QPSK}{quadrature phase-shift keying}
\newacronym{qrrls}{QR-RLS}{QR decomposition based recursive least squares}
\newacronym{quadriga}{QuaDRiGa}{QUAsi Deterministic RadIo channel GenerAtor}
\newacronym{r2d}{R2D}{reader-to-device}
\newacronym{ra}{RA}{Random Access}
\newacronym{ram}{RAM}{random-access memory}
\newacronym{ran}{RAN}{radio access network}
\newacronym{rar}{RAR}{Random Access Response}
\newacronym{rat}{RAT}{radio access technology}
\newacronym{raw}{RAW}{restricted access window}
\newacronym{rb}{Rb}{Rubidium}
\newacronym{rbs}{RBS}{radio base station}
\newacronym{rbw}{RBW}{resolution bandwidth}
\newacronym{rc}{RC}{raised-cosine}
\newacronym{rcs}{RCS}{radar cross section}
\newacronym{rdl}{RDL}{redistribution layer}
\newacronym{re}{RE}{radio element}
\newacronym{red}{RED}{radio equipment directiv}
\newacronym{redcap}{RedCap}{reduced capability}
\newacronym{relu}{ReLU}{rectified linear unit}
\newacronym{rf}{RF}{radio frequency}
\newacronym{rfeh}{RFEH}{radio frequency energy harvesting}
\newacronym{rfic}{RFIC}{radio-frequency integrated circuit}
\newacronym{rfid}{RFID}{radio frequency identification}
\newacronym{rfpt}{RFPT}{radio frequency power transfer}
\newacronym{rfsoc}{RFSoC}{Radio Frequency System-on-Chip}
\newacronym{rfwpt}{RF-WPT}{radio frequency wireless power transfer}
\newacronym{rir}{RIR}{room impulse response}
\newacronym{ris}{RIS}{reflective intelligent surface}%
\newacronym{rl}{RL}{reinforcement learning}
\newacronym{rlc}{RLC}{Radio Link Control}
\newacronym{rllmtc}{RLLMTC}{reliable low latency machine type communication}
\newacronym{rls}{RLS}{recursive least squares}
\newacronym{rms}{RMS}{root-mean-square}
\newacronym{rmse}{RMSE}{root-mean-square error}
\newacronym{rmt}{RMT}{random matrix theory}
\newacronym{rof}{RoF}{radio-over-fiber}
\newacronym{ros}{ROS}{robot operating system}
\newacronym{rpi}{RPi}{Raspberry Pi}
\newacronym{rpl}{RPL}{Routing Protocol for Low power and Lossy Networks}
\newacronym{rrc}{RRC}{Radio Resource Connection}
\newacronym{rreq}{RREQ}{route request packet}
\newacronym{rsrp}{RSRP}{Reference Signals Received Power}
\newacronym{rsrq}{RSRQ}{Reference Signal Received Quality}
\newacronym{rss}{RSS}{received signal strength}
\newacronym{rssi}{RSSI}{received signal strength indicator}
\newacronym{rtc}{RTC}{real time clock}
\newacronym{rtf}{RTF}{reader talks first}
\newacronym{rtk}{RTK}{real time kinematics}
\newacronym{rts}{RTS}{ray tracing simulator}
\newacronym{rtwt}{rTWT}{restricted TWT}
\newacronym{ru}{RU}{Radio Unit}
\newacronym{ruc}{rUC}{representative Use Cases}
\newacronym{rv}{RV}{random variable}
\newacronym{rw}{RW}{RadioWeaves}
\newacronym{rx}{RX}{receiver}
\newacronym{rzf}{RZF}{regularized zero forcing}
\newacronym{s-parameter}{S-parameter}{scattering parameter}
\newacronym{sa}{SA}{synchronization anchor}
\newacronym{sa5}{SA}{Stand Alone}
\newacronym{sar}{SAR}{specific absorption rate}
\newacronym{sbl}{SBL}{sparse Bayesian learning}
\newacronym{sc}{SC}{single carrier}
\newacronym{scfde}{SC-FDE}{single-carrier modulation with frequency-domain-equalization}
\newacronym{scfdma}{SCFDMA}{single-carrier frequency division multiple access}
\newacronym{scomp}{SC}{Split Computing}
\newacronym{scs}{SCS}{sub-carrier spacing}
\newacronym{sdap}{SDAP}{service data adaptation protocol}
\newacronym{sdg}{SDG}{Sustainable Development Goal}
\newacronym{sdm}{SDM}{sigma-delta modulator}
\newacronym{sdma}{SDMA}{spatial-division multiple access}
\newacronym{sdn}{SDN}{software-defined network}
\newacronym{sdo}{SDO}{Standards Development Organization}
\newacronym{sdof}{SDoF}{sigma-delta over fiber}
\newacronym{sdr}{SDR}{software-defined radio}
\newacronym{se}{SE}{spectral efficiency}
\newacronym{sei}{SEI}{specific emitter identification}
\newacronym{ser}{SER}{symbol-error rate}
\newacronym{sf}{SF}{spreading factor}
\newacronym{sfn}{SFN}{single frequency network}
\newacronym{sfo}{SFO}{sampling frequency offset}
\newacronym{sfp}{SFP}{small form-factor pluggable}
\newacronym{sha}{SHA}{Secure Hash Algorithm}
\newacronym{SigMF}{SigMF}{Signal Metadata Format}
\newacronym{simo}{SIMO}{single-input multiple-output}
\newacronym{sinr}{SINR}{signal-to-interference-plus-noise ratio}
\newacronym{siso}{SISO}{single-input single-output}
\newacronym{slam}{SLAM}{simultaneous localization and mapping}
\newacronym{slc}{SLC}{spatial leakage suppression}
\newacronym{slerp}{SLERP}{spherical linear interpolation}
\newacronym{sma}{SMA}{SubMiniature version A}
\newacronym{smc}{SMC}{specular multipath component}
\newacronym{smps}{SMPS}{switched mode power supply}
\newacronym{smt}{SMT}{Surface Mount Technology}
\newacronym{sndr}{SNDR}{signal-to-noise-and-distortion ratio}
\newacronym{snidr}{SNIDR}{signal-to-noise-and-interference-and-distortion ratio}
\newacronym{snir}{SNIR}{signal-to-interference-plus-noise ratio}
\newacronym{snr}{SNR}{signal-to-noise ratio}
\newacronym{soc}{SoC}{state of charge}
\newacronym{SoC}{SOC}{System on Chip}
\newacronym{sota}{SotA}{state of the art}
\newacronym{sp}{SP}{service point}
\newacronym{spdt}{SPDT}{single pole double throw}
\newacronym{spi}{SPI}{Serial Peripheral Interface}
\newacronym{spst}{SPST}{single pole single throw}
\newacronym{sram}{SRAM}{static random-access memory}
\newacronym{srd}{SRD}{short-range device}
\newacronym{srls}{SRLS}{standard recursive least squares}
\newacronym{srs}{SRS}{Sounding Reference Signal}
\newacronym{ssb}{SSB}{synchronisation signal block}
\newacronym{ssd}{SSD}{solid state drive}
\newacronym{ssq}{SSQ}{simulator sickness questionnaire}
\newacronym{sta}{STA}{station}
\newacronym{steam}{STEAM}{science, technology, engineering, the arts, and mathematics}
\newacronym{svd}{SVD}{singular value decomposition}
\newacronym{svm}{SVM}{support vector machine}
\newacronym{sw}{SW}{spherical wavefront}
\newacronym{swipt}{SWIPT}{simultaneous wireless information and power transfer}
\newacronym{synce}{SyncE}{Synchronous Ethernet}
\newacronym{t2t}{T2T}{tag-to-tag}
\newacronym{ta}{TA}{timing advance}
\newacronym{tau}{TAU}{tracking area update}
\newacronym{tcer}{TCER}{transported to consumed energy ratio}
\newacronym{tcp}{TCP}{Transmission Control Protocol}
\newacronym{tcxo}{TCXO}{temperature compensated crystal oscillator}
\newacronym{tdce}{TD-CE}{time-domain compression and expansion}
\newacronym{tdd}{TDD}{time division duplexing}
\newacronym{tdma}{TDMA}{time division-multiple access}
\newacronym{tdoa}{TDOA}{time-difference-of-arrival}
\newacronym{teg}{TEG}{Thermoelectric Generator}
\newacronym{teng}{TENG}{Triboelectric nanogenerators}
\newacronym{thz}{THz}{Terahertz}
\newacronym{tinyml}{TinyML}{Tiny Machine Learning}
\newacronym{tinyol}{TinyOL}{Tiny Online Learning}
\newacronym{tinyrl}{TinyRL}{Tiny Reinforcement Learning}
\newacronym{tinytl}{TinyTL}{Tiny Transfer Learning}
\newacronym{tl}{TL}{Transfer Learning}
\newacronym{to}{TO}{timing offset}
\newacronym{toa}{TOA}{time-of-arrival}
\newacronym{tof}{ToF}{time-of-flight}
\newacronym{tosm}{TOSM}{through-open-short-match}
\newacronym{tpms}{TPMS}{Tire-Pressure Monitoring System}
\newacronym{trl}{TRL}{technology readyness level}
\newacronym{trp}{TRP}{Transmission Reception Point}
\newacronym{tsg}{TSG}{Technical Specification Group}
\newacronym{tsn}{TSN}{time-sensitive networking}
\newacronym{ttf}{TTF}{tag talks first}
\newacronym{ttff}{TTFF}{Time To First Fix}
\newacronym{tti}{TTI}{transmission time interval}
\newacronym{ttm}{TTM}{time to market}
\newacronym{ttn}{TTN}{The Things Network}
\newacronym{twt}{TWT}{Target Wake Time}
\newacronym{tx}{TX}{transmitter}
\newacronym{uart}{UART}{Universal Asynchronous Receiver/Transmitter}
\newacronym{uav}{UAV}{unmanned aerial vehicle}
\newacronym{uc}{UC}{use case}
\newacronym{ucie}{UCIe}{Universal Chiplet Interconnect Express}
\newacronym{udp}{UDP}{User Datagram Protocol}
\newacronym{ue}{UE}{user equipment}
\newacronym{ugv}{UGV}{unmanned ground vehicle}
\newacronym{uhd}{UHD}{USRP hardware driver}
\newacronym{uhf}{UHF}{ultra-high frequency}
\newacronym{ul}{UL}{uplink}
\newacronym{ula}{ULA}{uniform linear array}
\newacronym{ule}{ULE}{ultra low energy}
\newacronym{ulp}{ULP}{ultra-low power}
\newacronym{uMIMO}{$\mu$-MIMO}{ultra-massive Multiple-Input Multiple-Output}
\newacronym{ummtc}{umMTC}{ultra massive machine type communication}
\newacronym{un}{UN}{United Nations}
\newacronym{unow}{uNOW}{unified non-orthogonal waveform}
\newacronym{up}{UP}{ultra passive}
\newacronym{upa}{UPA}{uniform planar array}
\newacronym{ura}{URA}{uniform rectangular array}
\newacronym{urllc}{URLLC}{ultra-reliable low-latency communications}
\newacronym{usrp}{USRP}{universal software radio peripheral}
\newacronym{uv}{UV}{unmanned vehicle}
\newacronym{uw}{UW}{unique word}
\newacronym{uwb}{UWB}{ultrawideband}
\newacronym{uwofdm}{UW-DFT-s-OFDM}{unique word DFT-s-OFDM}
\newacronym{v2v}{V2V}{vehicle-to-vehicle}
\newacronym{vco}{VCO}{voltage-controlled oscillator}
\newacronym{vep}{VEP}{virtual edge platform}
\newacronym{vf}{VF}{Vertical Farming}
\newacronym{vlc}{VLC}{visible light communication}
\newacronym{vlp}{VLP}{visible light positioning}
\newacronym{vna}{VNA}{vector network analyzer}
\newacronym{voc}{VOC}{Voltatile Organic Compound}
\newacronym{votable}{VOTable}{Virtual Observatory Table}
\newacronym{vr}{VR}{virtual reality}
\newacronym{vswr}{VSWR}{Voltage Standing Wave Ratio}
\newacronym{vuca}{VUCA}{volatile, uncertain, complex and ambiguous}
\newacronym{wb}{WB}{wideband}
\newacronym{wban}{WBAN}{wireless body area network}
\newacronym{wcma}{WCMA}{weather conditioned moving average}
\newacronym{wd}{WD}{Wireless distance}
\newacronym{wifi}{Wi-Fi}{Wireless Fidelity}
\newacronym{wimax}{WiMAX}{Worldwide Interoperability for Microwave Access}
\newacronym{wirelesshart}{WirelessHART}{Wireless Highway Addressable Remote Transducer Protocol}
\newacronym{wisp}{WISP}{Wireless Identification and Sensing Platform}
\newacronym{wlan}{WLAN}{wireless LAN}
\newacronym[plural=WPs,firstplural=work packages (WPs)]{wp}{WP}{work package}
\newacronym{wpc}{WPC}{Wireless Power Consortium}
\newacronym{wpt}{WPT}{wireless power transfer}
\newacronym{wr}{WR}{White Rabbit}
\newacronym{wrsn}{WRSN}{wireless rechargeable sensor network}
\newacronym{wsn}{WSN}{wireless sensor network}
\newacronym{wur}{WuR}{wake-up radio}
\newacronym{wus}{WuS}{wake-up signal}
\newacronym{xets}{XETS}{cross exponentially tapered slot}
\newacronym{xlmimo}{XL-MIMO}{extremely large-scale MIMO}
\newacronym{xr}{XR}{extended reality}
\newacronym{z3ro}{Z3RO}{zero third-order distortion}
\newacronym{zed}{ZED}{Zero-Energy device}
\newacronym{zf}{ZF}{zero-forcing}
\newacronym{zmcscg}{ZMCSCG}{zero mean circularly symmetric complex Gaussian}
\newacronym{zmq}{ZMQ}{ZeroMQ}
\newacronym{ztofdm}{ZT-DFT-s-OFDM}{zero-tail DFT-s-OFDM}

\newcommand*\emptycirc[1][1ex]{\tikz\draw (0,0) circle (#1);} 
\newcommand*\halfcirc[1][1ex]{%
  \begin{tikzpicture}
  \draw[fill] (0,0)-- (90:#1) arc (90:270:#1) -- cycle ;
  \draw (0,0) circle (#1);
  \end{tikzpicture}}
\newcommand*\fullcirc[1][1ex]{\tikz\filldraw (0,0) circle (#1);} 
\newcommand{\low}{\emptycirc\emptycirc\emptycirc}
\newcommand{\lowhigh}{\halfcirc\emptycirc\emptycirc}

\newcommand{\medium}{\fullcirc\halfcirc\emptycirc}

\newcommand{\high}{\fullcirc\fullcirc\fullcirc}

\usepackage{authblk}

\input{auto_names}

\usepackage[backend=biber, language=english, autolang=other, style=ieee]{biblatex}
 \AtBeginBibliography{\footnotesize}

\defineauthors[false]{tijl, liesbet, gilles}

\begin{document}
  \title{Long-Range Backscatter: A Bottom-Up Approach}

  \author{
  \IEEEauthorblockN{%
  Tijl Schepens\,\orcidlink{0009-0007-7177-4141},
  Gilles Callebaut\,\orcidlink{0000-0003-2413-986X},
  Liesbet Van der Perre\,\orcidlink{0000-0002-9158-9628},
  }
  
  \IEEEauthorblockA{Department of Electrical Engineering, KU Leuven, Belgium}}%

  \maketitle

  \begin{abstract}
    Continued progress towards energy-neutral \acrfull{iot} nodes expose the wireless communication link as the dominant energy bottleneck. While \acrfull{lpwan} technologies achieve long-range communication with multiple years of battery life, their active radios hinder reaching full energy neutrality. Long-range backscatter communication emerged as a key enabler, reaching one to three order of magnitude lower power consumption. New advancements leverage concepts from active radio systems such as \acrfull{css} modulation and integrate them on a low-power backscatter tag.
    
    This paper presents a comprehensive survey of long-range backscatter communication, using a bottom-up analysis spanning system topologies, hardware architecture, modulation techniques and medium access. Backscatter communication requires different topologies compared to active radios to reach longer communication distances. Different hardware architectures support backscattering a modulated signal with differing complexity, power consumption and spectral efficiency. At the physical layer binary switch-based modulation are well known and provide an easy form of modulation while \acrfull{css}-based modulation gain traction due to their robustness. \Acrfull{mac} techniques are examined with a focus on synchronization, concurrency and lightweight feedback mechanisms requiring low-power, low-complexity hardware.

    Building on these established solutions the paper evaluates the feasibility of long-range backscatter communication in different energy-neutral \acrfull{iot} applications. Starting from the available energy budget, harvested through solar, \acrfull{rf} or capacitive harvesting, feasible hardware, modulation and \acrfull{mac} solutions are explored.
  \end{abstract}


  \section{Introduction}
\IEEEPARstart{T}{he} Internet of Things landscape evolved rapidly over the recent years. The surge of connected devices puts a stress on increasing their lifetime. \Gls{iot} applications face autonomy constraints because of their batteries. This calls for new research to extend lifetimes lowering battery replacements. While long-range communication can achieve battery lifetimes above 20 years for very sporadic messaging (50 bytes per day), the lifetime drops below one year for more frequent updates (50 bytes per \SI{100}{\second}) \cite{morin2017comparison}. Increasing battery capacity extends the lifetime and enables faster update rates \cite{singh2020energy} but come with additional cost and volume. This calls for techniques to lower the nodes power consumption. Battery replacements contribute to the increasing e-waste and extraction of rare earth materials \cite{modarress2022threats} stressing the need for increased autonomy even further. Recent studies even extend the autonomy to the device lifetime by making devices energy-neutral~\cite{rossi2017energy,brunelli2019energy}.

Many newly developed technologies are created to move towards energy neutrality. Backscatter communication is one such technology. Instead of transmitting its own wireless signal it leverages existing signals to communicate. To do this it either absorbs or reflects the incoming wave. The concept of backscattering has long been used in different fields. \Gls{rfid} emerged as a main application in the RF field. \Gls{rfid} tags are fully passive by harvesting incoming RF energy. They are deployed in many areas such as retail, logistics, access control. This widespread usage results from their low complexity design, making them both low-cost and lightweight. Despite these benefits, \gls{rfid} adoption is limited by its limited range (10s of metres) and the need for dedicated readers.

A backscatter system consists of at least a tag and a transceiver. The tag backscatters the wireless signal coming from the transceiver. The transceiver can be split into a dedicated transmitter and a receiver in between which the tag is located. A local active device can replace the dedicated transmitter and activate nearby tags. The transmitter or local device can use a carrier signal which is a continuous tone onto which the tag modulates its data. The tag can also use ambient signals for backscattering which is the ongoing communication between a local device and a transceiver.

To achieve low-power operation, tags are kept low complexity limiting its communication range. Active radios boost their range through amplifiers. These are too power hungry for low-power backscatter tags. One major advancement is the introduction of \gls{lora} communication for backscattering by \textcite{talla2017lora}. \Gls{lora} excels in communication range due to its Chirp Spread Spectrum modulation. Implementing this modulation on a low-power, low complexity tag poses a true challenge researchers are tackling today. This review paper presents the evolution towards long-range backscatter communication. It elaborates on the different research related questions and compares different techniques being used. The practical feasibility of proposed solutions is assessed.

  \section{Paper Structure}
  This paper gives a bottom-up structured overview of long-range backscattering techniques, shows how to apply them in different application scenarios and provides a discussion on open challenges and technological gaps. It is structured as follows. \Cref{sec:topologies} introduces the concept of how tags modulate data using backscatter communication. It compares the communication topologies and system architectures used in \gls{lpwan} transmissions to those in backscatter.

  Different hardware architectures used to implement backscatter transmitters are discussed in \cref{sec:hardware}. These range from fixed-load modulation to more advanced IQ-modulators. Their strengths and weaknesses are highlighted and techniques to improve range such as tunnel diodes and carrier cancellation are discussed.

  Modulation forms a fundamental part in wireless communication and determines its inherent characteristics. \Cref{sec:active-modulation} briefly summarizes active long-range modulation schemes employed in \gls{lpwan} to establish a baseline for comparison with backscatter-based approaches. \Cref{sec:modulation} then provides an in-depth discussion of backscatter modulation techniques starting with simple binary switch-based modulation and transitioning to more advanced \gls{css}-based schemes.

  The low-complexity hardware of backscatter tags pose a challenge to implement \gls{mac} protocols in order to support synchronization, concurrency and feedback. \Cref{sec:mac} looks into the protocols implemented in active radio and explores how these are adapted for backscatter-based systems.

  Backscattering lends itself as an enabler for \gls{end} due to its low-power operation. A fundamental aspect to truly reach energy neutrality lies in the available energy budget harvested from the environment and how it is stored. \Cref{sec:autonomy} examines different harvesting and storage technologies and their impact on feasible communication strategies. Utilizing this information \cref{sec:assessment} evaluates different practical application scenarios and assess the suitability of the different hardware, modulation and \gls{mac} solutions based on the available energy budget.

  To conclude \cref{sec:discussion} covers topics such as concurrency, security, data rate and localization and looks into the available techniques while highlighting gaps for improvement. \Cref{sec:conclusion} concludes the paper with a summary of the key insights and outlines the open research questions.

  \section{Topologies / Architectures}\label{sec:topologies}
This section outlines the various topologies employed in both active transmission and backscatter communication. For backscattering, monostatic, bistatic, and ambient configurations exist, whereas active transmission typically adopts a star topology, functionally analogous to the monostatic setup in backscatter systems.

\subsection{Active Transmission (\acrshort{lpwan})}
The range and energy consumption of active transmissions in \glspl{lpwan} are fundamentally dictated by their underlying topology and architectural design~\cite{callebaut2021art}. \Glspl{lpwan} commonly implement a star topology, in which end devices communicate directly with a central gateway or base station over extended distances. This avoids energy-intensive multi-hop relaying~\cite{leenders2023energy} but shifts the full burden of long-range transmission to the energy-constrained end node. Given that communication is the dominant energy sink in such nodes, \cite{callebaut2021art} this architectural choice directly affects battery longevity. The base stations are often always in receive mode, allowing end devices to transmit packets in a \textit{send-and-forget} fashion, reducing the energy consumption otherwise required for synchronization and multiple access schemes.

\begin{figure*}[!t]
    \centering
    \hfill
    \begin{subfigure}[t]
        {0.3\linewidth}
        \centering
        \includegraphics[width=\textwidth]{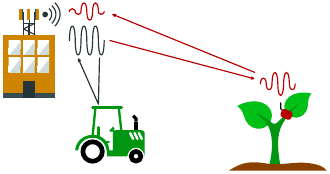}
        \caption{In a monostatic setup the carrier gets reflected by nearby objects resulting in the large reflected signal drowning out the backscattered signal.}
        \label{fig:monostatic-topo}
    \end{subfigure}\hfill%
    \begin{subfigure}[t]
        {0.3\linewidth}
        \centering
        \includegraphics[width=\textwidth]{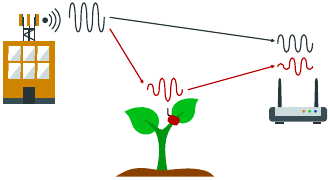}
        \caption{In a bistatic setup the carrier and backscattered signal suffer the same path loss.}
    \end{subfigure}\hfill%
    \begin{subfigure}[t]
        {0.3\linewidth}
        \centering
        \includegraphics[width=\textwidth]{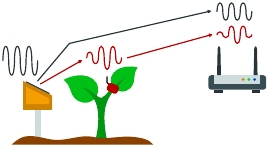}
        \caption{A tag utilizing ambient ongoing communication for backscattering.}
    \end{subfigure}
    \caption{The three main backscatter topologies.}
    \hfill \label{fig:topologies}
\end{figure*}

\subsection{Backscatter Principle}
A backscatter device reflects incoming \gls{rf} signals to carry information. The tag modulates data onto these signals by changing how much of the incoming signal it reflects back. This is determined by the reflection coefficient $\Gamma$. A tag changes its reflection coefficient by adjusting the load impedance connected to the antenna, as depicted in \cref{fig:backscatter-modulation}. The matching between the antenna and load impedance enables the tag to fully absorb ($|\Gamma| = 0$), fully reflect ($|\Gamma| = 1$) or partially reflect \cref{eq:reflection-coeff} the incoming signal. In \cref{eq:reflection-coeff}, $Z_{L}$ is the load impedance connected to the antenna and $Z_{A}$ is the antenna impedance.
\begin{equation}
    |\Gamma| = \frac{|Z_{L}- Z_{A}|}{|Z_{L}+ Z_{A}|}\label{eq:reflection-coeff}
\end{equation}
At its core a backscatter device toggles or switches between loads using an \gls{rf} switch. The resulting impedance mismatch attenuates the incoming signal by a factor $|\Gamma|$ and introduces a phase shift $\theta$.
\begin{align}
    A_{out}= |\Gamma| A_{in}                                        \\
    \theta_{out}= \theta_{in}+ \theta \label{eq:backscatter-effect}
\end{align}
The way the RF switch is controlled determines what modulation is performed (\gls{ook}, \gls{fsk}, \gls{psk},\dots) (\cref{sec:modulation}).

The load can consist of purely resistive elements, reactive elements or a combination of both. In case only resistive elements are used only the real part of the reflection coefficient gets altered resulting in purely amplitude changes of the backscattered signal. Only using reactive elements changes the imaginary part resulting in a phase change of the backscattered signal. Combining both load types makes both amplitude and phase changes possible.

\subsection{Ambient Backscatter}\label{sec:topology-ambient} 
Classic backscatter systems such as \gls{rfid} deploy a dedicated \gls{rf} source as the transmitter. These transmit a continuous carrier wave, on which the tags backscatter. However, constantly transmitting a single tone is power hungry. The transmitter cannot perform other communication tasks, such as communicating with active nodes, during these intervals. Ambient backscatter solves these problems by re-using commodity signals, such as WiFi, \gls{lte}, \gls{ble} and \gls{lora} which are already present. These signals either come from a base station or a device performing active transmissions in the field.

Ambient backscattering can also utilize both the monostatic and bistatic topology. Which topology presents itself as the most optimal depends on the application. An active node is generally closer to a backscatter tag than a base station. This favours the use of bistatic backscattering in ambient setups to maximize communication range. Active nodes transmit at lower power levels than a base station, making it crucial that the device is close to the tag. One active device can support multiple backscatter tags at close proximity. This lends itself to applications deploying active devices together with a multiple of tags backscattering on the uplink transmission. The distance between the tag and the active device determines the achievable backscatter range.

\begin{figure}
    \centering
    \includegraphics[width=\linewidth]{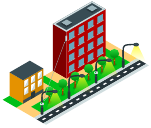}
    \caption{Active node relaying backscatter communication to the gateway.}
    \label{fig:backscatter-relay}
\end{figure}

Active devices can act as a relay point for backscatter tags when a base station is too far away (\cref{fig:backscatter-relay}). The tags communicate their data towards an active device relaying it to the base station much further away. They backscatter their data using either the uplink or downlink communication. This approach increases complexity and energy consumption at the active nodes end. A device requires the capability to receive and decode backscattered data on top of the active data. If the backscattering takes place during the uplink transmission a second transmission is required to relay the tags data. This becomes a major disadvantage in uplink centred IoT networks.

In order to backscatter on active communication, a tag has to be aware of it. Demodulation of and synchronization with ongoing transmissions increases a tag's complexity and its power consumption. The power consumption at the transmitter side does not increase, as no additional transmissions are required.

Backscattering on ambient signals is only suitable in medium to high traffic regions. Sufficient active transmissions are crucial for a tag to transmit its data.

As the backscatter communication relies on ambient transmissions, these tags cannot transmit at fixed intervals. The application must lend itself to intermittent traffic at unpredictable intervals.

The existing communication also poses a challenge when it comes to interference. When a tag shifts the incoming signal in frequency, the resulting signal might end up in another communication channel. Ongoing communication degrades the \gls{snr} and can block the backscatter communication entirely.

\subsection{Monostatic and Bistatic Backscattering}
In backscatter communication, a distinction is made between monostatic, bistatic, and ambient topologies. Monostatic and bistatic configurations rely on a dedicated transmitter to emit a continuous carrier signal, which is then modulated by the backscatter device. In contrast, ambient backscattering (\cref{sec:topology-ambient}) leverages existing ambient \gls{rf} transmissions, originating from other devices or infrastructure, as the carrier, enabling passive modulation without requiring a dedicated source.

As depicted in \cref{fig:topologies}, in monostatic backscattering the \gls{rf} source and receiver are co-located while in a bistatic setup the transmitter and receiver are placed in different locations.

Popular backscatter systems, such as \gls{rfid}, use a monostatic topology. A single transceiver generates the carrier and receives the backscattered signal. This simple setup holds a clear advantage in simplifying practical installation. The signal travels the same distance twice from transmitter to the tag and then returns to the receiver. The resulting propagation losses limit the possible communication range. A single transmission/reception point causes the unmodulated carrier to leak into the receiver due to limited isolation or reflections (\cref{fig:monostatic-topo}). The local carrier drowns out the return signal because it is significantly larger. This large in-band interference requires special suppression techniques to limit the resulting range reduction~\cite{brauner2009novel}.

Bistatic topologies reduce propagation losses and in-band interference, extending the communication range. The backscatter tag is placed close to the \gls{rf} source, allowing the receiver to be placed much further away (or vice versa). In this case, the unmodulated carrier and backscattered signal experience similar propagation conditions, improving the \gls{snr} with respect to monostatic systems. This alleviates the need for complex carrier suppression techniques, but the transmitter could still potentially saturate the receiver~\cite{10155565}.
However, a monostatic topology reduces deployment complexity, as a single transceiver base station can be used. Making these systems long-range is challenging due to the aforementioned problems. Bistatic systems solve these problems but are bound to a short transmitter-to-tag distance. These short distances complicate deployment, as not every application lends itself to placing tags close to transmitters.

  \section{Backscatter Hardware}\label{sec:hardware}
Different hardware approaches exist to generate backscatter signals. While some achieve better spectral efficiency others achieve a lower \gls{bom} cost. This section compares all the different techniques and explores their advantages.

\subsection{Fixed-load Modulation}\label{sec:fixed-load-modulation}
The simplest backscatter hardware consists of a single \gls{rf} switch alternating between two impedance states \cite{menon2023wisp}. This can be either a short and open, an open and a matched load or anything in between. The former case produces a constant envelope backscatter signal but with a \SI{180}{\degree} phase shift at the state transitions ($\Gamma=+1/-1$). This lends itself to produce frequency/phase shifting modulation schemes (\gls{bpsk}, \gls{css},\ldots). Because the signal remains constant envelope the reflected power is higher resulting in a higher \gls{snr} at the receiver \cite{karthaus2003rfid}. The reflection coefficient for this scenario can be Fourier expanded resulting in \cref{eq:constant-envelope-backscatter} \cite{ding2020harmonic}. The incoming wave is a constant carrier signal with frequency $f_0$.
\begin{align}
    \Gamma_{bs}(t) &= \frac{4}{\pi} \sum_{n=1,3,5}^{\infty} \frac{1}{n} \sin(2 n \pi f_{bs} t) \label{eq:constant-envelope-backscatter} \\
    S_{bs} &= \cos(2 \pi f_0 t) \cdot \Gamma_{bs}(t) \\
    S_{bs} &= \frac{4}{\pi} \sum_{n=1,3,5}^{\infty} \frac{1}{n} \cos(2 \pi f_0 t) \cdot \sin(2 n \pi f_{bs} t) \\
    S_{bs} &= \frac{2}{\pi} \sum_{n=1,3,5}^{\infty} \sin(2 \pi (f_0 + n f_{bs})t) - \sin(2 \pi (f_0 - n f_{bs})t) \label{eq:constant-envelope-backscatter-result}
\end{align}
The resulting backscatter signal (\cref{eq:constant-envelope-backscatter-result}) shows that this method gets hampered by low spectral efficiency as it produces odd harmonics ($f_0 + nf_{bs} (n=3,5,\ldots)$ and a mirror image ($f_{0} - nf_{bs} (n=3,5,\ldots)$. In existing communication bands there is a high chance of interfering with adjacent communication channels.
\begin{figure}[!t]
    \centering
    \includegraphics[width=\linewidth]{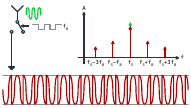}
    \caption{Square wave modulation in digital solutions generates significant harmonics.}
    \label{fig:single-load-backscatter}
\end{figure}

The same applies for the second scenario switching between full reflection and absorption ($\Gamma=+1/0$) or switching between two real impedances in between open and matched and matched and open (\cref{fig:single-load-backscatter}). The backscattered signal now supports amplitude based modulation schemes (\gls{ask}, \gls{ook}, \ldots).

Adding additional loads allows to construct a circle with constant \gls{vswr} on a smith chart. Following these points makes the phase transitions of the backscattered signal more fluent and reduces produced harmonics. In the ideal scenario the hardware is capable of synthesizing every reflection coefficient on a constant \gls{vswr} circle with magnitude 1. In that case no harmonics are produced and the backscattered signal power is maximized. In practice a set of discrete points is constructed on a constant \gls{vswr} circle through a set of fixed loads. These are connected to the antenna using a multi-throw switch network. Using 8 impedances allows to cancel the mirror image and third and fifth harmonics \cite{talla2017lora}. Adding additional intermediate levels reduces even higher harmonics \cite{li2020xorlora}.
\begin{figure}[!t]
    \centering
    \includegraphics[width=\linewidth]{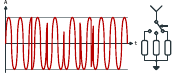}
    \caption{Principle schematic of a modulator using multiple backscatter loads to minimize the produced harmonics.}
    \label{fig:multi-load-backscatter}
\end{figure}

While this approach increases the spectral efficiency it comes at a greater complexity. Solutions built using \gls{cots} components suffer also from a cost and area increase due to the significant increase in \gls{bom} count.

\subsection{IQ Modulator}\label{sec:iq-modulator}
Transistors are capable of producing real-valued reflection coefficients over a certain range. This reduces tag complexity and increases modulation flexibility. Combining two transistors in an IQ-modulator allows to create a set of complex reflection coefficients on the smith chart \cite{guo2022rf,belo2019iq,correia2019chirp}. This supports any modulation type ranging from \gls{bpsk} to \gls{qam}. In practice the supported modulation types are limited by the tags hardware and the I/Q-imbalance which is quite severe for low-complexity hardware.

While IQ-modulators form the basis for modulation in many transmissions systems, adopting them for low-complexity backscatter tags forms a challenging task. Two \glspl{dac} control the transistor gate voltages in discrete steps. To achieve quadrature operation, the incoming carrier signal is split through a Wilkinson power splitter and shifted by \SI{45}{\degree} through a delay line. The delayed signal and the carrier signal are fed to the two transistors. In the delayed line the incident wave gets reflected and experiences a second \SI{45}{\degree} delay causing the signals to be in quadrature when they get combined again. This emulates the local oscillator of a standard IQ-modulator. The transistor gate control voltage is adjusted to achieve an in-phase and quadrature version of the baseband signal completing the IQ-modulator.

\begin{figure}
    \centering
    \includegraphics[width=0.5\linewidth]{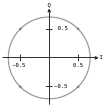}
    \caption{Two 1-bit \gls{rf} switches transition between fully reflecting and absorbing ($\Gamma=+1/-1$) creating four points on the IQ-plane.}
    \label{fig:digital-iq}
\end{figure}

Increasing the resolution of a \gls{dac} comes with a significant cost making it less attractive for low-cost tags. Controlling the transistor gate bias with a \gls{pwm} signal omits the \glspl{dac} completely \cite{zhang2016hitchhike,zhu2025inductor,liNovelLoadFreeSSB2025}. This creates four points on the IQ-plane as both switches transition between two impedance values, $Z_{L}= \inf$ and $Z_{L}= 0$ (\cref{fig:digital-iq}). In order to create additional \emph{virtual} impedance states the transistor gate control needs to be oversampled. The \gls{pwm} signal switches the transistors faster than the sampling clock of the receiver making it imperceivable. The tag adjusts the amplitude of the backscatter signal by changing the \gls{pwm} duty-cycle. At the receiver the discrete steps get filtered by the input bandpass filter creating a signal with smooth amplitude transitions. The phase gets manipulated by simultaneously adjusting the in-phase and quadrature amplitude.
The same effect can be achieved using a \gls{spdt} switch network with four complex impedances ($1+j, 1-j, -1+j, -1-j$). Switching at a multiple of the desired frequency shift achieves the same operation as the above circuit. The major downside of this approach is the increased \gls{bom} and increased complexity for monolithic integration.

\subsection{Load Control}
Solutions making use of fixed loads for modulation (\cref{sec:fixed-load-modulation}) contain one or more \gls{rf} switches which need to be controlled. An analogue solution makes use of a \gls{vco} to control the switches. The \gls{vco} changes its output frequency based on a \glspl{dac} output voltage \cite{talla2017lora} which is controlled by a \gls{mcu}. The load configuration determines which modulation technique is used while the \gls{vco} frequency determines the switching frequency.

\begin{figure}[!t]
    \centering
    \includegraphics[width=\linewidth]{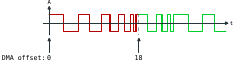}
    \caption{\gls{dma} outputs a waveform directly onto a \gls{gpio} to realize digital backscattering.}
    \label{fig:dma-backscatter}
\end{figure}

Implementing this circuitry fully analogue decreases temperature and supply voltage stability. For this reason \gls{dds} supersedes these circuits for signal generation offering a more flexible and stable solution. In backscatter circuits the digital output of the \gls{dds} circuit controls the load switches directly. \Gls{dds} circuitry is available in standard \glspl{ic} or can be implemented into an \gls{fpga} \cite{katanbaf2021simplifying} or \gls{mcu} \cite{tang2021self,tangPrototypeImplementationExperimental2025}. An \gls{mcu} utilizes its \gls{dma} hardware to do this in a low-power manner. Before transmission, the \gls{mcu} performs all necessary steps (scrambling, error correction coding, Gray coding,\dots) to convert the data into symbols. The image stored in flash memory is then loaded into \gls{ram}. Once this preparation is done the \gls{dma} controller outputs the waveform by copying the \gls{ram} contents to a \gls{gpio}. This output pin then controls the \gls{rf} switch to modulate the data. Changing the \gls{dma} source address allows to transmit a different symbol (\cref{fig:dma-backscatter}). This approach offers the lowest hardware complexity presenting a good solution for low-power tags.

\subsection{Tunnel Diodes}
Backscattered signals experience significant path loss resulting in hard to recover weak signals at the receiver. Reflection amplifiers overcome this limitation by amplifying the backscattered signal. Transistor based designs \cite{lazaro2025home} require significant energy ($~\SI{580}{\micro\watt}$). Amplifiers based on a tunnel diode \cite{amato2015long,varshneyTunnelScatterLowPower2019} maintain a much low power consumption (\SIrange{30}{150}{\micro\watt}) while still achieving significant gains (\SIrange{10}{40}{\dB}).

Some designs leverage the tunnel diode also as an oscillator \cite{varshneyTunnelEmitterTunnel2020}. This creates a a low-power oscillator capable of transmitting a carrier wave in case one is absent while also amplifying backscattered signals. This enables tags to transmit a carrier signal on their own \cite{varshneyTunnelScatterLowPower2019} or have a nearby node transmit a carrier signal while amplifying its backscattered signal \cite{varshneyTunnelEmitterTunnel2020}. Adjusting the tunnel diode oscillator's frequency enables cross-band operation on the tag \cite{guoEnablingCrossBandBackscatter2025}.

\subsection{Carrier Cancellation}
To retain proper receiver sensitivity in a monostatic setup carrier cancellation is mandatory. The transmitter remains active while the tag backscatters onto the carrier causing the strong carrier to be directly received again at the receive port. A directional coupler or circulator connected to the gateways antenna isolates both transmitting and receiving ports. However, as their isolation capabilities are finite a carrier cancellation circuit brings additional improvement.
\begin{figure}
    \centering
    \includegraphics[width=\linewidth]{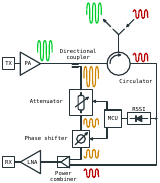}
    \caption{Principle schematic of a conventional carrier cancellation circuit \cite{zhonghua2019carrier} creating a \SI{180}{\degree} shifted copy of the incoming carrier at the receive port in order to cancel it.}
    \label{fig:carrier-cancellation}
\end{figure}

A conventional cancellation circuit uses a coupler to extract an attenuated and phase shifted version of the transmitted carrier \cite{zhonghua2019carrier} (\cref{fig:carrier-cancellation}). This signal copy gets added to the signal coming from the circulator in an attempt to fully cancel the remaining incoming carrier. The attenuator and phase shifter adjust the added signals amplitude and phase based on the \gls{rssi} value measured at the circulator output. Adding a \SI{180}{\degree} shifted copy with the same amplitude completely cancels the remaining carrier signal. In practice a full cancellation cannot be achieved and many other methods exist to further improve the cancellation operation.

  \section{Active Long-Range Modulation}\label{sec:active-modulation}
\begin{figure}
    \centering
    \begin{subfigure}{\linewidth}
        \centering
        \includegraphics[width=\textwidth]{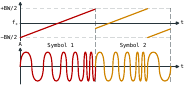}
        \caption{\gls{css} modulation uses linear chirps to encode data. The chirp starting frequency determines the used symbol.}
        \label{fig:css-modulation}
    \end{subfigure}
    
    \begin{subfigure}{\linewidth}
        \centering
        \includegraphics[width=\textwidth]{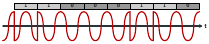}
        \caption{\Acrlong{dbpsk} adds \SI{90}{\degree} to the signals phase when transmitting a \texttt{1} and \SI{0}{\degree} when transmitting a \texttt{0}.}
        \label{fig:dbpsk-modulation}
    \end{subfigure}
    \begin{subfigure}{\linewidth}
        \centering
        \includegraphics[width=\textwidth]{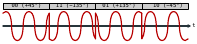}
        \caption{\Acrlong{qpsk} modulation uses 4 fixed phase shifts for its four symbols each carrying two bits.}
        \label{fig:qpsk-modulation}
    \end{subfigure}
    \caption{Different modulation techniques used in active long-range communication.}\label{fig:long-range-modulation}
\end{figure}
\Gls{lpwan} technologies are purpose-built to enable low-power, long-range messaging with minimal signalling overhead--ideally suited for sporadic transmissions of small payloads. At the \gls{phy} layer, these systems utilize highly robust modulation schemes that favour link budget and sensitivity over spectral efficiency~\cite{callebaut2021art}. Examples include \gls{css} (\cref{fig:css-modulation}) in \gls{lorawan}, ultra-narrowband \gls{dbpsk} (\cref{fig:dbpsk-modulation}) in Sigfox, and single-tone \gls{bpsk}/\gls{qpsk} (\cref{fig:qpsk-modulation}) in \gls{nbiot}. These modulation formats allow reception at extremely low \gls{snr}, often below \SI{0}{dB}, which is critical for extending coverage while using constrained transmission power. Importantly, many of these schemes--such as constant-envelope signalling--enable efficient power amplifier operation, a key factor in minimizing energy drain during transmission.

\section{Backscatter Modulation}\label{sec:modulation}
For long-range backscatter two main categories exist for the modulation techniques. The first consists of the binary switch-based modulations which are known from \gls{rfid}. These switch a binary number of fixed loads (\cref{sec:fixed-load-modulation}) to achieve modulation types such as \gls{ook} and \gls{fsk}. The second group consists of all modulation techniques making use of \gls{css} in any form. Within this category there are three main groups. A first group consists of tags generating chirp signals utilizing an incoming fixed carrier. The second group alter incoming chirps (either fixed or ambient) by applying \gls{ook} based modulation onto them to achieve long-range communication. A last group does the same but applies \gls{fsk} based modulation. Each category has its own specific trade-offs in complexity, spectral efficiency and compatibility with commodity receivers. \Cref{tab:modulation_comparison} gives an overview of the different modulation techniques with their main characteristics.

\begin{table*}[ht]
    \centering
    \renewcommand{\arraystretch}{1.2}
    \caption{Comparison of Modulation Techniques in Long-Range Backscatter Systems.}
    \label{tab:modulation_comparison}
    \begin{NiceTabular}{@{}lcccccclll@{}}
        \toprule
        \textbf{Modulation} & \textbf{Data Rate} & \textbf{Range} & \textbf{Power} & \thead{Spectral\\Eff.} & \thead{Ambient\\Carrier} & \thead{Commodity\\RX} & \textbf{Use Case} & \textbf{Section} & \textbf{Fig.} \\
        \midrule
        \multicolumn{10}{c}{TX carrier: Binary-based} \\
        \midrule
        OOK                 & \medium    & \lowhigh    & \low       & \emptycirc    & \emptycirc    & \fullcirc     & RFID, basic sensing                       & \cref{sec:bb-ook}             & \cref{fig:ook-modulation} \\
        FSK                 & \medium    & \high       & \low       & \emptycirc    & \emptycirc    & \emptycirc    & Environmental monitoring                  & \cref{sec:bb-fsk}             & \cref{fig:freq-modulation} \\
        \midrule
        \multicolumn{10}{c}{TX carrier: Chirp-based} \\
        \midrule
        CSS                 & \lowhigh   & \high       & \low       & \emptycirc    & \emptycirc    & \fullcirc     & Environmental monitoring                  & \cref{sec:css-backscatter}    & \cref{fig:css-modulation} \\
        OOK                 & \high      & \medium     & \lowhigh   & \emptycirc    & \emptycirc    & \emptycirc    & Sound monitoring, smart shelves           & \cref{sec:cb-ook}             & \cref{fig:harmonic-backscatter} \\        
        \enspace Ambient    & \medium    & \medium     & \high      & \fullcirc     & \fullcirc     & \emptycirc    & Livestock monitoring, flood sensors       & \cref{sec:cb-ambient}         & \cref{fig:aloba} \\
        \enspace CIM        & \lowhigh   & \high       &            & \emptycirc    & \fullcirc     & \emptycirc    & Warehouse monitoring                      & \cref{sec:cb-cim}             & \cref{fig:cim-modulation} \\
        FSK                 & \lowhigh   & \high       & \medium    & \emptycirc    & \fullcirc     & \emptycirc    & Logistics                                 & \cref{sec:cb-fsk}             & \cref{fig:p2lora-demodulation} \\
        \enspace CWT        & \lowhigh   & \medium     &            & \emptycirc    & \fullcirc     & \fullcirc     & Agriculture, implantable devices          & \cref{sec:cb-cwt}             & \cref{fig:codeword-translation} \\
        \enspace Offset CSS &            & \lowhigh    & \low       & \fullcirc     & \fullcirc     & \fullcirc     & Identification, home security             & \cref{sec:offset-css}         & \cref{fig:fsk-tag-id} \\
        \bottomrule
    \end{NiceTabular}
\end{table*}

Some modulation schemes incorporate a frequency shift to modulate their data. This pushes the backscattered signal out of band avoiding interference with the large carrier signal. Shifting to an adjacent channel increases the \gls{snr} at the receiver and as a consequence the communication range at the cost of additional required bandwidth. Other modulation techniques by default operate in the same band as the carrier signal. In these schemes the shift can be integrated through and additional switching operation. For example, in an \gls{ook} scheme this comes down to backscattering a frequency shifted waveform during the \texttt{on} state by toggling the load.

\subsection{Binary Switch-Based Modulation}
Binary switch-based modulation includes simple schemes such as \gls{ook} and \gls{fsk}. These techniques toggle between distinct impedance states to encode data.
\begin{figure}[!t]
    \centering
    \begin{subfigure}
        {\linewidth}
        \centering
        \includegraphics[width=\textwidth]{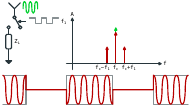}
        \caption{\gls{ook} either reflects or absorbs the incoming carrier generating two sidebands shifted by the on-off switching frequency.}
        \label{fig:ook-modulation}
        \vspace{5pt}
    \end{subfigure}
    \begin{subfigure}
        {\linewidth}
        \centering
        \includegraphics[width=\textwidth]{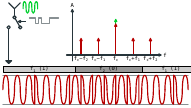}
        \caption{\gls{fsk} changes the switching frequency to encode data. The principle schematic shows 2-\gls{fsk} switching between an open and short load ($\Gamma=+1/0$). This creates two sidebands at both switching frequencies.}
        \label{fig:freq-modulation}
    \end{subfigure}
    \caption{Different modulation techniques using a square wave and a fixed load.}
    \label{fig:backscatter-modulation}
\end{figure}

\subsubsection{On-Off Keying}\label{sec:bb-ook}
\label{sec:ook} In backscattering, \gls{ook} translates into fully reflecting or absorbing the incoming signal. This results in transmitting a \texttt{1} (\texttt{on}) or \texttt{0} (\texttt{off}) respectively. The receiver needs to distinguish between only receiving the carrier or receiving both the carrier and a reflected signal. Classic RFID systems use a fixed carrier at around \SI{900}{\mega\hertz}~\cite{bharadia2015backfi} which they cancel out. The modulated signal remains and can be demodulated through standard practices.

This is not straightforward in ambient systems as the carrier signal is now a modulated signal (opposed to a continuous signal) and the carrier waveform is unknown to the receiver, complicating the demodulation of the ambient backscattered signal. Ambient systems resort to other demodulation techniques.

\subsubsection{Frequency Shift Keying}\label{sec:bb-fsk}
In binary switch-based modulations \gls{fsk} supersedes \gls{ook} in popularity due to its better robustness against fading and noise \cite{wang2022allspark}. This results in better \gls{ber} for the same \gls{snr} yielding a larger communication distance.

A tag generates frequency based modulation by toggling between reflection and absorption at a certain frequency. This superimposes a signal on top of the incoming carrier (see~\cref{fig:freq-modulation}). Generating two different frequencies achieves 2-\gls{fsk} modulation where each frequency represents a single bit (either \texttt{1} or \texttt{0}). Increasing the amount of used frequencies allows to transmit more bits per symbol and increases the data rate. Transmitting $2^{n}$ bits in one symbol requires n frequencies.

\begin{figure}
    \centering
    \includegraphics[width=0.6\linewidth]{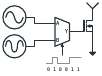}
    \caption{Principle schematic how 2-\gls{fsk} is generated on a low-power tag. The load switch is switched at a fixed frequency which depends on the data being transmitted.}
    \label{fig:fsk-modulation}
\end{figure}

2-\Gls{fsk} modulation has proven to be a viable solution for long-range backscattering \cite{varshney2017lorea,wang2022allspark}. A transmitter sends out a fixed carrier on which the tags backscatter. Two oscillators generate a square wave at a different frequency. A multiplexer decides which oscillator switches the antenna load on and off depending on the bit to be transmitted (\cref{fig:fsk-modulation}).

\subsection{Chirp Spread Spectrum}\label{sec:css-backscatter}
\Gls{css} forms a good basis for long-range backscattering as it is known for its robustness against interference and can be decoded below the noise floor. Generating chirps on a tag requires generating a linearly increasing frequency signal. It shifts the chirp in the time domain to generate different symbols. The tag is able to backscatter standard \gls{lora} packets when a fixed carrier is available \cite{talla2017lora,ding2020harmonic,zhu2025inductor,liNovelLoadFreeSSB2025}.

Instead of generating a fixed carrier ambient \gls{fm} radio broadcasts can be re-used \cite{daskalakis2022new}. These signals are broadcasted all over the world and allow the re-use of existing infrastructure. Many commodity devices incorporate an FM-demodulator opening the possibility for widespread demodulation. This combines the benefits of long-range communication with widespread usage. The tag backscatters chirps directly onto the incoming radio signals specifically onto the $\mathrm{L} - \mathrm{R}$ stereo audio channel. Stereo receivers decode both audio channels making the backscattered data available by subtracting the left and right audio channel.

\subsection{Chirp-Based OOK}\label{sec:cb-ook}
Chirp-based modulation techniques utilize chirps as the base waveform and either encode the data in the chirps or apply another form of modulation on top to encode data. In the latter case the chirps function as the carrier waveform which can come from ambient ongoing communication or a gateway sending dedicated chirps for backscattering. This method retains the advantage of robustness and long-range capabilities of the chirps and combines it with other benefits of the second modulation. As either a gateway or active node generates the chirps easier modulation techniques such as \gls{ook} can be used, simplifying tag design.

\begin{figure}
    \centering
    \includegraphics[width=\linewidth]{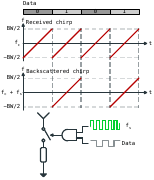}
    \caption{Chirp-based \gls{ook} relies on the gateway for generating the linear chirps. The tag either shifts the incoming chirp when transmitting a \texttt{1} or does nothing when transmitting a \texttt{0}.}
    \label{fig:harmonic-backscatter}
\end{figure}

Harmonic backscatter \cite{huang2021freeback,lin2024harmonic} implements chirp-based \gls{ook} with a transmitter generating fixed linear chirp signals on which the tags backscatter. The tags generate a fixed frequency signal which switches the antenna load on and off. The switch actuates while transmitting a 1-bit and gets disabled when transmitting a 0. This results in the incoming chirp signal being shifted to a an adjacent channel when transmitting a \texttt{1} (\texttt{on}) and no signal being present in the channel while transmitting a \texttt{0} (\texttt{off}) (\cref{fig:harmonic-backscatter}). The receiver listens on the adjacent channel to demodulate the incoming data. Depending on the demodulation, tags modulate either one \cite{lin2024harmonic} or multiple bits \cite{huang2021freeback} per chirp.

\subsubsection{Ambient Chirps}\label{sec:cb-ambient}
Instead of using fixed linear chirps as carrier, ambient \gls{lora} chirps can be used. This reduces the gateways energy consumption as no dedicated transmissions are required for backscatter. Deployments require a nearby active \gls{lora} node with active uplink transmission. Demodulation now relies on interference between the backscattered and active transmission at the receiver. The \gls{ook} modulation translates to the tag fully reflecting (\texttt{on}) or absorbing (\texttt{off}) the incoming symbol. The receiver captures the superposition of the reflected symbol and the ambient \gls{lora} signal (in this case the carrier). In the two extreme cases, which practically almost impossible to occur, both signals are either received phase-aligned or with opposite phases. In the fromer case, constructive interference induces an amplitude increase due to the backscattered \texttt{on} symbol. The latter extreme, with opposite phases (\SI{180}{degree} out of phase) results in destructive interference reducing the received signal amplitude. Any other phase alignment results in phase and amplitude variations. When the incoming symbol is absorbed by the tag to transmit a 0 (\texttt{off}) there is no amplitude or phase variation caused by the backscattered signal.

\begin{figure}
    \centering
    \includegraphics[width=\linewidth]{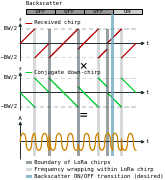}
    \caption{Decoding \gls{ook} modulated backscatter chirps using ambient communication. The backscattered chirps are combined with the ambient transmitted chirps in the same channel. After conjugate down-chirp multiplication a sinusoidal tone with phase jumps remains, caused by frequency wrapping in the chirp or the transition from starting or stopping backscatter.}
    \label{fig:aloba}
\end{figure}

In order to demodulate the backscattered symbols the signal at the receiver is multiplied by its conjugate \cite{guo2020aloba} (\cref{fig:aloba}). This results in a constant sine wave (per \gls{lora} symbol) containing phase jumps. The transitions between \gls{lora} symbols introduce these jumps as well as the phase wrapping of a chirp within a \gls{lora} symbol (indicated in \cref{fig:aloba} as gray vertical lines). Eliminating these phase jumps \cite{guo2020aloba} allows to keep only the jumps created by the backscattering symbols, i.e., backscattered \texttt{on}. Comparing the phase of the constant sine wave with a fixed carrier, transitioning between a backscattered \texttt{on} and \texttt{off} induce a different phase jump, allowing the symbols to be extracted by clustering techniques. As each tag induces a different phase shift based on e.g., its location they will induce other clusters allowing multiple tags to be multiplexed.

\subsubsection{Chirp-Interval Modulation}\label{sec:cb-cim}
Standard chirp-based \gls{ook} encodes its data in the presence or absence of a backscattered chirp. Another option is to encode the data in the time between two backscattered chirps \cite{peng2022ambient}. Defining $n$ possible time delays allows to transmit $2^{n}$ bits during the \texttt{off} symbol. This significantly increases achievable data rates while maintaining good long-range capabilities. Throughput gains of \SIrange{4.7}{8.6}{\times} are possible when compared to previous solutions
\cite{peng2018plora}.
\begin{figure}[!t]
    \centering
    \includegraphics[width=\linewidth]{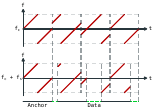}
    \caption{Chirp-interval modulation backscatters ambient chirps in an adjacent channel. The data is not encoded in the presence/absence of a chirp, but in the time between backscattered chirps.}
    \label{fig:cim-modulation}
\end{figure}

The \texttt{on} symbol contains no information and functions as an anchor chirp. They are fixed in length and stem from ambient chirps which are shifted in frequency (\cref{fig:cim-modulation}). This frequency shift avoids interference with the incoming ambient signal and simplifies demodulation if $f_{s}> BW / 2$. A fixed oscillator generates the shifting waveform omitting the need for a \gls{vco} and simplifying the hardware. Keeping the time between anchor chirps short allows to increase the achievable data rate.
\begin{figure}[!t]
    \centering
    \includegraphics[width=\linewidth]{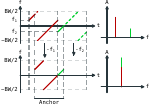}
    \caption{Misalignment between the start of the chirp and the shifting operation causes the frequency content to be spread out. Twin chirp cancellation solves this problem by shifting the chirps based on the original chirps starting frequency.}
    \label{fig:twin-chirp-cancellation}
\end{figure}

Shifted chirps are misaligned with the ambient chirps (\cref{fig:cim-modulation}). This introduces frequency discontinuities in the backscattered signal. These split the transmitted energy in multiple frequencies lowering the transmission range. Twin-chirp cancellation~\cite{peng2022ambient} counters this issue. The receiver uses the chirps coming from the ambient source to shift the backscattered chirps. If the ambient chirp has a starting frequency of $f$ the anchor chirp is shifted by $-f$. This shift concentrates all the chirp energy in one single \gls{fft} bin (\cref{fig:twin-chirp-cancellation}).

As a downside standard \gls{lora} receivers can no longer be used to decode the backscatter information. A new receiver algorithm utilizes a sliding window to locate anchor symbols and measures the time between them to decode the data.

\subsection{Chirp-Based FSK}\label{sec:cb-fsk}
Chirp-based \gls{fsk} relies on shifting incoming chirps at different frequencies. This can be done in different ways shifting the complexity more towards the tag or the gateway.

\begin{figure}
    \centering
    \includegraphics[width=\linewidth]{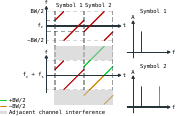}
    \caption{Blind chirp modulation shifts incoming chirps twice and splices them together resulting in a continuous \gls{lora} chirp.}
    \label{fig:blind-chirp-modulation}
\end{figure}
A first approach, called blind chirp modulation, utilizes two shifting frequencies to modulate incoming ambient chirps~\cite{peng2018plora}. Two RF switches perform the shifting operation. One shifts the incoming chirp by $BW / 2$ while the other shifts it by $-BW / 2$ (\cref{fig:blind-chirp-modulation}). An \gls{fpga} splices the two shifted chirps into a new backscattered chirp. The tag only performs this splicing operation while transmitting a \texttt{1}. When transmitting a \texttt{0} the incoming chirp is left unaltered. An additional frequency shift $f_{shift}$ shifts the backscattered chirps out of the active communication band to improve the \gls{snr} and simplify the demodulation.

For demodulation, the receiver multiplies both the active transmission and the backscattered signal with the same down chirp. An \gls{fft} analysis reveals whether the resulting peaks fall into the same or different bins. A spliced chirp falls in a different bin than the original one resulting in a transmitted \texttt{1} while a shifted chirps falls in the same bin yielding a \texttt{0}. This modulation scheme suffers from low data rates since only one bit can be encoded per chirp.

\begin{figure}
    \centering
    \includegraphics[width=0.8\linewidth]{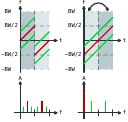}
    \caption{\gls{fsk} modulated backscatter chirps contain frequency continuities. The receiver uses a switch and splice method to create a base up-chirp to concentrate the frequency content in two bins.}
    \label{fig:p2lora-demodulation}
\end{figure}

The complexity of splicing the chirp can be offloaded to the receiver \cite{jiang2021long}. The tag utilizes FSK modulation to shift incoming chirps with a certain frequency. This shifting process generates a lower and upper sideband backscatter signal (\cref{fig:p2lora-demodulation}). When a chirp contains a frequency discontinuity this results in the energy being split into two \gls{fft} peaks after applying a de-chirp. The energy is split again for the backscatter signal due to the generation of an upper and lower sideband signal, resulting in four \gls{fft} peaks.

In order to concentrate the energy into two \gls{fft} peaks the chirps are spliced into a base up-chirp at the receiver. By calculating the starting frequency from the incoming chirp and determining the wrapping point the chirp can be split into two parts. Swapping these two parts and splicing them back together results in a base up-chirp concentrating the \gls{fft} energy into two peaks.

As the shifted chirps interfere with the incoming ones, carrier cancellation is necessary. A reconstruction of the excitation signal in the time domain cancels the carrier from the hybrid signal.

\subsubsection{Codeword Translation}\label{sec:cb-cwt}
Ambient \gls{lora} backscatter systems apply several techniques to generate valid \gls{lora} symbols. This allows the re-use of existing receivers without any modifications. Codeword translation \cite{zhang2017freerider} achieves this goal by translating an incoming symbol into another valid symbol.

Every modulation technique has a codebook with a fixed amount of codewords. A codeword represents a certain symbol on the physical layer and is used to transmit a number of bits. With N the amount of codewords the amount of bits per codeword becomes $\log_{2}(N)$. For example 2-\gls{fsk} utilizes two codewords in which each is a sinusoidal tone with a different frequency on the physical layer. As there are only two symbols every symbol transmits one bit $\log_{2}(2) = 1$. Now applying codeword translation on this example requires changing the frequency of the incoming symbol to represent the other. Applying this only when transmitting a 1-bit and leaving the symbol as is for a 0-bit results in the logic table in \cref{tab:codeword-translation}. Two separate receivers decode the original and manipulated data. From the table it becomes clear that a simple XOR operation between the two data streams yields the data transmitted by the tag. The backscattered signal is shifted out of band to ease the receiver design.
\begin{table}
    \centering
    \caption{Logic table for 2-\gls{fsk} codeword translation.}
    \begin{tabular}{lllll}
        \toprule \multicolumn{2}{c}{\textbf{Symbol}} & \multicolumn{3}{c}{\textbf{Data}} \\
        \textbf{Incoming}                            & \textbf{Backscattered}           & \textbf{Original} & \textbf{Backscattered} & \textbf{Tag} \\
        \midrule $f_{1}$                             & $f_{2}$                          & 1                 & 0                      & 1            \\
        $f_{2}$                                      & $f_{1}$                          & 0                 & 1                      & 1            \\
        $f_{1}$                                      & $f_{1}$                          & 1                 & 1                      & 0            \\
        $f_{2}$                                      & $f_{2}$                          & 0                 & 0                      & 0            \\
        \bottomrule
    \end{tabular}
    \label{tab:codeword-translation}
\end{table}

\begin{figure}
    \centering
    \includegraphics[width=\linewidth]{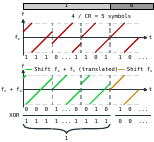}
    \caption{Codeword translation needs to manipulate the right amount of
    symbols to avoid disrupting the output data stream after decoding and de-interleaving.}
    \label{fig:codeword-translation}
\end{figure}

In \gls{lora} the $\text{SF}$ determines the amount of codewords in the codebook. As the spreading factor represents the number of bits transmitted per symbol the amount of symbols becomes $2^\text{SF}$. Applying codeword translation is not straightforward as \gls{lora} encodes its data before transmission. Tags manipulating the data on the physical layer are actually altering the encoded data. This prevents the tag from directly manipulating the data bits at the output of the receiver. The tag instead alters a set of symbols respecting the applied encoding. To decoded the tags data, the receiver detects if a string of bits corresponding to the manipulated symbols has been altered or not. It applies an XOR operation between the original and backscattered data. To transmit a \texttt{1} the tag manipulates the right amount of symbols by shifting them $f_{s}+ f_{t}$ flipping the data bits. After the XOR this yields a string of 1 bits (\cref{fig:codeword-translation}). When transmitting a \texttt{0} the tag shifts the same amount of symbols by $f_{s}$ which results in all 0 bits at the receiver.

To manipulate the right symbols the tag must be aware of how the data is encoded on the physical layer \cite{li2020xorlora}. \gls{lora} adds redundancy to perform \gls{fec} and make the data more robust \cite{joachim2019complete}. It applies a Hamming code and interleaving. The Hamming code adds $\frac{4}{CR}$ additional bits and the interleaving shuffles them. $CR$ represents the coding rate ranging from $\frac{4}{5}$~to~$\frac{4}{8}$. Afterwards the data is interleaved in blocks of $SF \times \frac{4}{CR}$ bits. In order for the manipulated data to decode correctly the tag manipulates this set of bits at once. The amount of symbols corresponding with these bits is determined by the spreading factor $SF$. This results in $\frac{4}{CR}$ symbols being manipulated at once to transmit a single bit. This lowers the data rate by $4 SF$ compared to standard \gls{lora} (\cref{eq:lora-cwt-dr}).
\begin{equation}
    R_{b}= \frac{4\ \text{BW}}{\text{CR}\ 2^{\text{SF}}}\label{eq:lora-cwt-dr}
\end{equation}

In order for the tag to know what symbol period to use, the transmitter sends the modulation parameters upfront. A low-power envelope detector realizes the demodulation on the tag. This detector senses the presence or absence of a \gls{lora} packet to achieve \gls{ook}. A Barker Code at the beginning of the frame triggers the data reception after which the data is decoded.

\subsubsection{Offset CSS}
\label{sec:offset-css} 
\begin{figure}
    \centering
    \includegraphics[width=0.8\linewidth]{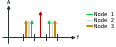}
    \caption{Every tag shifts incoming \gls{lora} packets with a slightly different
    offset into the adjacent channel. The offset is used to identify the tag.}
    \label{fig:fsk-tag-id}
\end{figure}

Previous approaches allow to transmit data through \gls{css} modulated packets. When a tag only requires identification it suffices to make it capable of shifting packets in frequency instead \cite{lazaro2021room} of modulating data. Every tag shifts the incoming packets to an adjacent channel using a unique fixed frequency offset (\cref{fig:fsk-tag-id}). This doesn't require packet detection and other complex techniques resulting in a complete analogue solution. This results in low-latency and low-power communication as there are no digital parts introducing delays and additional processing. Utilizing ambient \gls{css} packets for backscattering leverages their long-range capability.
Making all tags backscatter to the same channel enables a single receiver to continuously monitor it. Standard \gls{lora} transceivers include frequency correction to compensate oscillator drift. Through correlation the receiver synchronizes the incoming signal with the preamble of the packet. The calculated frequency deviation is available in a register on these transceivers. The tags shift the incoming packet into the same channel with slightly different frequency offsets. This offset can then be determined through this register to identify each tag. All tags utilize the same channel minimizing the spectrum usage. As a downside this does not allow tags to transmit concurrently as collisions become indistinguishable.

  \section{Medium Access and Network Coordination (MAC Layer)}\label{sec:mac}
The MAC layer in a communication system provides flow control and multiplexing. It plays an important role in low-power applications as it increases the complexity and consequently the power consumption of a node. A well designed MAC layer must be tailored to its application domain.

\subsection{Active Radio}
At the \gls{mac} layer, \glspl{lpwan} minimize signalling overhead by employing device-initiated, asynchronous uplink access, often based on pure ALOHA or slotted ALOHA~\cite{callebaut2021art}. These protocols omit handshakes, scheduling, or frequent synchronization, which would otherwise increase idle listening and signalling burden. For example, \gls{lorawan} Class~A devices open receive windows only after transmission, and Sigfox nodes operate in a strict uplink-centric model with optional and sparse downlink opportunities. Even cellular \gls{nbiot}, which uses more complex connection-oriented access, supports extreme sleep durations and delay-tolerant operation via \gls{psm} and \gls{edrx} mechanisms. The signalling strategies across these protocols are thus optimized not for throughput or interactivity, but for rare, lightweight messaging at minimal energy cost, which is an essential requirement for autonomous, long-lived IoT deployments.

\subsection{Challenges in Synchronization and Association}
Scaling up ambient backscatter networks requires addressing synchronization challenges, especially with heterogeneous and mobile tags. The near-far problem, channel reciprocity asymmetries, and variable delay propagation complicate association. Distributed \glspl{mac} or AI-based scheduling may help adapt dynamically to topology and interference conditions.

\subsection{Backscatter}\label{sec:concurrency}
The low-complexity hardware of backscatter tags make use of uncoordinated random access protocols like slotted ALOHA \cite{guo2022saiyan,guo2024low}. Tags transmit in predefined slots after detecting energy beacons, reducing collisions while preserving simplicity. Subcarrier diversity and frequency hopping mitigate congestion and exploit the ambient spectrum \cite{guo2022saiyan,guo2020aloba}.

Query-based \glspl{mac} support large amount of concurrent transmissions by enabling tags to transmit after a query beacon~\cite{hessar2019netscatter, jiang2021long}. Symbol allocation is performed by assigning unique cyclic shifts, frequency bins, or time slots to each tag, allowing concurrent decoding at the receiver. This centralized coordination supports scalability but requires tight synchronization.

Assigning a unique cyclic shift per device allows the same amount of devices to transmit concurrently as symbols used in the system $N$ \cite{hessar2019netscatter}. This increases the concurrency at the cost of data rate. A standard \gls{lora} system transmits $log_{2}(N)$ bits per symbol while now every device transmits one bit using its unique symbol.
\begin{figure}[!t]
    \centering
    \begin{subfigure}
        {\linewidth}
        \centering
        \includegraphics[width=\textwidth]{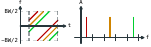}
        \caption{Different \gls{lora} symbol transmitted perfectly aligned with 1 \gls{fft} bin of spacing.}
        \vspace{5pt}
    \end{subfigure}
    \begin{subfigure}
        {\linewidth}
        \centering
        \includegraphics[width=\textwidth]{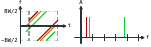}
        \caption{Misalignment causes signals to fall in the same bin and become
        indistinguishable.}
    \end{subfigure}
    \caption{Enable concurrent transmissions by assigning a unique \gls{lora} symbol
    to every tag.}
    \label{fig:netscatter}
\end{figure}
A query message, sent by the transmitter, triggers the tags to all start transmitting simultaneously (\cref{fig:netscatter}). Without timing synchronization incoming symbols overlap making them indistinguishable at the receiver. Some timing mismatches, stemming from hardware delays and propagation delays, remain. Empty \gls{fft} bins account for these. Not using these \gls{fft} bins allows chirps from adjacent bins to drift without wrong decoding as a result.

As devices are distributed in the field, they experience different propagation losses. A tag closer to the transmitter overpowers devices further away. Tags estimate their \gls{snr} at the transmitter using the \gls{rss} of the query message. Additionally, the transmitter assigns further spaced chirps to devices close the transmitter than to tags further away.

Devices require association to enter the network and get a unique cyclic shift assigned. Dedicated cyclic shifts $N_assoc$ are reserved for this process. The gateway uses one slot for the association with high-\gls{snr} and one other slot for low-\gls{snr} devices solving the near-far problem. Conflicts during the association process can be solved with protocols like ALOHA.

This solution relies on custom receiver designs for demodulation. CPLoRa \cite{liCPLoRaParallelLoRa2025} focuses on utilizing standard \gls{lora} modulation and achieving parallel operation by assigning unique frequency offsets per tag. This maintains the present infrastructure with commodity receivers but at the cost of deployment complexity and decreases spectral efficiency.

The downside of this system is the need of a dedicated carrier transmitter. P2LoRa presents a solution utilizing ambient \gls{lora} signals~\cite{jiang2021long}. The tag modulates incoming chirps using \gls{ook} where every tag utilizes different shifting frequencies. After reception the receiver observes different \gls{fft} peaks for the different transmitting tags. Again empty \gls{fft} bins avoid wrong demodulation results due to timing mismatches. As P2LoRa tags don't utilize feedback signals much more empty \gls{fft} bins are required. The different frequency shifts assigned to every tag create sidelobes which impact adjacent \gls{fft} peaks. A Hanning window reduces phase discontinuities lowering inter-tag interference.

Another approach is the use of non-linear chirps \cite{li2022curvinglora}. Generating non-linear chirps on a low-power low-cost tags poses a main challenge \cite{ren2023prism}. In CurvingLoRa, a receiver demodulates the the chirps using a sliding window. A window, with a non-linear down-chirp, slides over the received signal. When the window aligns with a chirp its \gls{fft} peak will be the highest allowing it to be decoded.

\begin{figure}[!t]
    \centering
    \begin{subfigure}
        {\linewidth}
        \centering
        \includegraphics[width=\textwidth]{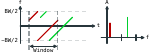}
        \caption{The tag closest to the receiver drowns-out the tags further
        away.}
        \label{fig:curving-lora-interference}
        \vspace{5pt}
    \end{subfigure}
    \begin{subfigure}
        {\linewidth}
        \centering
        \includegraphics[width=\textwidth]{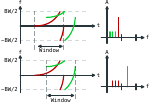}
        \caption{The energy of non-aligned curved chirps spreads out in the
        frequency domain.}
        \label{fig:curving-lora-symbols}
    \end{subfigure}
    \caption{Comparison between using linear and curved \gls{lora} chirps.}
    \label{fig:curving-lora}
\end{figure}
Now when standard chirps collide the chirp closest to the receiver\footnote{Near-far problem.} will generate the largest \gls{fft} peak (\cref{fig:curving-lora-interference}). Even when the sliding window is aligned with the signal of the furthest node, the largest chirp will generate a larger \gls{fft} peak. This causes only the chirp of the closest node to be demodulated and the other chirps to be regarded as noise. No demodulation may even take place if both have the same strength as they drown each other out. With non-linear chirps, the energy of non-aligned chirps spreads out over the \gls{fft} bins (\cref{fig:curving-lora-symbols}). The \gls{fft} peak of the closest node is largely reduced and no longer interferes with the demodulation of nodes further away. This approach allows demodulation of colliding chirps when the transmissions are not aligned. When transmissions are aligned utilizing different types of non-linear chirps allows demodulation and boosts concurrency. However this increases deployment complexity as deployment arrangements need to be known upfront.

\begin{figure}[!t]
    \centering
    \includegraphics[width=\linewidth]{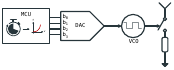}
    \caption{An \gls{mcu} controls a \gls{dac}/\gls{vco} to generate non-linear chirps.}
    \label{fig:prism-diagram}
\end{figure}
Prism \cite{ren2023prism} adapts this approach for ambient backscattering. The tag converts incoming linear chirps to non-linear ones. It utilizes a \gls{mcu}, \gls{dac} and \gls{vco} to generate the chirps (figure~\ref{fig:prism-diagram}). The \gls{mcu} runs a timer to change the frequency over time. Every time the timer interrupt triggers the \gls{dac} output is changed to output the next frequency. The \gls{vco} generates this frequency, based on the \gls{dac} voltage, and controls an RF switch to toggle the antenna impedance.

The modulation scheme of Aloba (\cref{sec:cb-ambient}) \cite{guo2020aloba} inherently supports concurrent transmissions. The signal at the receiver comes the superposition of $M$ backscatter signals and the carrier. After signal transformation and reconstruction $2^{M}$ points appear on the constellation diagram. Parallel decoding algorithms are then able to decode the backscatter signals.

\subsection{Synchronization and Feedback Mechanisms}
Tags require the ability to detect and synchronize with ongoing transmissions in order to use them for backscattering. Their tight energy budgets prevent the use of conventional radio front-ends requiring new low-power techniques. These come with many challenges and each have their own limitations. On top of enabling ambient backscatter, these receivers also allow tags to receive feedback from gateways. Reliable operation in dynamic wireless environments requires some form of bidirectional signalling to acknowledge transmissions, adapt data rates, and coordinate retransmissions. This section explores recent advancements in low-power receivers for backscatter tags and how these allow tags to use ambient transmissions for backscattering and implement feedback mechanisms.

\subsubsection{Ambient Packet Detection}
Ambient backscatter communication requires synchronization with the ongoing communication. The design challenge lies in achieving precise timing synchronization with minimal energy cost. To achieve this packet detection often relies on energy-based techniques like \gls{rssi} monitoring, envelope detection, or correlation against preambles from incumbent protocols. As a downside these techniques trade energy savings for range. Energy hungry receiver components --- \gls{lna}, \gls{fft}, Mixer --- are avoided limiting the tag range to tens of meters.

\begin{figure}
    \centering
    \includegraphics[width=\linewidth]{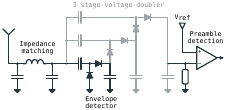}
    \caption{The ambient signal detection circuits used by \cite{peng2018plora}.
    \cite{jiang2021long} adds the 3-stage voltage doubler to increase the range.}
    \label{fig:ambient-signal-detection}
\end{figure}

To synchronize with the intermittent communication tags often resort to cross-correlation \cite{peng2018plora}. An envelope detector first converts the incoming signal to baseband (\cref{fig:ambient-signal-detection}). This signal tracks the \gls{rssi} of the incoming \gls{lora} packet and outputs ten equally spaced pulses while receiving the packets preamble \cite{guo2020aloba}. A low-power \gls{adc} then samples this signal and passes the information to an \gls{fpga} or \gls{mcu}. Correlating this unique pattern to a pre-stored preamble avoids the need for decoding the incoming signal. This reduces the required sampling rate and power consumption. Replacing the \gls{adc} with a voltage comparator allows to reduce \cite{jiang2021long,guo2022saiyan} the power consumption even further.

Adding a three-stage voltage doubler to the envelope detector increases its output signal strength. This doubles the achievable detection range from \SIrange{50}{100}{\meter}.

\subsubsection{Feedback Signals}\label{sec:feedback}
Modern communication systems use feedback signalling to optimize the \gls{qos}. It enables the communicating devices to switch channels, adapt data rate, change modulation strategy, bandwidth or other parameters. The use of feedback signals also enables package acknowledgments. It avoids blind packet transmissions and the associated wasted energy.

To enable this technology, a tag must be able to demodulate incoming signals. Previous solutions \cite{peng2018plora}, \cite{guo2020aloba} synchronize with incoming \gls{lora} packets to enable ambient backscattering. These however are not able to demodulate the data and act accordingly. \cite{guo2022saiyan,guo2024low} are the first to design a \gls{lora} backscatter tag able to demodulate incoming packets. Standard demodulation techniques are too power hungry to use on a backscatter tag. More low-power techniques exist but at the cost of reduced reception range.

\begin{figure}[!t]
    \centering
    \includegraphics[width=\linewidth]{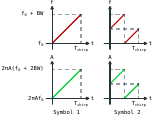}
    \caption{Derivative of a \gls{lora} chirp translates the frequency change into an
    amplitude change.}
    \label{fig:saiyan-amplitude}
\end{figure}

A \gls{lora} symbol can be represented by \cref{eq:lora-symbol} \cite{guo2022saiyan,guo2024low}. The signal has an amplitude of A and its frequency increases linearly over time according to $f(t) = f_{0}+ kt$. The initial frequency is represented by $f_{0}$ and changes at rate $k$.
\begin{equation}
    s(t) = A \sin\left(2 \pi f(t) t\right) \label{eq:lora-symbol}
\end{equation}
Differentiating a \gls{lora} chirp results in a signal with an amplitude proportional to the chirps frequency (\cref{eq:lora-diff}). Decoding the symbols is now a matter of tracking the amplitude and determining when it reaches its peak (\cref{fig:saiyan-amplitude}).
\begin{equation}
    \dfrac{\mathrm{d}s(t)}{\mathrm{d}t}= \underbrace{2 \pi A (f_0 + 2kt)}_{\mathrm{Amplitude}}
    \cos(2 \pi f(t) t) \label{eq:lora-diff}
\end{equation}
One possibility to achieve the differential operation is through the use of an SAW filter. These exhibit a very sharp frequency response, closely matching the narrow bandwidth of a \gls{lora} signal. After passing through the filter, the chirps are down-converted to baseband through an envelope detector. A low-power voltage comparator detects the peaks and feeds this information to an \gls{mcu}. This method avoids the usage of a power hungry \gls{adc}.

Two case studies demonstrate that acknowledgments and channel hopping improve the communication reliability. Utilizing acknowledgments boosts the packet reception rate from \SIrange{45.6}{70.1}{\percent}. Increasing the number of acknowledgments to two or three times make this increase further to \SI{83.3}{\percent} and \SI{95.5}{\percent} respectively. This increase in reliability comes at the cost of complexity and current consumption. While these solutions consume orders of magnitude less power compared to standard \gls{lora}, they still increase power consumption of a tag. This complicates the creation of an energy neutral device.

A tag becomes complex and power hungry when it supports full-duplex communication in order to receive feedback signals. Reducing the feedback to only data rate adoption avoids the need for communication. It allows a tag to perform channel estimation through the \gls{rssi} \cite{huang2021freeback}. This relies on the use of a monostatic setup where the transmission and reception channel are symmetric. Instead of power hungry \glspl{adc} simple circuitry monitors the \gls{rssi} resulting in a low power consumption. After calculating the signal strength the tag adopts its data rate. It informs the receiver in the preamble what data rate it used to transmit the data.

  \section{Application Autonomy}\label{sec:autonomy}
In low-power or \acrlongpl{end} the communication is tightly coupled to the energy budget as it is the most power hungry part. The amount of energy that can be harvested and stored determines what technologies can be used or need to improve in order to become feasible. To explore which \gls{end} applications benefit from a backscatter implementation, the most prominent harvesting and storage solutions are explored.

\subsection{Energy Harvesting}
\label{sec:energy-harvesting}
\subsubsection{Solar}\label{sec:solar-harvesting}
Solar became the main energy harvesting source in long-range backscattering \cite{peng2018plora, guo2020aloba, guo2022saiyan,guo2024low}. As backscatter tags demand a small footprint, to reduce cost and environmental impact, they require a small-area\footnote{Small-area is defined here as a maximum surface area of \SI{4}{\centi\meter\squared}.} solar panel with a high power density. In an indoor setting the incident light spectrum narrows compared to the outdoor spectrum. This causes the output of silicon cells to drop below \SI{10}{\micro\watt\per\centi\meter\squared} \cite{raj2018power}. Available commercial cells of the Panasonic Amorton series\footnote{Taking the AM-1456 \cite{PanasonicAM1456} and AM1437 \cite{PanasonicAM1437} as reference.} achieve around \SI{3.6}{\micro\watt\per\centi\meter\squared}. Other cell technologies consist of organic, dye-sensitised and perovskite. These yield higher power densities but lack commercial availability and are more expensive. An illuminance of \SI{200}{\lux} forms the baseline when looking at available devices. Cells evaluated at light levels above \SI{500}{\lux} are not representative for a normal lighting level in an office/residential building \cite{EN12464-1}.

The main downside of solar panels is the dependence on the application environment. A steady power source requires the illuminance to be guaranteed. In an outdoor scenario this is never guaranteed due to the day-night cycle. Night cycles take up the largest part of the cycle in Nordic countries decreasing the harvested power even further. This introduces the need for sufficient energy storage to bridge these illuminance gaps. Outdoor panels achieve a much higher power density of around \SI{26}{\milli\watt\per\centi\meter\squared} under solar irradiance. This allows the tag to store much more energy during light hours.

In an indoor setting the tag must harvest and store energy while the light is on. This can be purely incandescent light or a combination of indirect sunlight and incandescent light. As the power density of these panels lies far below the outdoor versions the tag harvests much less energy. This shows that the low-power operation of backscatter tags forms an especially interesting technology for indoor applications.

\subsubsection{\Acrfull{rf}}
\Gls{rf} energy harvesting comes forward as the most convenient solution. It has no dependency on the environment and can even penetrate obstacles. On the downside it has an extremely low efficiency limiting the applicable range. Several factors influence the power received by a device being the antenna gains $G_{T}, G_{R}$, transmit power $P_{T}$, frequency $\lambda = \frac{c}{f}$ and the distance from the transmitter $d$ (\cref{eq:rfeh-power}) \cite{la2019strategies}.
\begin{align}
    P_{A} & = P_{T}\frac{G_{T}G_{R}\lambda^{2}}{(4 \pi d)^{2}}\tau \label{eq:rfeh-power}          \\
    d     & = \sqrt{\frac{P_{T}G_{T}G_{R}}{P_{A}}}\frac{\lambda}{4 \pi}\tau \label{eq:rfeh-range}
\end{align}

In practice this results in only the low-power \gls{asic} designs being able to be powered using \gls{rf} harvesting. Theoretically an \gls{asic} design consuming \SI{9}{\micro\watt} of power \cite{talla2017lora} can be powered at a range of \SI{5.9}{\meter} when transmitting at \SI{20}{\dBm} and utilizing two omnidirectional antennas with a gain of \SI{3.25}{\dBi} (\cref{eq:rfeh-range}). Practical limitations reduce this range even further, for example a system with a theoretical range of \SI{20}{\meter} gets limited to \SI{16}{\meter} when using a transmit power of \SI{30}{\dBm} at a frequency of \SI{470}{\mega\hertz} \cite{tangPrototypeImplementationExperimental2025}.

Extending the harvesting range requires the use of directional antennas. A horn antenna with a moderate gain of \SI{15}{\dBi} extends the range to \SI{23}{\meter}. Utilizing directional antennas limits the application domain as the position of the tag in accordance with the transmitter needs to be well defined.

The limited ranges shown here lend themselves to ambient solutions in a bistatic setup. The tags are placed rather close (\SIrange{5}{10}{\meter}) to the ambient nodes in order to achieve sufficient backscatter range. This allows the tag to harvest sufficient energy from the ambient node and backscatter on its ongoing communication.

Recent studies \cite{vanmuldersKeepingEnergyNeutralDevices2024} focus on multi-antenna systems to increase the transferred power using \gls{rf} waves. The use of constructive interference allows to deliver higher energy levels to a tag by creating a power spot. \Gls{csi}-free methods are favourable to avoid increasing tag complexity.

\subsubsection{Capacitive}\label{sec:cap-harvesting}
Many household appliances are permanently connected to the mains power lines or connect to it frequently. The power lines emit magnetic and electric fields allowing nodes to harvest from. As the neutral and hot line run closely together in the same cable their magnetic fields will cancel each other. The magnetic fields depend on how much current runs through the cable making it an unreliable harvesting source. Electric fields around the cable are the result of the equipotential difference between them. As the cables run close together the stray electric field available to harvest from will be small. This field however forms a stable harvesting source as it only depends on the voltage across the cable.

\begin{figure}
    \centering
    \includegraphics[width=0.6\linewidth]{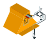}
    \caption{The current flows in the capacitive harvesting system~\cite{gulati2018capharvester}.}
    \label{fig:capharvester}
\end{figure}

A copper foil wrapped around the power lines will form a capacitive coupling to harvest the stray electric fields. The circuit requires a reference connection to extract the energy from these stray capacitances (\cref{fig:capharvester}). Earth ground forms this reference as the neutral conductor is connected to it at the distribution side of the transformer. Metal appliances have their enclosures connected to earth ground. Utilizing this as a reference seems evident but is cumbersome in practice. Instead a copper electrode functioning as a local ground reference eases deployment \cite{gulati2018capharvester}. One shortcoming is that this solution needs a good physical contact with the environment --- the ground, a shelf, a wall --- in order to operate. This disallows the solution to be dangling in mid air.

The harvester stores its energy in a TAJD477K004RNJ tantalum capacitor from AVX \cite{KYOCERA_AVX_TAJ}. It specifies a maximum leakage current of \SI{18.8}{\micro\ampere} which is in line with the selection limit of \SI{10}{\nano\ampere\per\micro\farad\volt} specified in \cite{Franklin_SolTantLeakage}. This limit is guaranteed at high temperatures and is a bad indicator for the actual leakage at room temperature. As the capacitor leakage has not been accounted for the harvesting capabilities will lie higher than what has been measured. With the tantalum capacitor the harvested energy becomes \SI{270.6}{\micro\joule} in 12 minutes over \SI{14}{\centi\meter} of cable. It demonstrates that it can work with low leakage storage solutions, while additional measurements would quantify the actual limit.

\subsection{Energy Storage}
Every backscatter tag needs some form of energy storage. The power coming directly from the harvesting source is unpredictable and unstable requiring sufficient buffering. Additionally every application needs some autonomy to perform sensing operations. The buffer size depends on the power consumption and how much autonomy is required by the application.

Backscattering enables the use of smaller and more sustainable storage solutions thanks to its low power consumption. This drives applications towards the use of capacitor based technologies. The \gls{edlc} type is most suited for short and quick charge-discharge cycles. It bridges a couple of days with its stored energy at most due to the inherently high self-discharge rates. After 30 days an \gls{edlc} only retains \SI{79.6}{\percent} of its initial charge after charging it fully \cite{callebaut2021art}. A technology such as \gls{lic} is able to power applications much longer. It retains \SI{94}{\percent} of its charge and comes with a higher energy density. These are also called hybrid capacitors as they combine the architecture of an \gls{edlc} with the anode of a Li-Ion battery.

In \glspl{edlc} the internal \gls{esr} depends a lot on the construction of the capacitor. Smaller versions tend to have a much higher \gls{esr} reaching up to tens of ohms \cite{kemet_fc_series_2023}. For active transmissions this poses a problem as the radio draws tens of milliamperes in a short time period. Backscattering solves this problem through its low-power consumption staying in the microampere range.

Batteries offer higher energy densities and low self-discharge rates making them suitable for energy storage beyond a year. Backscatter nodes harvest energy from their environment requiring rechargeable battery solutions for which \gls{lco} and \gls{lipo} are the most used today \cite{callebaut2021art}.

  \section{Assessment of Available Solutions}\label{sec:assessment}
Integrating backscattering solutions into an application gets dictated by the available energy budget. The used harvesting technique and the application environment determine the energy availability while the \gls{mcu}, \gls{rf} front-end and used sensors/actuators define the consumption. Achieving the proper balance achieves energy neutrality. Here we only consider the consumption of the backscatter circuit and the \gls{mcu} to assess the usage in different applications.

Chirp-based modulation techniques easily compare in terms of power consumption due to the shared specification of receiver sensitivity. Other researched modulation types often don't specify the sensitivity of the receiver used. Calculating the required energy\footnote{This comparison is based on the power consumption reported in their corresponding papers. While this offers a good indicative value for comparison reasons, much depends on the hardware implementation and optimizations will change the outcome.} to transmit a single bit enables a fair comparison and aids the selection during application design.
\begin{equation}
    R_{b}= SF \cdot \frac{4 BW}{(4 + CR) 2^{SF}} \cdot 1000 \label{eq:lora-bitrate}
\end{equation}
For proper comparison the parameters of the \gls{lora} packets are fixed to a coding rate of $4/8$ and bandwidth of \SI{125}{\kilo\hertz}. The energy per bit gets plotted against the maximum achievable range. This depends on many different factors with transmission power, receiver sensitivity and path loss being dominant. The transmitter adopts a transmission power up to \SI{20}{\dBm} which is the maximum allowed by European regulations. The receiver sensitivity for \gls{lora} based systems can be found in the datasheet of the Semtech \gls{lora} receivers (SX1276/77/78/89) \cite{callebaut2019characterization}. The selected spreading factor is proportional with the achievable range and inversely proportional with the data rate. Lastly the path loss is defined by the environment of the application. \textcite{callebaut2019characterization} list the path loss for three scenarios based on a measured dataset. For this comparison the urban setting is selected.

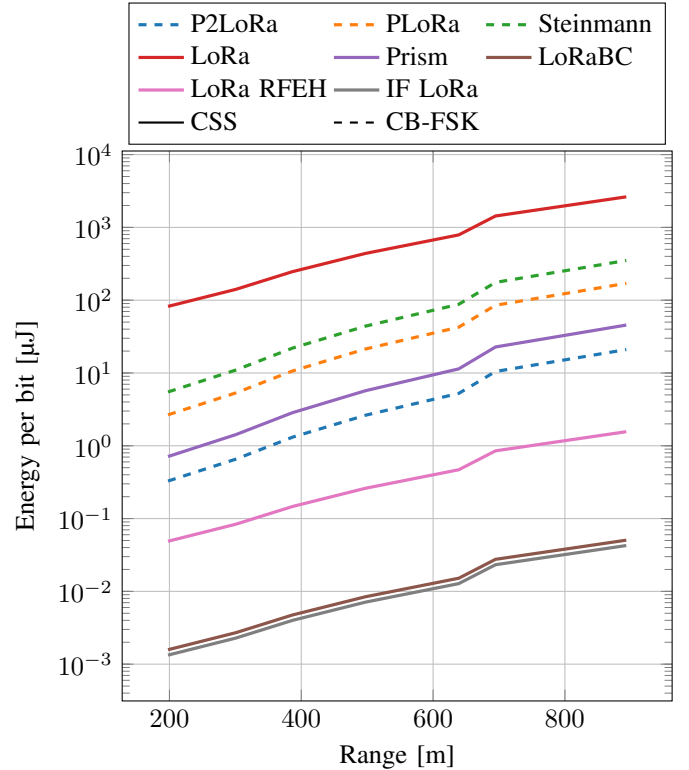
\begin{figure}
    \centering
    \begin{tikzpicture}
      \begin{axis}[
        xlabel={Range [m]},
        ylabel={Energy per bit [\unit{\micro\joule}]},
        ymode=log,
        grid=major,
        width=\linewidth,
        height=\linewidth,
        legend columns=3,
        legend cell align=left,
        legend style={
          at={(0.5,1.27)},
          anchor=north
        }
      ]
      
        \addplot+[dashed] table[x=range_m, y=energy_per_bit_uJ_P2LoRa, col sep=comma] {figures/comparison-calc-dist.csv};
        \addlegendentry{P2LoRa}
        
        \addplot+[dashed] table[x=range_m, y=energy_per_bit_uJ_PLoRa, col sep=comma] {figures/comparison-calc-dist.csv};
        \addlegendentry{PLoRa}
        
        \addplot+[dashed] table[x=range_m, y=energy_per_bit_uJ_Steinmann, col sep=comma] {figures/comparison-calc-dist.csv};
        \addlegendentry{Steinmann}
        
        \addplot table[x=range_m, y=energy_per_bit_uJ_LoRa, col sep=comma] {figures/comparison-calc-dist.csv};
        \addlegendentry{LoRa}
        
        \addplot table[x=range_m, y=energy_per_bit_uJ_Prism, col sep=comma] {figures/comparison-calc-dist.csv};
        \addlegendentry{Prism}
        
        \addplot table[x=range_m, y=energy_per_bit_uJ_LoRaBC, col sep=comma] {figures/comparison-calc-dist.csv};
        \addlegendentry{LoRaBC}
        
        \addplot table[x=range_m, y=energy_per_bit_uJ_LoRa_RFEH, col sep=comma] {figures/comparison-calc-dist.csv};
        \addlegendentry{LoRa RFEH}
        
        \addplot table[x=range_m, y=energy_per_bit_uJ_IF_LoRa, col sep=comma] {figures/comparison-calc-dist.csv};
        \addlegendentry{IF LoRa}

        \addlegendimage{solid, thick, white}
        \addlegendentry{}

        \addlegendimage{solid, thick}
        \addlegendentry{CSS}

        \addlegendimage{dashed, thick}
        \addlegendentry{CB-FSK}
      \end{axis}
    \end{tikzpicture}
    \caption{Comparison of the different technologies utilizing chirp-based modulation in terms of energy used to transmit a single bit. Transmit power is set to \SI{20}{\dBm}.}
    \label{fig:energy-per-bit}
\end{figure}

The theoretically maximum achievable range based on these three parameters gets defined as
\begin{equation}
    S_{\mathrm{RX}}\geq P_{\mathrm{TX}}+ \mathrm{PL}. \label{eq:link-budget}
\end{equation}
with $S_{\mathrm{RX}}$ the receiver sensitivity, $P_\mathrm{TX}$ the transmitter power and $\mathrm{PL}$ the path loss between the transmitter and receiver. As long as the signal power is above the receivers sensitivity it can be decoded. Note that this agnostic of the user topology as a monostatic setup has double the path loss of a bistatic setup for the same tx-tag-rx distance. This simplification however does not take effects such as carrier interference in monostatic setups into account. Combining this with the formula to calculate the energy per bit (\cref{eq:lora-bitrate}) results in \cref{fig:energy-per-bit}.

From this comparison (\cref{fig:energy-per-bit}) it becomes clear that a combination of low-power and good data rate are important. Solutions with a low data rate \cite{ren2023prism, peng2018plora, jiang2021long, steinmann2025} en up with a higher energy consumption per transmitted bit due to the increased air time. These solutions are one to two orders of magnitude more efficient than active transmissions. This, while a higher data rate boasts three orders of magnitude reduction \cite{tang2021self}. To go even more low-power and approach \gls{rfid} solutions (\SIrange{1}{7}{\micro\watt}) \cite{ashry2009compact, shen2012design} an \gls{asic} needs to be developed \cite{talla2017lora,zhu2025inductor}.

\begin{table*}
    \centering
    \caption{Summary on the applicability of the different technologies.}
    \begin{tabular}{ccccccc}
        \toprule \textbf{Tx to Tag}                     & \textbf{Topology}       & \textbf{Ambient}          & \textbf{Tag to Rx}                         & \textbf{Harvesting}     & \textbf{Technology}       & \textbf{Applications} \\
        \midrule \multirow{3}*{\SIrange{1}{10}{\meter}} & \multirow{3}*{Bistatic} & \multirow{3}*{\checkmark} & \multirow{3}*{\SIrange{500}{1000}{\meter}} & \multirow{2}*{\gls{rf}} & \multirow{2}*{\gls{asic}} & Agriculture           \\
                                                        &                         &                           &                                            &                         &                           & E-waste monitoring    \\
                                                        &                         &                           &                                            & Solar                   & \gls{cots}                & Agriculture           \\
        \SIrange{1}{20}{\meter}                         & Monostatic              &                           & \SIrange{1}{20}{\meter}                    & \gls{rf}                & \gls{asic}                & E-waste monitoring    \\
        \SIrange{10}{50}{\meter}                        & Bistatic                & \checkmark                & \SIrange{100}{500}{\meter}                 & Solar                   & \gls{cots}                & Agriculture           \\
        \SIrange{50}{200}{\meter}                       & Monostatic              &                           & \SIrange{50}{200}{\meter}                  & Solar                   & \gls{cots}                & Agriculture           \\
        \bottomrule
    \end{tabular}
\end{table*}

\subsection{E-Waste}
Continuous technological advancements cause a rapid increase in global e-waste. Old equipment gets replaced and ends up in landfills. These obsoleted devices contain a wide range of precious materials such as metals, glass and plastic. As these resources are finite, recycling them is of utmost importance. While attempts are made to collect and recycle e-waste, only \SI{17.4}{\percent} gets collected and recycled globally \cite{balde2024ewastemonitor}. Annually the collection and recycling rate increases by 0.5 billion kg on average getting outpaced by a steady increase in e-waste of 2.3 billion kg per year. By 2022, the world generated 62 billion kg of e-waste stressing the need for proper collection and handling of these devices.

\subsubsection{E-Waste Monitoring}\label{sec:ewaste-mon}
Keeping track of e-waste flows and monitoring e-waste quantities remain difficult challenges. Proper data collection forms the basis for decision making to protect the environment and public health. This shows the need for technology to track e-waste in order to stop incinerating or disposing precious materials. As this technology serves to lower e-waste adding additional batteries is undesirable. Backscatter communication serves as a key technology to achieve this goal by providing wireless communication with an extremely low energy consumption.

Vapes popped up as a major e-waste contributor in recent years. It is estimated that 42 million kg of e-cigarettes were sold in 2022 alone \cite{balde2024ewastemonitor}. These contain lithium-ion batteries, heating elements and circuits boards which end up as waste in landfills. Most of these end up incinerated as they are designed as single use products. Identifying these and saving them from landfills would have allowed to recycle 130 thousand kg of lithium in 2022 alone. This shows that adding a low cost tag for localisation can bring significant gains in tackling e-waste.

As tags are deployed in extremely large scales \gls{rfid}-like \gls{rf} energy harvesting becomes the only viable energy source. To achieve an acceptable range a directional antenna is required (\cref{sec:energy-harvesting}) allowing the reader to scan for e-waste devices in a waste stream\footnote{} up to \SI{20}{\meter} away. As the energy budget is only several micro watts at this range \gls{asic} solutions \cite{talla2017lora,zhu2025inductor} are necessary. Energy storage is limited to temporal storage in a small capacitive element as the tag is only active while receiving \gls{rf} power. Keeping the tag low-cost pushes the modulation to low-complexity binary-based solutions (\cref{sec:bb-ook}/\cref{sec:bb-fsk}). These can be implemented in hardware using a simple fixed-load modulator toggling a switch between open and closed state. The penalty in spectral efficiency doesn't pose an issue as the communication range is extremely limited.

The world generates \num{2010} billion kg of municipal solid waste yearly \cite{kaza2018waste}. In contrast the yearly generation of 62 billion kg of e-waste is \SI{3}{\percent} of waste generated. This low density of electronic equipment in the waste stream relaxes the requirement for concurrent transmissions in the \gls{mac}. The need for feedback can be alleviated by allowing the tags to continuously transmit at random time intervals.

\subsubsection{Hoarded Equipment}\label{sec:hoarded}
Obsolete or damaged electronic equipment often ends up in cupboards or warehouses. An average household owns 74 electric and electronic devices of which 13 are hoarded \cite{balde2022weee}. When they contain batteries over time they form a safety and environmental hazard. The batteries can start leaking hazardous materials into the environment such as lead and lithium. In case of fire batteries become a fuel source which cause explosions. Detecting and reporting that a device became obsolete facilitates recycling before it becomes a hazard. The same holds true for other devices such as fridges. These contain hazardous refrigerants which must be recycled before they leak out.

The long-range required to contact remote gateways brings chirp-based modulation (\cref{sec:css-backscatter}/\cref{sec:cb-ook}/\cref{sec:cb-fsk}) forward as preferred technology. In urban settings the path loss increases significantly \cite{callebaut2019characterization} hampering proper coverage with fixed infrastructure. Indoor transmitters functioning as carrier generating beacons can bridge this gap. The bistatic setup increases the achievable range to hundreds of metres. As the spectrum in these scenarios is crowded with ongoing communication. Multi-load or IQ-modulator hardware is a must to limit out-of-band interference. At the \gls{mac} level proper acknowledgements ensure no messages are missed.

Tags monitor equipment for inactivity over longer periods and alert gateways for proper collection. Capacitive and solar harvesting (\cref{sec:cap-harvesting} and \cref{sec:solar-harvesting}) come forward as the best solution for energy provisioning. The solar option has the clear benefit that it can harvest energy while the device is inactive. The capacitive solution requires a large storage element to bridge a period of several months. As the harvested energy levels are low (\SI{270.6}{\micro\joule} in 12~minutes) the storage element may not be too large to limit leakage losses.

Selecting the proper technologies starts with the available energy budgets being \SI{230}{\milli\joule} and \SI{32}{\milli\joule} per day for solar and capacitive harvesting respectively. The solar case uses a \SI{3.5}{\centi\meter\squared} panel harvesting during 5 light hours per day. From these budgets a part gets consumed by the standby power of the \gls{mcu}. Taking the STM32U031 series from ST Microelectronics as an example gives a standby power of \SI{288}{\nano\watt}\footnote{\SI{160}{\nano\ampere} at \SI{1.8}{\volt}} with the \gls{rtc} enabled. A full day requires \SI{25}{\milli\joule} showing that the capacitive solution has only \SI{7}{\milli\joule} left to charge its storage element. For the solar solution \SI{142}{\milli\joule} remains available to power the backscatter hardware.

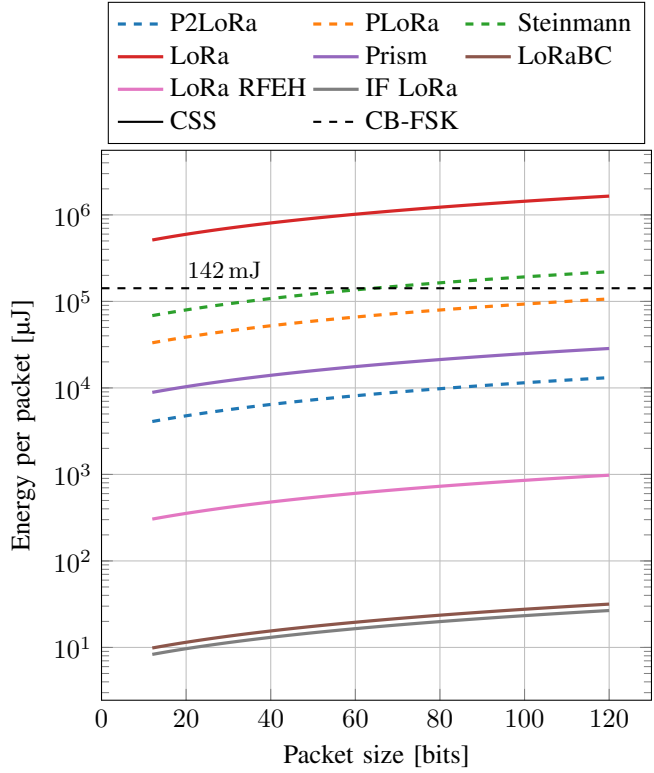
\begin{figure}
    \centering
    \begin{tikzpicture}
      \begin{axis}[
        xlabel={Packet size [bits]},
        ylabel={Energy per packet [\unit{\micro\joule}]},
        ymode=log,
        xmin=0,
        xmax=130,
        grid=major,
        width=\linewidth,
        height=\linewidth,
        legend columns=3,
        legend cell align=left,
        legend style={
          at={(0.5,1.27)},
          anchor=north
        }
      ]
      
        \addplot+[dashed] table[x=packet_size_bits, y=energy_per_packet_uJ_P2LoRa, col sep=comma] {figures/e-waste-packet-cons.csv};
        \addlegendentry{P2LoRa}
        
        \addplot+[dashed] table[x=packet_size_bits, y=energy_per_packet_uJ_PLoRa, col sep=comma] {figures/e-waste-packet-cons.csv};
        \addlegendentry{PLoRa}
        
        \addplot+[dashed] table[x=packet_size_bits, y=energy_per_packet_uJ_Steinmann, col sep=comma] {figures/e-waste-packet-cons.csv};
        \addlegendentry{Steinmann}
        
        \addplot table[x=packet_size_bits, y=energy_per_packet_uJ_LoRa, col sep=comma] {figures/e-waste-packet-cons.csv};
        \addlegendentry{LoRa}
        
        \addplot table[x=packet_size_bits, y=energy_per_packet_uJ_Prism, col sep=comma] {figures/e-waste-packet-cons.csv};
        \addlegendentry{Prism}
        
        \addplot table[x=packet_size_bits, y=energy_per_packet_uJ_LoRaBC, col sep=comma] {figures/e-waste-packet-cons.csv};
        \addlegendentry{LoRaBC}
        
        \addplot table[x=packet_size_bits, y=energy_per_packet_uJ_LoRa_RFEH, col sep=comma] {figures/e-waste-packet-cons.csv};
        \addlegendentry{LoRa RFEH}
        
        \addplot table[x=packet_size_bits, y=energy_per_packet_uJ_IF_LoRa, col sep=comma] {figures/e-waste-packet-cons.csv};
        \addlegendentry{IF LoRa}
        
        \addlegendimage{solid, thick, white}
        \addlegendentry{}

        \addlegendimage{solid, thick}
        \addlegendentry{CSS}

        \addlegendimage{dashed, thick}
        \addlegendentry{CB-FSK}
        
        \addplot[dashed, thick, black, domain=0:150]{142000} node[pos=0.12, above right, font=\small]{\SI{142}{\milli\joule}};
      \end{axis}
    \end{tikzpicture}
    \caption{Energy required to backscatter a packet with varying data sizes over a distance of \SI{900}{\meter}.}
    \label{fig:e-waste-options}
\end{figure}

\cref{fig:e-waste-options} compares the different chirp-based modulations in terms of energy required to transmit a single packet with varying length over a range of \SI{900}{\meter}. It shows that only the codeword translation solution from \textcite{steinmann2025} becomes infeasible beyond 64-bit data packets for solar harvesting. Active transmissions require almost $4\times$ more energy at 12-bits of data making it infeasible for this application.

\begin{figure}
    \centering
    \begin{tikzpicture}
      \begin{groupplot}[
        group style={
            group name=plots1,
            group size=1 by 2,
            x descriptions at=edge bottom,
            vertical sep=10pt,
        },
        xlabel={Storage duration [days]},
        grid=major,
        width=\linewidth,
        height=0.5\linewidth,
        legend cell align=left,
        legend pos=north west,
      ]
        \nextgroupplot[ylabel={Leakage energy [J]}]
            \addplot table[x=duration_days, y=EDLC_leakage_energy_J, col sep=comma] {figures/overdimensioning-leakage.csv};
            \addlegendentry{EDLC}
            
            \addplot table[x=duration_days, y=LiC_leakage_energy_J, col sep=comma] {figures/overdimensioning-leakage.csv};
            \addlegendentry{LiC}

        \nextgroupplot[ylabel={Leakage energy [mJ]}]
            \addplot[mplgreen] table[x=duration_days, y=Li_ion_leakage_energy_mJ, col sep=comma] {figures/overdimensioning-leakage.csv};
            \addlegendentry{Li-Ion}
      \end{groupplot}
    \end{tikzpicture}
    \caption{Energy needed to compensate for leakage in different storage media depending on the storage duration. \Glspl{edlc} \cite{abraconADCM-S07R5S}, \SI{1}{\micro\ampere} leakage \gls{lic} \cite{abracon2021ahcr}, \SI{3}{\percent} self-discharge Li-ion battery.}
    \label{fig:storage-leakage-energy}
\end{figure}
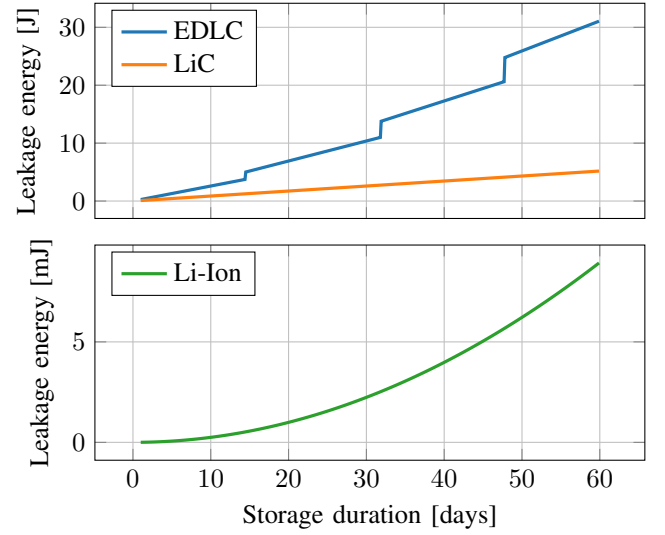

Other storage solutions posses too high leakage currents for example, \glspl{edlc} have a maximum leakage current of \SI{4}{\micro\ampere} \cite{abraconADCM-S07R5S}.
The capacitive solution needs to store the energy during the inactivity period of several months. Enough energy is required to power the standby \gls{mcu} as to facilitate a backscatter transmission at the end of the sleep period. On top of that the storage element leaks away part of the energy internally. This self-discharge has to be compensated in additional storage size. \cref{fig:storage-leakage-energy} compares \glspl{edlc}, \glspl{lic} and Li-ion batteries in terms of energy leakage. \Glspl{edlc} require up to \SI{10}{\farad} of additional storage to compensate leakage making it unfeasible as this increases leakage even further. \Glspl{lic} offer a much lower leakage requiring only \SIrange{1}{2}{\farad} for compensation. As these posses a leakage of only \SI{1}{\micro\ampere} for large storage sizes of \SI{10}{\farad} \cite{abracon2021ahcr} this overhead is manageable. Li-ion batteries are superior in term of leakage and are preferred when there are strict space requirements as these only require \SI{10}{\milli\joule} of additional energy to bridge 60 days. 

\subsection{Agriculture}
The world population keeps expanding and is expected to peak at 10.3 billion people in the mid-2080s \cite{un2024prospects}. Feeding all these people poses a large challenge for the agricultural sector. The \gls{iot} plays a major role in transforming the sector to increase crop yield, improving economical gain and environmental impact \cite{thilakarathne2025internet}. Soil and crop sensors spread across a field monitor crop health, identify areas of stress and disease and monitor overall crop yield.

\begin{figure*}
    \centering
    \begin{subfigure}
        [b]{0.24\textwidth}
        \centering
        \includegraphics[width=\textwidth]{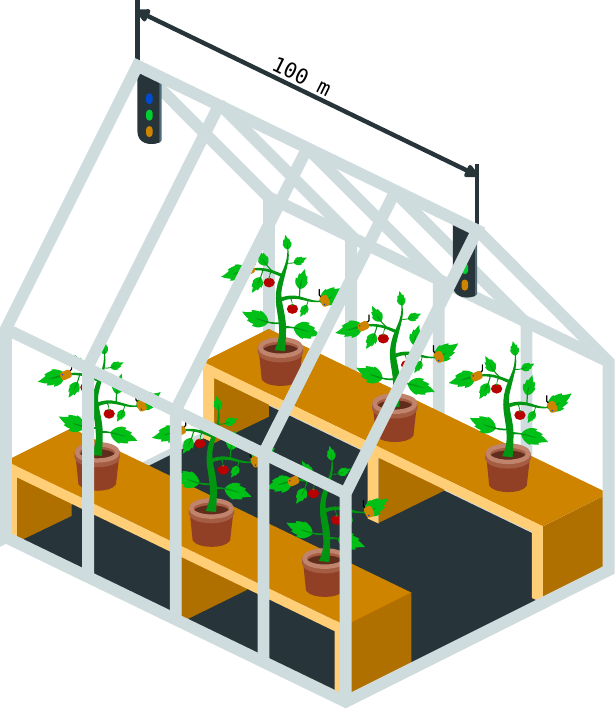}
        \caption{Plants in a greenhouse.}
        \label{fig:greenhouse}
    \end{subfigure}
    \begin{subfigure}
        [b]{0.24\textwidth}
        \centering
        \includegraphics[width=\textwidth]{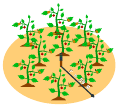}
        \caption{Plants in an open field.}
        \label{fig:open-field}
    \end{subfigure}
    \begin{subfigure}
        [b]{0.24\textwidth}
        \centering
        \includegraphics[width=\textwidth]{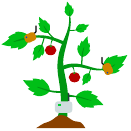}
        \caption{Active node powering tags.}
        \label{fig:single-plant}
    \end{subfigure}
    \begin{subfigure}
        [b]{0.24\textwidth}
        \centering
        \includegraphics[width=\textwidth]{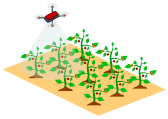}
        \caption{Reading out tags with a drone.}
        \label{fig:drone-read}
    \end{subfigure}
    \caption{Different topologies that can be used to monitor plants.}
    \label{fig:agriculture}
\end{figure*}

Monitoring crops takes place in greenhouses or outside in an open field (\cref{fig:agriculture}). Measuring environmental parameters next to crop growth enhances the monitoring capabilities and enables the use of a bistatic setup. Active nodes spread out between the crops monitor the environment and periodically transmit this data to the gateway. Crops get monitored by attaching sensors to its stem or leaves. To avoid disturbing plant growth these have to be extremely lightweight making backscatter an ideal solution. When the active nodes transmit their data the tags start backscattering onto this communication to report their measurements. Active nodes situated very close to the backscatter tags can provide them with power through \gls{rf} harvesting. Otherwise solar comes forward as the most prominent harvesting source. As this setup uses ambient backscattering the active nodes determine the update rate of the backscattering nodes.

In an open field this translates to active nodes placed in the field providing communication for tags within a range of \SI{50}{\meter} (\cref{fig:open-field}). Inside a greenhouse the same setup can be created by placing gateways at a fixed interval of \SI{100}{\meter} (\cref{fig:greenhouse}). One active node and four backscatter nodes can be placed every \SI{250}{\meter\squared} resulting in 80 devices needing battery replacements on a small farm of \SI{20000}{\meter\squared}. Comparing this to 400 devices with an all active solution shows how backscattering brings a significant improvement. This becomes more prominent when combining additional tags per active node.

As outdoor solar panels posses higher much higher efficiencies, and outdoor daylight is always present during part of the day, energy is more prominently available. This enables the use of \gls{cots} hardware solutions. For the modulation solutions compatible with ambient \gls{lora} communication are required. This limits the selection to ambient \gls{ook} \cite{guo2020aloba}, chirp-interval modulation \cite{peng2022ambient}, chirp-based \gls{fsk} \cite{peng2018plora,jiang2021long} or codeword translation based \gls{fsk} \cite{zhang2017freerider,steinmann2025}. While ambient \gls{ook} achieves optimal spectral efficiency and easily enables concurrency, chirp-interval modulation offers higher data rates which allows to transmit aggregated data at a lower energy budget. For ambient solutions this comes as a big benefit as the transmission interval is determined by the active node and thus unpredictable. The loss in spectral efficiency can be compensated by selecting an iq-modulator for the hardware.

For the \gls{mac} concurrency poses the main challenge. Some modulations achieve this inherently such as ambient \gls{ook} \cite{guo2020aloba} while others need additional measures such as assigning different shifting frequencies per tag. Other solutions such as assigning unique symbols or applying non-linear chirps (\cref{sec:concurrency}) become infeasible due to the use of ambient chirps. The most optimal solution is determined by the allowed complexity during deployment.

\begin{figure}
    \centering
    \begin{tikzpicture}
      \begin{axis}[
        xlabel={Payload size [bits]},
        ylabel={Energy per day [\unit{\joule}]},
        grid=major,
        xmin=0,
        xmax=136,
        width=\linewidth,
        height=\linewidth,
        legend columns=3,
        legend cell align=left,
        legend pos=north west,
      ]
      
        \addplot table[x=payload_bits, y expr=\thisrow{energy_uJ_1x_per_day} / 1e6, col sep=comma] {figures/p2lora-agriculture-packet-cons.csv};
        \addlegendentry{1/day}
        
        \addplot table[x=payload_bits, y expr=\thisrow{energy_uJ_5x_per_day} / 1e6, col sep=comma] {figures/p2lora-agriculture-packet-cons.csv};
        \addlegendentry{5/day}
        
        \addplot table[x=payload_bits, y expr=\thisrow{energy_uJ_20x_per_day} / 1e6, col sep=comma] {figures/p2lora-agriculture-packet-cons.csv};
        \addlegendentry{20/day}
        
        \addplot table[x=payload_bits, y expr=\thisrow{energy_uJ_50x_per_day} / 1e6, col sep=comma] {figures/p2lora-agriculture-packet-cons.csv};
        \addlegendentry{50/day}
      \end{axis}
    \end{tikzpicture}
    \caption{Energy required daily to backscatter onto ambient communication using P2\gls{lora} and using the packet detection circuit of Aloba for a different amount of transmissions per day and a varying payload length.}
    \label{fig:agriculture-duty-cycled}
\end{figure}
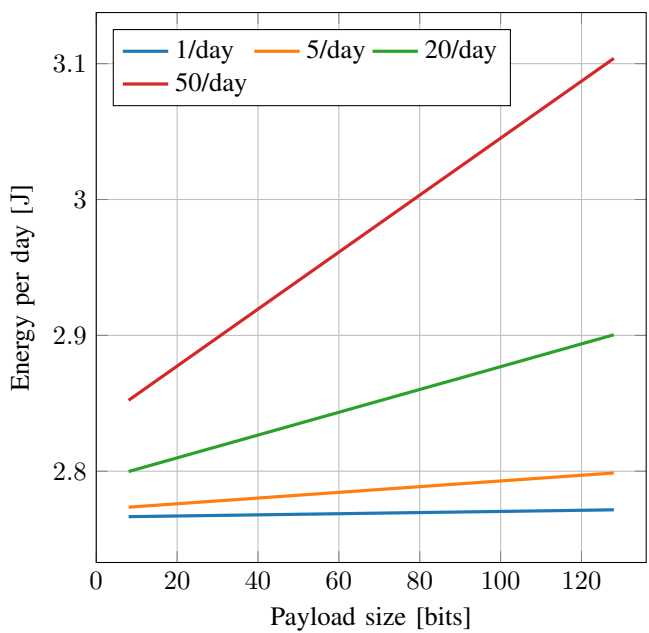

Solar harvesting provides the largest energy budget with \SI{3.9}{\joule} of energy on an overcast day (\SI{2000}{\lux} for 6~hours) and a \SI{3.4}{\centi\meter\squared} AM-5815\footnote{The operating current was extrapolated to \SI{40}{\micro\ampere} based on the simulated curve in the datasheet.} panel from Panasonic \cite{PanasonicAM5815}. Simulating the daily power requirement of a solution using P2\gls{lora} for ambient backscattering, the detection circuit is estimated to consume \SI{1.4}{\micro\watt} and a STM32L011 \gls{mcu} in low-power run mode consumes \SI{30.6}{\micro\ampere} resulting in \cref{fig:agriculture-duty-cycled}. It shows how the required energy varies depending on the required transmissions per day and the payload size. Below 5 transmissions per day the required energy is largely independent of the payload. Requiring 5 transmissions daily with a 64-bit payload yields a consumption of \SI{2.78}{\joule} leaving \SI{1.12}{\joule} to power sensors, energy management circuitry and compensate storage capacitor leakage. Residing to \gls{lic} for storage consumes \SI{173}{\milli\joule} of additional energy. The remaining \SI{947}{\milli\joule} should suffice for the additional circuitry but forms a challenging optimization task.

Single plant monitoring allows the usage of \gls{rf} energy harvesting as the active node is very close to the tags (\cref{fig:single-plant}). However this requires \gls{asic} solutions as the harvested energy is extremely low. The tags no longer aggregate data but only transmit data while being powered by the ambient transmission. This poses a challenging task as it must perform both the measurement and transmission while maintaining tight synchronization with the ambient packet.

An alternative to avoid the usage of ambient communication is the use of drones (\cref{fig:drone-read}). These fly close to the tags and can be equipped with directional antennas directed downwards. This shifts the need for maintenance from the active nodes to the drone itself. The advantage lies in the possibility to automate the flight pattern and charging of the drone reducing the need for human intervention. Even large-scale agriculture deploys drones using specifically developed deployment strategies \cite{liang2019drone}. Drones equipped with spraying equipment allow to perform a double role. They spray/water the crops and read out sensors at the same time. \textcite{katanbaf2021simplifying} demonstrate that a drone flying at \SI{18}{\meter} altitude is capable of covering an area of \SI{2400}{\meter\squared}. \textcite{renAeroEchoAgriculturalLowpower2025} place fixed gateways in the field and excite the tags using a drone. Tags are grouped in cells which are excited all at once utilizing multi-tag decoding in the same frequency channel optimizing spectral usage. Two route planning strategies optimize the coverage and energy efficiency of the overall system.

\subsection{Logistics}
\begin{figure*}
    \centering
    \begin{subfigure}
        [b]{0.48\textwidth}
        \centering
        \includegraphics[width=\textwidth]{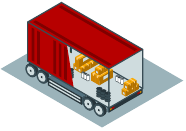}
        \caption{Inside metal truck.}
        \label{fig:truck}
    \end{subfigure}
    \hspace{2em}
    \begin{subfigure}
        [b]{0.24\textwidth}
        \centering
        \includegraphics[width=\textwidth]{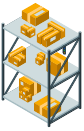}
        \caption{On a shelf in a warehouse.}
        \label{fig:warehouse}
    \end{subfigure}
    \caption{Tracking goods in the logistics chain has its main challenge in the environment. Tags are placed close to gateways enabling \gls{rf} energy harvesting but many obstacles block the path attenuating the received power.}
    \label{fig:logistics}
\end{figure*}
Inventory management plays a key role in many businesses. It avoids insufficient or excessive stock levels and enables tracking goods from supplier to customer. In the logistics domain it is crucial to track goods in order to optimize the process, construct proper planning, keep track of progress and ensure timely deliveries to customers. Continuous tracking of goods allows to locate them anywhere in the logistics chain at any time. This provides more fine grained tracking opposed to discrete tracking where goods are scanned at fixed locations. Currently active \gls{rfid} forms the key enabling technology due to its increased range \cite{casella2022evolution}. As the location of these goods can range from being somewhere in a warehouse to being shipped on a boat this increased range is necessary to support all these locations. The downside of this approach is the need for batteries which comes with an increased cost and e-waste. This pushes most logistic chains towards using discrete tracking still today.

With long-range backscattering a monostatic setup creates the possibility to mount gateways in the different steps of the logistics chain. These track the goods as they are transported with a truck or stored in a warehouse (\cref{fig:logistics}). The achievable range depends on the path loss in the specific environment and determines how many gateways need to be installed to achieve full coverage. Standard trucks are limited in size and require only one beacon while a warehouse can be hundreds of square metres requiring beacons to be installed at a specific shelve interval. In warehouses the metal shelves used to store the goods pose a major obstacle for the \gls{rf} communication. Proper coverage requires precise planning, good measurements and additional gateways. Adding sensors to the backscatter tags allows to monitor the goods for temperature changes, fall damage and much more.

Assuming a deployment of 4 gateways per \SI{200}{\square\meter} a tag at the furthest point in the centre of this area. Assuming the area is squared this results in
\begin{equation}
    \frac{\sqrt{\sqrt{A}^2 + \sqrt{A}^2}}{2} = \frac{\sqrt{2} \sqrt{A}}{2}
\end{equation}
with $A$ the total covered area of \SI{200}{\square\meter} the tag-to-gateway distance is \SI{10}{\meter} worst case. This distance makes \gls{rf} based energy harvesting viable but pushes the hardware to \gls{asic} solutions to remain feasible. As the communication distance is limited the main challenges lie in the \gls{mac}-layer. Here the enablement of parallel transmissions and ensuring privacy pose the most critical objectives. Many goods are stored closely together and will start to transmit simultaneously. Many modulations rely on distinct features being programmed in every tag to allow concurrency (\cref{sec:concurrency}). In logistics it is impossible to plan up front which tags will end up together making this approach infeasible. Modulations such as Aloba \cite{guo2020aloba} inherently support concurrency while for others query-based \glspl{mac} allow concurrent transmission after a query beacon. The tight synchronization in query based systems pose an additional challenge in these environments. The dense gateway deployment enables indoor localisation using backscattering \cite{hou2024lobaca,jiang2023locra}. One remaining challenge lies in combining the localization with the concurrency requirement.

  \section{Discussion and Technological Gaps}\label{sec:discussion}
Long-range backscatter evolves rapidly to become an enabling technology for \glspl{end}. While most research focuses on power-consumption and range many challenges remain in other domains such as concurrency and security. Currently available technologies are discussed while highlighting where gaps exist for future researchers.

\subsection{Backscatter Options}
\renewcommand{\arraystretch}{1.2}
\begin{table*}
    \centering
    \caption{Comparison of long-range backscatter systems.}
    \label{tab:perf-comparison}
    \begin{tabular}{lccrrrrrcr}
        \toprule \textbf{Paper}                             & \textbf{Topology}        & \thead{Ambient \\ carrier}          & \textbf{$P_{TX}$}   & \textbf{$P_{RX}$}     & \multicolumn{2}{c}{Range} & \textbf{Power}     & \textbf{Hardware}      & \textbf{Frequency}  \\
        \cmidrule(l){6-7}                                   &                         &                           &                     &                       & Tx-Tag                    & Tag-Rx             & Active                   &                    &                    \\
                                                            &                         &                           & \unit{\dBm}         & \unit{\dBm}           & \unit{\meter}             & \unit{\meter}      & \unit{\micro\watt}       &                    & \unit{\mega\hertz} \\
        \midrule
        \multicolumn{10}{c}{\gls{css}} \\
        \midrule
        Prism \cite{ren2023prism}                           & Bistatic                & \fullcirc                 & 20.0                &                       & 5.0                      & 500.0              & 695.0                    & Non-linear         &    \\
        \multirow{2}*{\textcite{talla2017lora}}             & \multirow{2}*{Bistatic} & \emptycirc                & \multirow{2}*{30.0} & \multirow{2}*{-142.0} & 237.5                    & 237.5              & \multirow{2}*{9.3\textsuperscript{$\dagger$}} & \multirow{2}*{VCO/PLL} & \multirow{2}*{915} \\
                                                            &                         & \emptycirc                &                     &                       & 5.0                      & 2800.0             &                          &                    &     \\
        \textcite{katanbaf2021simplifying}                  & Monostatic              & \emptycirc                & 30.0                & -148.0                & 90.0                     & 90.0               & 9.3\textsuperscript{$\dagger$}                & VCO/PLL & 915 \\
        \textcite{guo2022rf}                                & Bistatic                & \emptycirc                & 30.0                &                       & 0.2                      & 643.5              & 339.2                    & IQ Mod.            & 2400  \\
        IF LoRa \cite{zhu2025inductor}                      & Bistatic                & \emptycirc                & 19.0                & -136.0                & 1.0                      & 1200.0             & 7.8\textsuperscript{$\dagger$}& Digital-IQ    & 900 \\
        \textcite{tang2021self}                             & Bistatic                & \emptycirc                & 30.0                &                       & 28.0                     & 381.0              & 286.0                    & Digital            & 433 \\
        \midrule
        \multicolumn{10}{c}{Binary-based FSK} \\
        \midrule
        AllSpark \cite{wang2022allspark}                    & Monostatic              & \emptycirc                & 33.0                &                       & 700.0                    & 700.0              & \num{244 200.0}          & FSK                & 900 \\
        \multirow{3}*{LoRea \cite{varshney2017lorea}}       & \multirow{3}*{Bistatic} & \emptycirc                & \multirow{2}*{28.0} & \multirow{3}*{-124.0} & 1.0                      & 3400.0             & \multirow{2}*{70.0}      & \multirow{3}*{FSK} & \multirow{2}*{868} \\
                                                            &                         & \emptycirc                &                     &                       & 75.0                     & 75.0               &                          &                    &     \\
                                                            &                         & \emptycirc                & 26.0                &                       & 1.0                      & 225.0              & 650.0                    &                    & 2400\\
        \midrule
        \multicolumn{10}{c}{Chirp-based OOK} \\
        \midrule
        Freeback \cite{huang2021freeback}                   & Monostatic              & \emptycirc                & 5.0                 &                       & 27.0                     & 27.0               &                          & OOK                & 900 \\
        \textcite{lin2024harmonic}                          & Monostatic              & \emptycirc                & 2.5                 &                       & 50.0                     & 50.0               & 31.5\textsuperscript{$\ddagger$}& OOK         & 925 \\
        Aloba \cite{guo2020aloba}                           & Bistatic                & \fullcirc                 & 20.0                &                       & 5.0                      & 300.0              &                          & OOK                & 900 \\
        \midrule
        \multicolumn{10}{c}{Chirp-based FSK} \\
        \midrule
        Pacim \cite{peng2022ambient}                        & Bistatic                & \fullcirc                 & 20.0                &                       & 1.0                      & 800.0              &                          & CIM                & 900 \\
        PLoRa \cite{peng2018plora}                          & Bistatic                & \fullcirc                 & 21.0                & -103.0                & 0.2                      & 1100.0             & \num{2600.0}             & FSK                & 900 \\
        \multirow{2}*{P2LoRa \cite{jiang2021long}}          & \multirow{2}*{Bistatic} & \multirow{2}*{\fullcirc}  & 20.0                & \multirow{2}*{-141.0} & 5.0                      & 500.0\textsuperscript{*} & \multirow{2}*{320.0}& \multirow{2}*{FSK}& \multirow{2}*{433} \\
                                                            &                         &                           & 30.0                &                       & 100.0                    & 100.0              &                          &                    &     \\
        XORLoRa \cite{li2020xorlora}                        & Bistatic                & \fullcirc                 & 10.0                &                       & 1.0                      & 30.0               &                          & CWT                & 900 \\
        \multirow{2}*{\textcite{steinmann2025}}             & \multirow{2}*{Bistatic} & \multirow{2}*{\fullcirc}  & \multirow{2}*{10.0} & \multirow{2}*{}       & 1.0                      & 500.0              & \multirow{2}*{670.0}     & \multirow{2}*{CWT} & \multirow{2}*{868} \\
                                                            &                         &                           &                     &                       & 60.0                     & 60.0               &                          &                    &     \\
        \textcite{lazaro2025home}                           & Bistatic                & \fullcirc                 & 10.0                &                       &                          &                    & 54.0                     & Offset CSS         & 433 \\
        \bottomrule
        \multicolumn{10}{l}{\textsuperscript{*}We ignore the result of \SI{2.2}{\kilo\meter} as the transmitter to tag distance is unspecified.} \\
        \multicolumn{10}{l}{\textsuperscript{$\dagger$}Simulated.} \\
        \multicolumn{10}{l}{\textsuperscript{$\ddagger$}Modulator.}
    \end{tabular}
\end{table*}

While many different solutions exist for long-range backscattering, comparing them remains a challenging task. The test conditions vary significantly, some solutions optimize the hardware for power consumption while others use standard development boards and frequency bands depend on the country where the experiments are conducted. \cref{tab:perf-comparison} attempts to give an overview of the available solutions and their test conditions. Fields left blank are either unclear or not specified. Every solution uses a modulation technique in combination with a certain hardware architecture. This results in a certain power consumption and performance which is specific to that solution. Other hardware topologies are also capable of generating the specific modulation but can lower the power consumption but at the cost of spectral efficiency for example.

Important to not is that ambient solutions are limited in range by their detection circuit. For example, beyond the listed equidistant range of \SI{100}{\meter}, P2LoRa \cite{jiang2021long} fails to detect ambient packets and the tag no longer gets woken up. Some solutions make use of directional antennas increasing the achievable range. Pacim \cite{peng2022ambient} utilizes a \SI{12}{\dBi} directional antenna at the receiver, \textcite{talla2017lora} and \textcite{katanbaf2021simplifying} use a directional \SI{6}{\dBi} and \SI{9}{\dBc} patch antenna respectively.

Nevertheless backscattering remains a trade-off between power consumption and range. The modulation technique (\cref{sec:modulation}) mostly determines the achievable range while the hardware (\cref{sec:hardware}) has a big impact on the spectral efficiency and the power consumption. IQ-modulator topologies (\cref{sec:iq-modulator}) support any type of modulation but at a greater complexity and thus power consumption while fixed-load modulation with a single \gls{rf} switch (\cref{sec:fixed-load-modulation}) offers the simplest solution but at the worst spectral efficiency. Making the right trade-off is application dependant where many factors such as environment, cost and range are taken into account.

All solutions are tested using \gls{cots} components in many cases leading to a sub-optimal result. Either this has to be optimized towards power consumption and cost for commercialization or monolithic integration needs to be adopted. It reduces both the manufacturing cost and power consumption of the tags enabling the use of printed manufacturing technologies. Their minimal environmental impact \cite{zheng2018life} form an additional benefit of these technologies. For extremely low-power applications utilizing \gls{rf} harvesting, such as logistics, this is the only viable solution.

Reflection amplifiers boost the reflected signal at the cost of additional consumption. A \SI{22.3}{\dB} gain can be achieved with \SI{580}{\micro\watt} of additional power \cite{lazaro2025home}. Using tunnel diodes yields similar results but at much lower power consumptions \cite{amato2015long,varshneyTunnelScatterLowPower2019}. A \gls{mac}-layer approach to increase the achievable range relies on feedback signals (\cref{sec:feedback}). Re-transmissions and adapting transmission parameters at the tags side \cite{guo2022saiyan,guo2024low} improves the overall \gls{qos}. The power consumption penalty is in the same range as for tunnel diodes (\SI{84}{\micro\watt}).

\subsection{Topology}
Proper gateway deployment remains a key challenge for backscatter solutions. In conventional \gls{lora} networks, gateways are centrally deployed to provide wide-area coverage. In backscattering this translates to monostatic architectures, limiting the communication distance to approximately \SI{100}{\meter} due to carrier interference. Bistatic architectures significantly extend this range up to \SI{1}{\kilo\meter}. However, this comes at the cost of increased deployment complexity and requires the tag to be located close to the transmitter. Adopting an equidistant setup the communication distance is close to that of the monostatic case (\SIrange{100}{200}{\meter}). This setup offers an attractive compromise by using active nodes in the field to provide the carrier through their ambient transmissions. This maximizes backscatter range while avoiding the need for deploying high-power transmitters over large areas.

Non-ambient backscatter solutions introduce additional constraints by relying on a dedicated carrier. To reduce power consumption gateways typically only transmit when tags need to be queried. For time critical applications such as home security systems \cite{lazaro2025home}, the carrier must be continuously available, wasting substantial amount of power when tags are inactive. Other designs introduce a wake-up radio to trigger uplink communication \cite{katanbaf2021simplifying,hessar2019netscatter}. While enabling on-demand transmissions it requires an always-on radio significantly increasing power consumption. Their low-complexity nature limit the achievable sensitivity (e.g. $>\SI{-55}{\dBm}$ in \cite{katanbaf2021simplifying}), which in practice results in a significant reduction in communication range.

Experimental results show that packet detection ranges of up to \SI{50}{\meter} are feasible when operating at a sampling rate $480\times$ lower than standard \gls{lora}. This highlights these receivers ultra-low-power operation capabilities. However, this performance comes with a penaltiy in detection robustness. For example, Aloba achieves a packet detection range of \SI{30}{\meter} at a packet detection error rate of \SI{1}{\percent}, while operating at a sampling rate of \SI{250}{\kilo\hertz} and a power consumption of \SI{300}{\micro\watt}. Similarly, P\gls{lora} \cite{peng2018plora} achieves \SI{50}{\meter} of detection range with a transmit power of \SI{21}{\dBm}. This illustrates that ambient packet detection enables substantial reduction in infrastructure energy consumption but at a limited achievable detection range limiting reliable backscatter communication in practical deployments.

\subsection{Concurrency}
Many applications densely populate areas with many tags. A local event triggers multiple tags which will contend for transmission time. While feedback signals allow re-transmissions in case of collisions they become ineffective in these large scale networks. The high rate of re-transmissions increases the power consumption and causes denial of service in the worst case.

Several solutions exist assigning different symbols \cite{hessar2019netscatter}, frequency channels \cite{jiang2021long} or different curved symbols \cite{ren2023prism} to each tag to avoid collisions. P2\gls{lora} goes up to 101 parallel transmissions with a bandwidth of \SI{500}{\kilo\hertz} while Prism boost this towards 700. NetScatter achieves 256 concurrent transmissions with a transmission bandwidth of \SI{500}{\kilo\hertz} while a bandwidth of \SI{2}{\mega\hertz} boosts this up to 1000. However they all share the same drawback requiring manual assignment of the unique property to each tag. This results in manually adjusting every tag and carefully planning the deployment upfront. To avoid this labour intensive task more complex techniques are required for collision avoidance. The main challenge lies in designing these for a low complexity tag with low-power operation in mind. Next to this solutions requiring different frequency channels per tag have a high spectral occupation which is undesirable. New solutions should minimize this impact as much as possible.

\textcite{kim2023lora} adjust the gateways transmission power to control the amount of contending devices. ALOHA resolves basic contentions but the goodput still drops significantly when many devices are competing. Lowering the transmission power limits the amount of devices able to perform a backscatter transmission. In order to avoid prioritizing devices close to the transmitter an algorithm is implemented which increases the priority of devices losing a transmission. This method has only been tested through simulations and still requires an assessment on complexity and power consumption impact. \textcite{renAeroEchoAgriculturalLowpower2025} lower the transmission power of the preamble waking up the tags to select the contending devices. Afterwards collisions resolve by assigning a random transmission delay per tag and using non-linear chirps.

\subsection{Security}
Long-range backscattering calls for security measures. Tags detecting a carrier instantaneously transmit their data. The tag doesn't know who is providing the carrier and to whom it is transmitting its data. Encrypting the data provides a way to stop eavesdropping but comes with increased hardware complexity. In applications such as logistics and asset management tracking moving tags remains possible due to their unique fingerprint. Authentication techniques \cite{juels2005strengthening,jiang2023backscatter} address issues such as impersonation and spoofing. 

AuthScatter \cite{zhangAuthScatterAccurateRobust2025} addresses multiple security issues through a physical-layer mutual authentication scheme. It leverages channel fading and random numbers as a one-time pad to protect the identity key exchange procedure. Physical-layer fingerprints construct shared identity keys providing efficient identification while a challenge-response authentication ensures secure exchange of the keys. The one-time pad protects against eavesdropping, spoofing, replay and counterfeiting attacks. The use of channel reciprocity and random number knowledge avoids the need for channel estimation and complex processing mechanisms such as encryption.

Implementing security features always comes with added complexity and power consumption. For long-range backscatter the need for an on-tag receiver remains the main challenge as it limits the communication distance.

\subsection{Enhancing Data Rate}
\Gls{lora} inherently operates at low data rates. Certain applications, video surveillance for example, collect significant amounts of data. Transmitting these at low data rates increases power consumption and latency. Increasing the chirp bandwidth poses a straightforward solution to increase the data rate. Going outside of the \gls{lora} specification allows to reach a symbol time of \SI{26}{\micro\second} with a bandwidth of \SI{5}{\mega\hertz} \cite{guo2024mighty}. This results in a data rate of \SI{218.8}{\kilo\bit\per\second} compared to \SI{21.8}{\kilo\bit\per\second} for a standard bandwidth of \SI{500}{\kilo\hertz}. With this increase in data rate comes a decrease in power consumption as the air time reduces ($\frac{T_{sym}}{10}$). As a downside this will reduce the achievable range and the amount of available channels.

\begin{figure}
    \centering
    \includegraphics[width=0.8\linewidth]{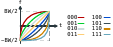}
    \caption{Different polynomial functions giving chirps different curvatures. Additional
    bits are encoded per curvature.}
    \label{fig:mighty-modulation}
\end{figure}

Instead of deploying non-linear chirps to enable concurrent transmissions (\cref{sec:concurrency}) \cite{li2022curvinglora} they can be utilized to encode additional bits per symbol \cite{guo2024mighty}. A chirp curves according to a certain polynomial function. By defining N different curvatures, $\log_{2}(n)$ additional bits can be encoded per symbol. For example defining 8 different chirp shapes allows 3 additional bits to be transmitted (\cref{fig:mighty-modulation}). Proper demodulation of these symbols requires sufficient spacing between them after the \gls{fft} operation. This limits the number of additional bits that can be encoded per symbol. Higher spreading factors will reduce the number of bits that can be encoded.

Deploying multiple antennas on the backscatter tag allows to multiplex different data streams. These can be separated in frequency resulting in a large spectrum usage and increased power consumption \cite{guo2024mighty}. Separating the streams through different modulation parameters allows a receiver to demodulate them even when colliding. The use of multiple antennas guarantees that each antenna receives the full signal strength and the tag won't suffer in range.

Practical experiments show that the above methods combined can reach data rates of \SIrange{1.6}{4.2}{\mega\bit\per\second} with a range up to \SI{150}{\meter} \cite{guo2024mighty}. The high data rate reduces the air time and results in a low energy consumption of \SI{25}{\nano\joule\per\bit}. As a downside the peak power consumption of the tag rises to \SI{70}{\milli\watt} requiring sufficient buffering. Also radio setup and shutdown will play an important factor at these power consumption levels.

Extending \gls{lora} backscatter based on \gls{dds} with amplitude modulation boosts the data rate from \SI{4}{\kilo\bit\per\second} to \SI{64}{\kilo\bit\per\second} when utilizing a 16QAM scheme \cite{shiRateenhancedSquarewaveLoRa2024}. However, amplitude modulation inherently limits the communication range due to its worse demodulation characteristics at longer ranges. Dividing a chirp into sub-chirps with an individual frequency offset achieves a similar throughput gain but with the same range characteristics as classic \gls{lora} backscatter \cite{baiSubLoRaHighThroughputLoRa2026}. The usage of ambient signals limits tag-to-transmitter distance but this can be circumvented by switching to a constant carrier wave.

Higher data rates are achievable by leveraging wider bandwidths (e.g., \gls{uwb}), higher-order modulations, and dynamic switching among multiple backscatter paths. Systems like \cite{parksTurbochargingAmbientBackscatter2014,trotterMultiantennaTechniquesEnabling2012a} employ multiple backscatter paths or beamforming to maximize throughput while maintaining ultra-low power operation.

In~\cite{song2021exploiting}, they present a polar code-based strategy that adapts the coding rate to the current channel. In order to do this, the receiver estimates the possible coding rate and sends the desired code rate to backscatter device. According to their measurements, the polar-code based solution largely outperforms P\gls{lora} (with hamming) both in terms of propagation distance and goodput.

\subsection{Localization}
Many applications require localisation. The required localisation range varies depending on the application. For indoor applications a range of \SIrange{10}{20}{\meter} suffices while outdoors ranges beyond \SI{100}{\meter} are the starting point. \gls{aoa} forms a widely used technology to facilitate localisation of devices \cite{hou2024lobaca}. Multiple receiving nodes are deployed in the field. These receive both the carrier signal and the backscattered signal. Every beacon calculates the \gls{aoa} and the combination allows to locate the device. Another approach calculates the travel time and distance of the backscatter signal to several receivers and utilizes this information to localize the tag \cite{jiang2023locra}.

These solutions require a dense gateway deployment where 3-4 gateways cover less than \SI{200}{\square\meter}. Solutions eliminating the need for multiple gateways \cite{nandakumar3DLocalizationSubCentimeter2018} rely on knowledge of the floor plan of a room. This limits their use to fixed indoor setups. Long-range solutions utilizing \gls{lora} modulation for localisation exist but are limited to active radios \cite{bansalOwLL2021}.

Backscatter localization solutions suffer from two main disadvantages. They lack long-range capabilities and require additional infrastructure investments. \textcite{hou2024lobaca} achieve very good accuracies of \SI{5}{\centi\meter} close range ($\leq \SI{5}{\meter}$) and \SI{71}{\centi\meter} long range ($\leq \SI{40}{\meter}$) while \textcite{jiang2023locra} achieve \SI{6.8}{\centi\meter} and \SI{88}{\centi\meter} at \SI{5}{\meter} and \SI{50}{\meter} from the base station respectively. Installing receivers in every room indoors allows room-level tracking through \gls{rssi} \cite{lazaro2021room}. This operates on the data backscattered by the tag and doesn't require designated signalling.

When only motion information is required \textcite{jiang2021sense,liu2024lomu,liu2024enable} are able to extract this based on the phase difference between the signals received from multiple backscatter tags. Multiple tags can be sensed simultaneously with a range of up to \SI{400}{\meter}.

\subsection{Sustainability}
The \gls{iot} brings large social benefits but they come at a large environmental cost. This is reflected in the 50~million metric tons of e-waste produced globally each year \cite{jakhar2023environmental}. Backscattering forms a key technology to push for more sustainability through elimination of the e-waste coming from battery replacements. Capacitors provide an alternative energy source with an energy density \SIrange{10}{20}{\times} lower than batteries. However, they endure millions of charge-discharge cycles while batteries can only do a couple of thousand \cite{meena2024green}. They charge extremely fast making them suitable for high-power intermittent harvesting sources. New materials are studied to replace the toxic ones being used today to fabricate the capacitor electrodes \cite{meena2024green}. This further lowers their environmental impact far below present battery technologies.

A typical \gls{iot} node contributes the most through its \gls{pcb} and \gls{ic} production to the \gls{gwp} \cite{cappelle2024iot}. Since backscatter tags utilize less complex technologies their required circuitry reduces. This lowers their \gls{pcb} size and environmental impact accordingly. When moving to \gls{asic} solutions the highest gain in \gls{gwp} can be achieved. The evolution of \gls{rfid} showed that microchips of \SI{0.16}{\milli\meter\squared} can provide the necessary circuitry for backscatter communication. Even if this surface area triples or quadruples it remains orders of magnitudes lower than the area achieved with multiple standard \glspl{ic}.

Solutions utilizing individual \gls{rf} transistors to represent impedances \cite{guo2022rf,belo2019iq,zhu2025inductor} offer a smaller footprint solution. Especially when attempting to reduce harmonics, the additional impedances required increase the footprint. Implementing these impedances at the \gls{ic} level proves to be area intensive and expensive. An \gls{asic} simulation gives a comparable area compared to commercial \gls{rfid} chips \cite{zhu2025inductor}.

Commercial \gls{rfid} tags make use of printed electronics instead of PCBs. This results in a more low cost solution and a smaller environmental impact. An comparative \gls{lca} \cite{zheng2018life} reveals that the production of a PCB antenna requires much more materials and has much larger environmental impact compared to a printed antenna. In the stage of raw material preparation the PCB solution already emits 100 times more. This shows that moving towards printed electronics has significant benefits for which backscattering is a crucial enabler.

\subsection{Backscatter and Active Synergy}
Large gaps exist between the receiver sensitivity and the \gls{rss} during data transmission. This gap can be leveraged to reduce the power consumption of radios by lowering their transmit power. Due to the low efficiencies of active radios the power saving turns out rather low. \textcite{rostami2018polymorphic} integrated an active radio and passive backscattering on the same device. Their approach leveraged the strong points of both to build an optimized solution while masking their weaknesses. Backscattering operates at much lower power levels than active radios and doesn't suffer from the low efficiency at lower transmit powers. Therefore it is used to transmit the data when there is a gap between the receiver sensitivity and the \gls{rss}. Through active-assisted backscatter they make the active radio step in when the \gls{rss} becomes too low and backscattering fails.

  \section{Conclusions and Open Questions}\label{sec:conclusion}
Long-range backscatter communication evolved into an enabling technology ultra-low-power and even energy-neutral \gls{iot} deployments. Reflecting instead of generating the radio signals reduces the required energy for wireless transmission by two to three orders of magnitude using \gls{cots} components and up to five order of magnitude when realized as \gls{asic}. This enormous reduction makes maintenance-free operation feasible boosting the sustainability of the \gls{iot} in a world with ever increasing e-waste.

The communication range in backscatter systems gets determined by a three-way interplay between topology, the available energy budget and tag complexity. Bistatic and ambient topologies emerge as the key enables for long-range operation, providing communication distances of several hundreds of meters and even beyond a kilometer when tags are placed sufficiently close to the transmitter. In contrast, monostatic architectures remain attractive for their deployment simplicity but face fundamental challenges related to carrier self-interference.

Selecting the most optimal backscatter solution depends on the specific application. Cost- and scale-driven applications such as logistics, waste monitoring and asset tracking reside to \gls{asic} based designs, \gls{rf} energy harvesting and simple binary modulation schemes. They trade the reduced hardware complexity and cost for limited spectral efficiency and range. Applications with higher range or reliability requirements benefit from higher energy budgets enabled by solar or capacitive harvesting. Fields such as agriculture make use of chirp-based modulation, ambient backscattering and more advanced \gls{mac}-layer techniques.

Advances in \gls{mac}-layer designs overcome challenges concerning concurrency and \gls{qos}. Feedback-based transmissions improve communication reliability but scale poorly under dense deployments due to excessive contention and energy overhead. Parallel decoding, symbol-domain multiplexing and non-linear chirps demonstrate that large-scale concurrent backscatter networks are possible, but at the expense of deployment flexibility, spectral efficiency or receiver complexity. Designing collision-resilient \gls{mac} protocols with feedback signalling that remain compatible with low-complexity tags remains an open challenge.

Extending backscattering to long-range application amplifies security and privacy risks. Eavesdropping, spoofing, tracking and jamming pose significant threats which can be tackled using cryptographic protection and mutual authentication mechanisms. However, their direct application to backscatter gets constrained by the low-power/low-complexity requirements of the tags. Recent physical‑layer authentication and lightweight security techniques are promising, but further work is required to balance robustness, range, and energy consumption. In many real‑world deployments security considerations may ultimately determine whether backscatter can be adopted at scale

Long-range backscatter does not form a drop-in replacement for active radios. Instead it has to be carefully matched to the application context and energy constraints to build wireless \gls{iot} nodes for previously impractical or impossible domains. By eliminating the need for batteries backscatter communication plays a key role in developing scalable, sustainable and long-lived \gls{iot} deployments. Ongoing developments focus on tackling challenges coming from concurrency, security and optimized integration needs in future 6G networks and beyond.

\section{Acknowledgments}
The research was supported by the EU under the projects SUSTAIN-6G (Grant Agreement~101191936) and Ambient-6G (Grant Agreement~101192113).

  \appendices

  \ifCLASSOPTIONcaptionsoff
  \newpage
  \fi

  \printbibliography

@inproceedings{amato2015long,
	title        = {{Long range and low powered RFID tags with tunnel diode}},
	author       = {Amato, Francesco and Peterson, Christopher W and Akbar, Muhammad B and Durgin, Gregory D},
	year         = {2015},
	booktitle    = {2015 IEEE International Conference on RFID Technology and Applications (RFID-TA)},
	pages        = {182--187},
	organization = {IEEE}
}

@inproceedings{varshneyTunnelScatterLowPower2019,
  title = {{{TunnelScatter}}: {{Low Power Communication}} for {{Sensor Tags}} Using {{Tunnel Diodes}}},
  shorttitle = {{{TunnelScatter}}},
  booktitle = {The 25th {{Annual International Conference}} on {{Mobile Computing}} and {{Networking}}},
  author = {Varshney, Ambuj and Soleiman, Andreas and Voigt, Thiemo},
  date = {2019-10-11},
  series = {{{MobiCom}} '19},
  pages = {1--17},
  publisher = {Association for Computing Machinery},
  location = {New York, NY, USA},
  doi = {10.1145/3300061.3345451},
  url = {https://dl.acm.org/doi/10.1145/3300061.3345451},
  urldate = {2026-02-13},
  isbn = {978-1-4503-6169-9},
}

@inproceedings{varshneyTunnelEmitterTunnel2020,
  title = {Tunnel Emitter: Tunnel Diode Based Low-Power Carrier Emitters for Backscatter Tags},
  shorttitle = {Tunnel Emitter},
  booktitle = {Proceedings of the 26th {{Annual International Conference}} on {{Mobile Computing}} and {{Networking}}},
  author = {Varshney, Ambuj and Corneo, Lorenzo},
  date = {2020-09-18},
  series = {{{MobiCom}} '20},
  pages = {1--14},
  publisher = {Association for Computing Machinery},
  location = {New York, NY, USA},
  doi = {10.1145/3372224.3419199},
  url = {https://dl.acm.org/doi/10.1145/3372224.3419199},
  urldate = {2026-02-13},
  isbn = {978-1-4503-7085-1},
}

@article{baiSubLoRaHighThroughputLoRa2026,
  title = {{{SubLoRa}}: {{High-Throughput LoRa Backscatter Communication}}},
  shorttitle = {{{SubLoRa}}},
  author = {Bai, Jingyi and Du, Caihui and Yu, Jihong and Ren, Ju and Zhang, Rongrong and Yao, Haipeng},
  date = {2026},
  journaltitle = {IEEE Transactions on Mobile Computing},
  pages = {1--13},
  issn = {1558-0660},
  doi = {10.1109/TMC.2026.3655394},
  url = {https://ieeexplore.ieee.org/abstract/document/11358788},
  urldate = {2026-02-13},
}

@inproceedings{liCPLoRaParallelLoRa2025,
  title = {{{CPLoRa}}: {{Parallel LoRa Backscatter Communications Compatible}} with {{Commodity LoRa Receivers}}},
  shorttitle = {{{CPLoRa}}},
  booktitle = {2025 {{IEEE}} 102nd {{Vehicular Technology Conference}} ({{VTC2025-Fall}})},
  author = {Li, Zelong and Gong, Shimin and Li, Lanhua and Lyu, Bin and Li, Feng and Niyato, Dusit},
  date = {2025-10},
  pages = {1--5},
  issn = {2577-2465},
  doi = {10.1109/VTC2025-Fall65116.2025.11310535},
  url = {https://ieeexplore.ieee.org/abstract/document/11310535},
  urldate = {2026-02-13},
  eventtitle = {2025 {{IEEE}} 102nd {{Vehicular Technology Conference}} ({{VTC2025-Fall}})},
}

@article{belo2019iq,
	title        = {{IQ impedance modulator front-end for low-power LoRa backscattering devices}},
	author       = {Belo, Daniel and Correia, Ricardo and Ding, Yuan and Daskalakis, Spyridon Nektarios and Goussetis, George and Georgiadis, Apostolos and Carvalho, Nuno Borges},
	year         = {2019},
	journal      = {IEEE Transactions on Microwave Theory and Techniques},
	publisher    = {IEEE},
	volume       = {67},
	number       = {12},
	pages        = {5307--5314}
}

@article{bharadia2015backfi,
	title        = {{Backfi: High throughput wifi backscatter}},
	author       = {Bharadia, Dinesh and Joshi, Kiran Raj and Kotaru, Manikanta and Katti, Sachin},
	year         = {2015},
	journal      = {ACM SIGCOMM Computer Communication Review},
	publisher    = {ACM New York, NY, USA},
	volume       = {45},
	number       = {4},
	pages        = {283--296}
}

@article{callebaut2021art,
	title        = {{The art of designing remote iot devices--technologies and strategies for a long battery life}},
	author       = {Callebaut, Gilles and Leenders, Guus and Van Mulders, Jarne and Ottoy, Geoffrey and De Strycker, Lieven and Van der Perre, Liesbet},
	year         = {2021},
	journal      = {Sensors},
	publisher    = {MDPI},
	volume       = {21},
	number       = {3},
	pages        = {913}
}

@article{daskalakis2022new,
	title        = {{The New Era of Long-Range ``Zero-Interception'' Ambient Backscattering Systems: 130 m with 130 nA Front-End Consumption}},
	author       = {Daskalakis, Spyridon Nektarios and Georgiadis, Apostolos and Tentzeris, Manos M and Goussetis, George and Deligeorgis, George},
	year         = {2022},
	journal      = {Sensors},
	publisher    = {MDPI},
	volume       = {22},
	number       = {11},
	pages        = {4151}
}

@article{ding2020harmonic,
	title        = {{Harmonic suppression in frequency shifted backscatter communications}},
	author       = {Ding, Yuan and Lihakanga, Romwald and Correia, Ricardo and Goussetis, George and Carvalho, Nuno Borges},
	year         = {2020},
	journal      = {IEEE Open Journal of the Communications Society},
	publisher    = {IEEE},
	volume       = {1},
	pages        = {990--999}
}

@article{gulati2018capharvester,
	title        = {{CapHarvester: A stick-on capacitive energy harvester using stray electric field from AC power lines}},
	author       = {Gulati, Manoj and Parizi, Farshid Salemi and Whitmire, Eric and Gupta, Sidhant and Ram, Shobha Sundar and Singh, Amarjeet and Patel, Shwetak N},
	year         = {2018},
	journal      = {Proceedings of the ACM on Interactive, Mobile, Wearable and Ubiquitous Technologies},
	publisher    = {ACM New York, NY, USA},
	volume       = {2},
	number       = {3},
	pages        = {1--20}
}

@inproceedings{guo2020aloba,
	title        = {{Aloba: Rethinking ON-OFF keying modulation for ambient LoRa backscatter}},
	author       = {Guo, Xiuzhen and Shangguan, Longfei and He, Yuan and Zhang, Jia and Jiang, Haotian and Siddiqi, Awais Ahmad and Liu, Yunhao},
	year         = {2020},
	booktitle    = {Proceedings of the 18th conference on embedded networked sensor systems},
	pages        = {192--204}
}

@inproceedings{guo2022rf,
	title        = {{RF-transformer: A unified backscatter radio hardware abstraction}},
	author       = {Guo, Xiuzhen and He, Yuan and Yu, Zihao and Zhang, Jiacheng and Liu, Yunhao and Shangguan, Longfei},
	year         = {2022},
	booktitle    = {Proceedings of the 28th annual international conference on mobile computing and networking},
	pages        = {446--458}
}

@inproceedings{guo2022saiyan,
	title        = {{Saiyan: Design and implementation of a low-power demodulator for $\{$LoRa$\}$ backscatter systems}},
	author       = {Guo, Xiuzhen and Shangguan, Longfei and He, Yuan and Jing, Nan and Zhang, Jiacheng and Jiang, Haotian and Liu, Yunhao},
	year         = {2022},
	booktitle    = {19th USENIX Symposium on Networked Systems Design and Implementation (NSDI 22)},
	pages        = {437--451}
}

@article{guo2024low,
  title={A low-power demodulator for LoRa backscatter systems with frequency-amplitude transformation},
  author={Guo, Xiuzhen and He, Yuan and Nan, Jing and Zhang, Jiacheng and Liu, Yunhao and Shangguan, Longfei},
  journal={IEEE/ACM Transactions on Networking},
  year={2024},
  publisher={IEEE}
}

@inproceedings{hessar2019netscatter,
	title        = {{$\{$NetScatter$\}$: Enabling $\{$Large-Scale$\}$ Backscatter Networks}},
	author       = {Hessar, Mehrdad and Najafi, Ali and Gollakota, Shyamnath},
	year         = {2019},
	booktitle    = {16th USENIX Symposium on Networked Systems Design and Implementation (NSDI 19)},
	pages        = {271--284}
}

@inproceedings{jakhar2023environmental,
	title        = {{The Environmental Impact of Electronic Waste: A Bibliometric Analysis}},
	author       = {Jakhar, Sunil Kumar and Batra, Raman and SB, Vinay Kumar},
	year         = {2023},
	booktitle    = {2023 International Conference on Power Energy, Environment \& Intelligent Control (PEEIC)},
	pages        = {1349--1353},
	organization = {IEEE}
}

@inproceedings{jiang2021long,
	title        = {{Long-range ambient LoRa backscatter with parallel decoding}},
	author       = {Jiang, Jinyan and Xu, Zhenqiang and Dang, Fan and Wang, Jiliang},
	year         = {2021},
	booktitle    = {Proceedings of the 27th Annual International Conference on Mobile Computing and Networking},
	pages        = {684--696}
}

@inproceedings{katanbaf2021simplifying,
	title        = {{Simplifying backscatter deployment:$\{$Full-Duplex$\}$$\{$LoRa$\}$ backscatter}},
	author       = {Katanbaf, Mohamad and Weinand, Anthony and Talla, Vamsi},
	year         = {2021},
	booktitle    = {18th USENIX Symposium on Networked Systems Design and Implementation (NSDI 21)},
	pages        = {955--972}
}

@article{la2019strategies,
	title        = {{Strategies and techniques for powering wireless sensor nodes through energy harvesting and wireless power transfer}},
	author       = {La Rosa, Roberto and Livreri, Patrizia and Trigona, Carlo and Di Donato, Loreto and Sorbello, Gino},
	year         = {2019},
	journal      = {Sensors},
	publisher    = {MDPI},
	volume       = {19},
	number       = {12},
	pages        = {2660}
}

@article{leenders2023energy,
	title        = {{An Energy-Efficient LoRa Multi-Hop Protocol through Preamble Sampling for Remote Sensing}},
	author       = {Leenders, Guus and Callebaut, Gilles and Ottoy, Geoffrey and Van der Perre, Liesbet and De Strycker, Lieven},
	year         = {2023},
	journal      = {Sensors},
	publisher    = {MDPI},
	volume       = {23},
	number       = {11},
	pages        = {4994}
}

@inproceedings{li2022curvinglora,
	title        = {{$\{$CurvingLoRa$\}$ to boost $\{$LoRa$\}$ network throughput via concurrent transmission}},
	author       = {Li, Chenning and Guo, Xiuzhen and Shangguan, Longfei and Cao, Zhichao and Jamieson, Kyle},
	year         = {2022},
	booktitle    = {19th USENIX Symposium on Networked Systems Design and Implementation (NSDI 22)},
	pages        = {879--895}
}

@inproceedings{lin2024harmonic,
	title        = {{Harmonic Long-Range Backscatter with Frequency-Shifted Lightweight Tag}},
	author       = {Lin, Junliang and Zhang, Xiannan and Xu, Rongtao and Wang, Gongpu and Quek, Tony QS},
	year         = {2024},
	booktitle    = {2024 IEEE 100th Vehicular Technology Conference (VTC2024-Fall)},
	pages        = {1--5},
	organization = {IEEE}
}

@article{meena2024green,
	title        = {{Green supercapacitors: review and perspectives on sustainable template-free synthesis of metal and metal oxide nanoparticles}},
	author       = {Meena, Jayaprakash and shankari Sivasubramaniam, Shapna and David, Ezhumalai and others},
	year         = {2024},
	journal      = {RSC Sustainability},
	publisher    = {Royal Society of Chemistry},
	volume       = {2},
	number       = {5},
	pages        = {1224--1245}
}

@inproceedings{peng2018plora,
	title        = {{PLoRa: A passive long-range data network from ambient LoRa transmissions}},
	author       = {Peng, Yao and Shangguan, Longfei and Hu, Yue and Qian, Yujie and Lin, Xianshang and Chen, Xiaojiang and Fang, Dingyi and Jamieson, Kyle},
	year         = {2018},
	booktitle    = {Proceedings of the 2018 conference of the ACM special interest group on data communication},
	pages        = {147--160}
}

@article{peng2022ambient,
	title        = {{Ambient LoRa backscatter system with chirp interval modulation}},
	author       = {Peng, Yuxiang and He, Shiyue and Zhang, Yu and Niu, Zhiang and Xiao, Lixia and Jiang, Tao},
	year         = {2022},
	journal      = {IEEE Transactions on Wireless Communications},
	publisher    = {IEEE},
	volume       = {22},
	number       = {2},
	pages        = {1328--1342}
}

@ARTICLE{10155565,
  author={Kaplan, Ahmet and Vieira, Joao and Larsson, Erik G.},
  journal={IEEE Transactions on Wireless Communications}, 
  title={Direct Link Interference Suppression for Bistatic Backscatter Communication in Distributed MIMO}, 
  year={2024},
  volume={23},
  number={2},
  pages={1024-1036},
  doi={10.1109/TWC.2023.3285250}}

@inproceedings{ren2023prism,
	title        = {{Prism: High-throughput LoRa backscatter with non-linear chirps}},
	author       = {Ren, Yidong and Cai, Puyu and Jiang, Jinyan and Du, Jialuo and Cao, Zhichao},
	year         = {2023},
	booktitle    = {IEEE INFOCOM 2023-IEEE Conference on Computer Communications},
	pages        = {1--10},
	organization = {IEEE}
}

@article{talla2017lora,
	title        = {{Lora backscatter: Enabling the vision of ubiquitous connectivity}},
	author       = {Talla, Vamsi and Hessar, Mehrdad and Kellogg, Bryce and Najafi, Ali and Smith, Joshua R and Gollakota, Shyamnath},
	year         = {2017},
	journal      = {Proceedings of the ACM on interactive, mobile, wearable and ubiquitous technologies},
	publisher    = {ACM New York, NY, USA},
	volume       = {1},
	number       = {3},
	pages        = {1--24}
}

@article{tang2021self,
	title        = {{Self-sustainable long-range backscattering communication using RF energy harvesting}},
	author       = {Tang, Xiaoqing and Xie, Guihui and Cui, Yongqiang},
	year         = {2021},
	journal      = {IEEE Internet of Things Journal},
	publisher    = {IEEE},
	volume       = {8},
	number       = {17},
	pages        = {13737--13749}
}

@inproceedings{varshney2017lorea,
	title        = {{LoRea: A backscatter architecture that achieves a long communication range}},
	author       = {Varshney, Ambuj and Harms, Oliver and P{\'e}rez-Penichet, Carlos and Rohner, Christian and Hermans, Frederik and Voigt, Thiemo},
	year         = {2017},
	booktitle    = {Proceedings of the 15th ACM Conference on Embedded Network Sensor Systems},
	pages        = {1--14}
}

@article{wang2022allspark,
	title        = {{AllSpark: Enabling long-range backscatter for vehicle-to-infrastructure communication}},
	author       = {Wang, Xuan and Kou, Xin and Li, Haoyu and Wang, Fuwei and Fang, Dingyi and Ma, Yunfei and Chen, Xiaojiang},
	year         = {2022},
	journal      = {IEEE Internet of Things Journal},
	publisher    = {IEEE},
	volume       = {9},
	number       = {24},
	pages        = {25525--25537}
}

@book{un2024prospects,
   author = "United Nations Department of Economic and Social Affairs",pages = "",
   title = "World Population Prospects 2024",
   publisher = "United Nations",
   year = "2024",
   url = "https://www.un-ilibrary.org/content/books/9789211065138",
}

@misc{balde2024ewastemonitor,
  title = {The Global E-waste Monitor 2024},
  author = {Baldé, Cornelis P. and Kuehr, Ruediger and Yamamoto, Tales and McDonald, Rosie and D’Angelo, Elena and Althaf, Shahana and Bel, Garam and Deubzer, Otmar and Fernandez-Cubillo, Elena and Forti, Vanessa and Gray, Vanessa and Herat, Sunil and Honda, Shunichi and Iattoni, Giulia and Khetriwal, Deepali S. and Luda di Cortemiglia, Vittoria and Lobuntsova, Yuliya and Nnorom, Innocent and Pralat, Noémie and Wagner Michelle},
  year = {2024},
  organization = {International Telecommunication Union (ITU) and United Nations Institute for Training and Research (UNITAR)},
  url = {https://ewastemonitor.info/wp-content/uploads/2024/12/GEM_2024_EN_11_NOV-web.pdf},
  note = {Accessed 15 May 2025},
}

@book{kaza2018waste,
  title={What a waste 2.0: a global snapshot of solid waste management to 2050},
  author={Kaza, Silpa and Yao, Lisa and Bhada-Tata, Perinaz and Van Woerden, Frank},
  year={2018},
  publisher={World Bank Publications}
}

@article{ashry2009compact,
  title={A compact low-power UHF RFID tag},
  author={Ashry, Ahmed and Sharaf, Khaled and Ibrahim, Magdi},
  journal={Microelectronics Journal},
  volume={40},
  number={11},
  pages={1504--1513},
  year={2009},
  publisher={Elsevier}
}

@article{shen2012design,
  title={Design and implementation of an ultra-low power passive UHF RFID tag},
  author={Shen, Jinpeng and Wang, Xin'an and Liu, Shan and Zong, Hongqiang and Huang, Jinfeng and Yang, Xin and Feng, Xiaoxing and Ge, Binjie},
  journal={Journal of Semiconductors},
  volume={33},
  number={11},
  pages={115011},
  year={2012},
  publisher={IOP Publishing}
}

@article{lazaro2025home,
  title={Home surveillance system based on LoRa backscattering},
  author={Lazaro, Marc and Lazaro, Antonio and Villarino, Ramon and Girbau, David},
  journal={Scientific Reports},
  volume={15},
  number={1},
  pages={12063},
  year={2025},
  publisher={Nature Publishing Group UK London}
}

@techreport{balde2022weee,
  title        = {Update of WEEE Collection Rates, Targets, Flows, and Hoarding –- 2021 in the EU-27, United Kingdom, Norway, Switzerland, and Iceland},
  author       = {C.P. Baldé and G. Iattoni and C. Xu and T. Yamamoto},
  year         = {2022},
  institution  = {United Nations Institute for Training and Research (UNITAR), SCYCLE Programme},
  address      = {Bonn, Germany},
  url          = {https://weee-forum.org/wp-content/uploads/2022/12/Update-of-WEEE-Collection_web_final_nov_29.pdf},
  note         = {Study conducted in collaboration with the WEEE Forum}
}

@article{callebaut2019characterization,
  title={Characterization of LoRa point-to-point path loss: Measurement campaigns and modeling considering censored data},
  author={Callebaut, Gilles and Van der Perre, Liesbet},
  journal={IEEE Internet of Things Journal},
  volume={7},
  number={3},
  pages={1910--1918},
  year={2019},
  publisher={IEEE}
}

@article{guo2024mighty,
  title={Mighty: Towards Long-Range and High-Throughput Backscatter for Drones},
  author={Guo, Xiuzhen and He, Yuan and Shangguan, Longfei and Chen, Yande and Gu, Chaojie and Shu, Yuanchao and Jamieson, Kyle and Chen, Jiming},
  journal={IEEE Transactions on Mobile Computing},
  year={2024},
  publisher={IEEE}
}

@article{song2021exploiting,
  title={Exploiting channel polarization for reliable wide-area backscatter networks},
  author={Song, Guochao and Wang, Wei and Yang, Hang and Zhang, Dongchen and Gao, Peng and Jiang, Tao},
  journal={IEEE Transactions on Mobile Computing},
  volume={21},
  number={12},
  pages={4338--4351},
  year={2021},
  publisher={IEEE}
}

@inproceedings{liang2019drone,
  title={Drone fleet deployment strategy for large scale agriculture and forestry surveying},
  author={Liang, Man and Delahaye, Daniel},
  booktitle={2019 IEEE Intelligent Transportation Systems Conference (ITSC)},
  pages={4495--4500},
  year={2019},
  organization={IEEE}
}

@article{thilakarathne2025internet,
  title={Internet of things enabled smart agriculture: Current status, latest advancements, challenges and countermeasures},
  author={Thilakarathne, Navod Neranjan and Bakar, Muhammad Saifullah Abu and Abas, Pg Emeroylariffion and Yassin, Hayati},
  journal={Heliyon},
  volume={11},
  number={3},
  year={2025},
  publisher={Elsevier}
}

@techreport{EN12464-1,
  title = {Light and lighting – Lighting of workplaces – Part 1: Indoor workplaces},
  type = {Standard},
  institution = {CEN},
  key = {EN 12464-1},
  year = {2021},
  url = {https://standards.iteh.ai/catalog/standards/cen/53fc4ff7-e7df-4ebd-a730-0d5f0ea888e0/en-12464-1-2021},
}

@article{raj2018power,
  title={Power sources for the internet of things},
  author={Raj, Abhi and Steingart, Dan},
  journal={Journal of The Electrochemical Society},
  volume={165},
  number={8},
  pages={B3130},
  year={2018},
  publisher={IOP Publishing}
}

@misc{kemet_fc_series_2023,
  title={FC Series Supercapacitors Datasheet},
  author={{KEMET Electronics Corporation}},
  year={2023},
  url={https://content.kemet.com/datasheets/KEM_S6011_FC.pdf},
  note={Document Number: KEM\_S6011\_FC},
  howpublished={\url{https://content.kemet.com/datasheets/KEM_S6011_FC.pdf}},
  institution={KEMET Electronics Corporation}
}

@article{casella2022evolution,
  title={The evolution of RFID technology in the logistics field: a review},
  author={Casella, Giorgia and Bigliardi, Barbara and Bottani, Eleonora},
  journal={Procedia Computer Science},
  volume={200},
  pages={1582--1592},
  year={2022},
  publisher={Elsevier}
}

@article{cappelle2024iot,
  title={IoT on the Road to Sustainability: Vehicle or Bandit?},
  author={Cappelle, Jona and Van der Perre, Liesbet and Fitzgerald, Emma and Ravyts, Simon and Gajda, Weronika and De Smedt, Valentijn and Cox, Bert and Callebaut, Gilles},
  journal={arXiv preprint arXiv:2405.20706},
  year={2024}
}

@inproceedings{hou2024lobaca,
  title={LoBaCa: Super-Resolution LoRa Backscatter Localization for Low-Cost Tags},
  author={Hou, Boxin and Wang, Jiliang},
  booktitle={IEEE INFOCOM 2024-IEEE Conference on Computer Communications},
  pages={1081--1090},
  year={2024},
  organization={IEEE}
}

@inproceedings{liu2024lomu,
  title={LoMu: Enable Long-Range Multi-Target Backscatter Sensing for Low-Cost Tags},
  author={Liu, Yihao and Jiang, Jinyan and Wang, Jiliang},
  booktitle={IEEE INFOCOM 2024-IEEE Conference on Computer Communications},
  pages={2069--2078},
  year={2024},
  organization={IEEE}
}

@article{liu2024enable,
  title={Enable Practical Long-Range Multi-Target Backscatter Sensing},
  author={Liu, Yihao and Jiang, Jinyan and Zhao, Jumin and Wang, Jiliang},
  journal={IEEE Transactions on Mobile Computing},
  year={2024},
  publisher={IEEE}
}

@inproceedings{jiang2023locra,
  title={LocRa: Enable practical long-range backscatter localization for low-cost tags},
  author={Jiang, Jinyan and Wang, Jiliang and Chen, Yijie and Liu, Yihao and Liu, Yunhao},
  booktitle={Proceedings of the 21st Annual International Conference on Mobile Systems, Applications and Services},
  pages={317--329},
  year={2023}
}

@inproceedings{li2020xorlora,
  title={XORLoRa: LoRa backscatter communication with commodity devices},
  author={Li, Hao and Tong, Xinyu and Li, Qianru and Tian, Xiaohua},
  booktitle={2020 IEEE 6th International Conference on Computer and Communications (ICCC)},
  pages={706--711},
  year={2020},
  organization={IEEE}
}

@inproceedings{zhang2017freerider,
  title={Freerider: Backscatter communication using commodity radios},
  author={Zhang, Pengyu and Josephson, Colleen and Bharadia, Dinesh and Katti, Sachin},
  booktitle={Proceedings of the 13th international conference on emerging networking experiments and technologies},
  pages={389--401},
  year={2017}
}

@inproceedings{zhang2016hitchhike,
  title={Hitchhike: Practical backscatter using commodity wifi},
  author={Zhang, Pengyu and Bharadia, Dinesh and Joshi, Kiran and Katti, Sachin},
  booktitle={Proceedings of the 14th ACM conference on embedded network sensor systems CD-ROM},
  pages={259--271},
  year={2016}
}

@article{joachim2019complete,
  title={Complete reverse engineering of LoRa PHY},
  author={Joachim, Tapparel},
  journal={Reverse\_Eng\_Report. pdf},
  year={2019}
}

@inproceedings{huang2021freeback,
  title={Freeback: Blind and distributed rate adaptation in LoRa-based backscatter networks},
  author={Huang, Gang and Yang, Panlong and Zhou, Hao and Yan, Yubo and He, Xin and Li, Xiangyang},
  booktitle={2021 IEEE Wireless Communications and Networking Conference (WCNC)},
  pages={1--6},
  year={2021},
  organization={IEEE}
}

@article{zhu2025inductor,
  title={Inductor-Free LoRa Backscatter},
  author={Zhu, Fengyuan and He, Jiaquan and Lin, Jiajun and Qin, Yifan and Wang, Bingbing and Di, Qilong and Jin, Meng and Tian, Xiaohua},
  journal={IEEE Transactions on Networking},
  year={2025},
  publisher={IEEE}
}

@inproceedings{juels2005strengthening,
  title={Strengthening EPC tags against cloning},
  author={Juels, Ari},
  booktitle={Proceedings of the 4th ACM workshop on Wireless security},
  pages={67--76},
  year={2005}
}

@article{zheng2018life,
  title={Life cycle assessment (LCA) for printed electronics},
  author={Zheng, Li-Rong and Tenhunen, Hannu and Zou, Zhuo},
  year={2018},
  publisher={Wiley Semiconductors}
}

@article{kim2023lora,
  title={LoRa Backscatter Network Efficient Data Transmission Using RF Source Range Control.},
  author={Kim, Dae-Young and Lee, SoYeon and Kim, Seokhoon},
  journal={Computers, Materials \& Continua},
  volume={74},
  number={2},
  year={2023}
}

@article{lazaro2021room,
  title={Room-level localization system based on LoRa backscatters},
  author={Lazaro, Antonio and Lazaro, Marc and Villarino, Ramon},
  journal={Ieee Access},
  volume={9},
  pages={16004--16018},
  year={2021},
  publisher={IEEE}
}

@inproceedings{rostami2018polymorphic,
  title={Polymorphic radios: A new design paradigm for ultra-low power communication},
  author={Rostami, Mohammad and Gummeson, Jeremy and Kiaghadi, Ali and Ganesan, Deepak},
  booktitle={Proceedings of the 2018 Conference of the ACM Special Interest Group on Data Communication},
  pages={446--460},
  year={2018}
}

@article{jiang2023backscatter,
  title={Backscatter communication meets practical battery-free Internet of Things: A survey and outlook},
  author={Jiang, Tao and Zhang, Yu and Ma, Wenyuan and Peng, Miaoran and Peng, Yuxiang and Feng, Mingjie and Liu, Guanghua},
  journal={IEEE Communications Surveys \& Tutorials},
  volume={25},
  number={3},
  pages={2021--2051},
  year={2023},
  publisher={IEEE}
}

@inproceedings{jiang2021sense,
  title={Sense me on the ride: Accurate mobile sensing over a LoRa backscatter channel},
  author={Jiang, Haotian and Zhang, Jiacheng and Guo, Xiuzhen and He, Yuan},
  booktitle={Proceedings of the 19th ACM Conference on Embedded Networked Sensor Systems},
  pages={125--137},
  year={2021}
}

@article{morin2017comparison,
  title={Comparison of the device lifetime in wireless networks for the internet of things},
  author={Morin, Elodie and Maman, Mickael and Guizzetti, Roberto and Duda, Andrzej},
  journal={IEEE Access},
  volume={5},
  pages={7097--7114},
  year={2017},
  publisher={IEEE}
}

@article{singh2020energy,
  title={Energy consumption analysis of LPWAN technologies and lifetime estimation for IoT application},
  author={Singh, Ritesh Kumar and Puluckul, Priyesh Pappinisseri and Berkvens, Rafael and Weyn, Maarten},
  journal={Sensors},
  volume={20},
  number={17},
  pages={4794},
  year={2020},
  publisher={MDPI}
}

@article{modarress2022threats,
  title={Threats of internet-of-thing on environmental sustainability by E-waste},
  author={Modarress Fathi, Batoul and Ansari, Alexander and Ansari, Al},
  journal={Sustainability},
  volume={14},
  number={16},
  pages={10161},
  year={2022},
  publisher={MDPI}
}

@inproceedings{rossi2017energy,
  title={Energy neutral design of an IoT system for pollution monitoring},
  author={Rossi, Maurizio and Tosato, Pietro},
  booktitle={2017 IEEE Workshop on Environmental, Energy, and Structural Monitoring Systems (EESMS)},
  pages={1--6},
  year={2017},
  organization={IEEE}
}

@article{brunelli2019energy,
  title={Energy neutral machine learning based iot device for pest detection in precision agriculture},
  author={Brunelli, Davide and Albanese, Andrea and d'Acunto, Donato and Nardello, Matteo},
  journal={IEEE Internet of Things Magazine},
  volume={2},
  number={4},
  pages={10--13},
  year={2019},
  publisher={IEEE}
}

@article{brauner2009novel,
  title={A novel carrier suppression method for RFID},
  author={Brauner, Thomas and Zhao, Xiongwen},
  journal={IEEE Microwave and Wireless Components Letters},
  volume={19},
  number={3},
  pages={128--130},
  year={2009},
  publisher={IEEE}
}

@inproceedings{nandakumar3DLocalizationSubCentimeter2018,
  title = {{{3D Localization}} for {{Sub-Centimeter Sized Devices}}},
  booktitle = {Proceedings of the 16th {{ACM Conference}} on {{Embedded Networked Sensor Systems}}},
  author = {Nandakumar, Rajalakshmi and Iyer, Vikram and Gollakota, Shyamnath},
  year = {2018},
  month = nov,
  series = {{{SenSys}} '18},
  pages = {108--119},
  publisher = {Association for Computing Machinery},
  address = {New York, NY, USA},
  doi = {10.1145/3274783.3274851},
  urldate = {2025-09-23},
  isbn = {978-1-4503-5952-8},
}

@incollection{bansalOwLL2021,
  title = {{{OwLL}}},
  booktitle = {Proceedings of the 20th {{International Conference}} on {{Information Processing}} in {{Sensor Networks}} (Co-Located with {{CPS-IoT Week}} 2021)},
  author = {Bansal, Atul and {View Profile} and Gadre, Akshay and {View Profile} and Singh, Vaibhav and {View Profile} and Rowe, Anthony and {View Profile} and Iannucci, Bob and {View Profile} and Kumar, Swarun and {View Profile}},
  year = {2021},
  month = may,
  series = {{{ACM Conferences}}},
  pages = {148--162},
  doi = {10.1145/3412382.3458263},
  urldate = {2025-09-23},
  isbn = {978-1-4503-8098-0},
}

@misc{abraconADCM-S07R5S,
  author       = {Abracon LLC},
  title        = {ADCM-S07R5S Supercapacitor Module Datasheet},
  year         = {2023},
  url          = {https://abracon.com/datasheets/ADCM-S07R5S.pdf},
  note         = {Accessed: 2025-10-03}
}

@misc{abracon2021ahcr,
  author       = {Abracon LLC},
  title        = {AHCR-S04R0S Series Lithium-Ion Capacitor Datasheet},
  year         = {2021},
  url          = {https://abracon.com/datasheets/AHCR-S04R0S.pdf},
  note         = {Accessed: 2025-10-08}
}

@article{liNovelLoadFreeSSB2025,
  title = {Novel {{Load-Free SSB Modulation}} for {{LoRa Backscatter Communication}}},
  author = {Li, Jianing and Yin, Guosheng and Wang, Jialong and Feng, Man and Chen, Hao and Xu, Wei and Lu, Weibing},
  year = {2025},
  journal = {IEEE Wireless Communications Letters},
  pages = {1--1},
  issn = {2162-2345},
  doi = {10.1109/LWC.2025.3610688},
  urldate = {2025-10-07},
}

@inproceedings{steinmann2025,
  title = {Development of an {{Ultra-Low-Power Bidirectional LoRa Backscatter Tag}}},
  booktitle = {2025 {{IEEE International Instrumentation}} and {{Measurement Technology Conference}} ({{I2MTC}})},
  author = {Steinmann, Till and Riedel, Frederik and Schaechtle, Thomas and Rupitsch, Stefan Johann},
  year = {2025},
  month = may,
  pages = {1--6},
  issn = {2642-2077},
  doi = {10.1109/I2MTC62753.2025.11078945},
  urldate = {2025-10-07},
}

@inproceedings{shiRateenhancedSquarewaveLoRa2024,
  title = {A {{Rate-enhanced Square-wave LoRa Backscatter Communication Scheme}}},
  booktitle = {2024 {{International Symposium}} on {{Antennas}} and {{Propagation}} ({{ISAP}})},
  author = {Shi, Wenxin and Su, Ming and Xu, Rui and Liu, Yuanan},
  year = {2024},
  month = nov,
  pages = {1--2},
  issn = {2995-987X},
  doi = {10.1109/ISAP62502.2024.10846241},
  urldate = {2025-10-10},
}

@article{guoEnablingCrossBandBackscatter2025,
  title = {Enabling {{Cross-Band Backscatter Communication With Twaltz}}},
  author = {Guo, Xiuzhen and Liu, Boya and Jing, Nan and Gu, Chaojie and Shu, Yuanchao and Chen, Jiming},
  year = {2025},
  month = nov,
  journal = {IEEE Transactions on Mobile Computing},
  volume = {24},
  number = {11},
  pages = {11323--11336},
  issn = {1558-0660},
  doi = {10.1109/TMC.2025.3581900},
  urldate = {2025-10-10},
}

@inproceedings{renAeroEchoAgriculturalLowpower2025,
  title = {{{AeroEcho}}: {{Towards Agricultural Low-power Wide-area Backscatter}} with {{Aerial Excitation Source}}},
  shorttitle = {{{AeroEcho}}},
  booktitle = {{{IEEE INFOCOM}} 2025 - {{IEEE Conference}} on {{Computer Communications}}},
  author = {Ren, Yidong and Li, Gen and Liu, Yimeng and Dong, Younsuk and Cao, Zhichao},
  year = {2025},
  month = may,
  pages = {1--10},
  issn = {2641-9874},
  doi = {10.1109/INFOCOM55648.2025.11044614},
  urldate = {2025-10-10},
}

@article{tangPrototypeImplementationExperimental2025,
  title = {Prototype {{Implementation}} and {{Experimental Evaluation}} for {{LoRa-Backscatter Communication Systems With RF Energy Harvesting}} and {{Low Power Management}}},
  author = {Tang, Xiaoqing and Liu, Xin and Xie, Guihui and Cui, Yongqiang and Li, Dong},
  year = {2025},
  month = jul,
  journal = {IEEE Transactions on Communications},
  volume = {73},
  number = {7},
  pages = {4811--4825},
  issn = {1558-0857},
  doi = {10.1109/TCOMM.2024.3522052},
  urldate = {2025-10-10},
}

@article{zhangAuthScatterAccurateRobust2025,
  title = {{{AuthScatter}}: {{Accurate}}, {{Robust}}, and {{Scalable Mutual Authentication}} in {{Physical Layer}} for {{Backscatter Communications}}},
  shorttitle = {{{AuthScatter}}},
  author = {Zhang, Yifan and Xie, Boxuan and Yang, Yishan and Yan, Zheng and Jäntti, Riku and Han, Zhu},
  date = {2025},
  journaltitle = {IEEE Transactions on Information Forensics and Security},
  volume = {20},
  pages = {6937--6952},
  issn = {1556-6021},
  doi = {10.1109/TIFS.2025.3585453},
  url = {https://ieeexplore.ieee.org/abstract/document/11071870},
  urldate = {2026-02-18},
}

@inproceedings{menon2023wisp,
  title = {Wireless {{Identification}} and {{Sensing Platform Version}} 6.0},
  booktitle = {Proceedings of the 20th {{ACM Conference}} on {{Embedded Networked Sensor Systems}}},
  author = {Menon, Rohan and Gujarathi, Rohit and Saffari, Ali and Smith, Joshua R.},
  date = {2023-01-24},
  series = {{{SenSys}} '22},
  pages = {899--905},
  publisher = {Association for Computing Machinery},
  location = {New York, NY, USA},
  doi = {10.1145/3560905.3568109},
  url = {https://dl.acm.org/doi/10.1145/3560905.3568109},
  urldate = {2026-03-03},
  isbn = {978-1-4503-9886-2},
}

@article{karthaus2003rfid,
  title = {Fully Integrated Passive {{UHF RFID}} Transponder {{IC}} with 16.7-{{\si{\micro\watt}}} Minimum {{RF}} Input Power},
  author = {Karthaus, U. and Fischer, M.},
  date = {2003-10},
  journaltitle = {IEEE Journal of Solid-State Circuits},
  volume = {38},
  number = {10},
  pages = {1602--1608},
  issn = {1558-173X},
  doi = {10.1109/JSSC.2003.817249},
  url = {https://ieeexplore.ieee.org/document/1233745},
  urldate = {2026-03-10},
}

@inproceedings{correia2019chirp,
  title = {Chirp {{Based Backscatter Modulation}}},
  booktitle = {2019 {{IEEE MTT-S International Microwave Symposium}} ({{IMS}})},
  author = {Correia, Ricardo and Ding, Yuan and Daskalakis, Spyridon Nektarios and Petridis, Panagiotis and Goussetis, George and Georgiadis, Apostolos and Carvalho, Nuno Borges},
  date = {2019-06},
  pages = {279--282},
  issn = {2576-7216},
  doi = {10.1109/MWSYM.2019.8700913},
  url = {https://ieeexplore.ieee.org/abstract/document/8700913},
  urldate = {2026-03-10},
  eventtitle = {2019 {{IEEE MTT-S International Microwave Symposium}} ({{IMS}})},
}

@inproceedings{vanmuldersKeepingEnergyNeutralDevices2024,
  title = {Keeping {{Energy-Neutral Devices Operational}}: A {{Coherent Massive Beamforming Approach}}},
  shorttitle = {Keeping {{Energy-Neutral Devices Operational}}},
  booktitle = {2024 {{IEEE}} 25th {{International Workshop}} on {{Signal Processing Advances}} in {{Wireless Communications}} ({{SPAWC}})},
  author = {Van Mulders, Jarne and Cox, Bert and Deutschmann, Benjamin J. B. and Callebaut, Gilles and De Strycker, Lieven and Van der Perre, Liesbet},
  date = {2024-09},
  pages = {6--10},
  issn = {1948-3252},
  doi = {10.1109/SPAWC60668.2024.10694523},
  url = {https://ieeexplore.ieee.org/abstract/document/10694523},
  urldate = {2026-03-17},
  eventtitle = {2024 {{IEEE}} 25th {{International Workshop}} on {{Signal Processing Advances}} in {{Wireless Communications}} ({{SPAWC}})},
}

@misc{PanasonicAM1456,
  title        = {Amorphous Silicon Solar Cell (Indoor Use) AM-1456 Datasheet},
  author       = {{Panasonic Solar Amorton Co., Ltd.}},
  organization = {Panasonic Solar Amorton Co., Ltd.},
  year         = {2019},
  month        = apr,
  url          = {https://panasonic.net/electricworks/amorton/assets/pdf/spec_PDF/indoor/AM-1456.pdf},
  note         = {Product specification datasheet}
}

@misc{PanasonicAM1437,
  title        = {Amorphous Silicon Solar Cell (Indoor Use) AM-1437 Datasheet},
  author       = {{Panasonic Solar Amorton Co., Ltd.}},
  organization = {Panasonic Solar Amorton Co., Ltd.},
  year         = {2019},
  month        = apr,
  url          = {https://panasonic.net/electricworks/amorton/assets/pdf/spec_PDF/indoor/AM-1437.pdf},
  note         = {Product specification datasheet}
}

@techreport{Franklin_SolTantLeakage,
  author      = {Franklin, R. W.},
  title       = {Analysis of Solid Tantalum Capacitor Leakage Current},
  institution = {KYOCERA AVX Components Corporation},
  address     = {Paignton, Devon, United Kingdom},
  type        = {Technical Paper},
  year        = {},
  url         = {https://www.kyocera-avx.com/docs/techinfo/Tantalum-NiobiumCapacitors/soltant.pdf},
  note        = {Accessed: 2026-04-08}
}

@misc{KYOCERA_AVX_TAJ,
  author       = {{KYOCERA AVX}},
  title        = {TAJ Series -- Standard and Low Profile Tantalum Capacitors},
  howpublished = {\url{https://datasheets.kyocera-avx.com/TAJ.pdf}},
  year         = {},
  note         = {Datasheet, accessed 2026-04-08}
}

@misc{PanasonicAM5815,
  author       = {{Panasonic Solar Amorton Co., Ltd.}},
  title        = {{Amorphous Silicon Solar Cells Specification: Model AM-5815}},
  year         = {2019},
  month        = jul,
  howpublished = {Technical datasheet (PDF)},
  url          = {https://panasonic.net/electricworks/amorton/assets/pdf/spec_PDF/outdoor/AM-5815.pdf},
  note         = {Accessed April 13, 2026}
}

@article{parksTurbochargingAmbientBackscatter2014,
  title = {Turbocharging Ambient Backscatter Communication},
  author = {Parks, Aaron N. and Liu, Angli and Gollakota, Shyamnath and Smith, Joshua R.},
  date = {2014-08-17},
  journaltitle = {SIGCOMM Comput. Commun. Rev.},
  volume = {44},
  number = {4},
  pages = {619--630},
  issn = {0146-4833},
  doi = {10.1145/2740070.2626312},
  url = {https://dl.acm.org/doi/10.1145/2740070.2626312},
  urldate = {2026-04-14},
}

@inproceedings{trotterMultiantennaTechniquesEnabling2012a,
  title = {Multi-Antenna Techniques for Enabling Passive {{RFID}} Tags and Sensors at Microwave Frequencies},
  booktitle = {2012 {{IEEE International Conference}} on {{RFID}} ({{RFID}})},
  author = {Trotter, Matthew S. and Valenta, Christopher R. and Koo, Gregory A. and Marshall, Blake R. and Durgin, Gregory D.},
  date = {2012-04},
  pages = {1--7},
  issn = {2374-0221},
  doi = {10.1109/RFID.2012.6193051},
  url = {https://ieeexplore.ieee.org/document/6193051},
  urldate = {2026-04-14},
  eventtitle = {2012 {{IEEE International Conference}} on {{RFID}} ({{RFID}})}
}

@article{zhonghua2019carrier,
  title={Carrier extraction cancellation circuit in RFID reader for improving the Tx-to-Rx isolation},
  author={Zhonghua, Ma and Yanfeng, Jiang},
  journal={IET Circuits, Devices \& Systems},
  volume={13},
  number={5},
  pages={622--629},
  year={2019},
  publisher={Wiley Online Library}
}

  \begin{IEEEbiography}
    [{\includegraphics[width=1in,height=1.25in,clip,keepaspectratio]{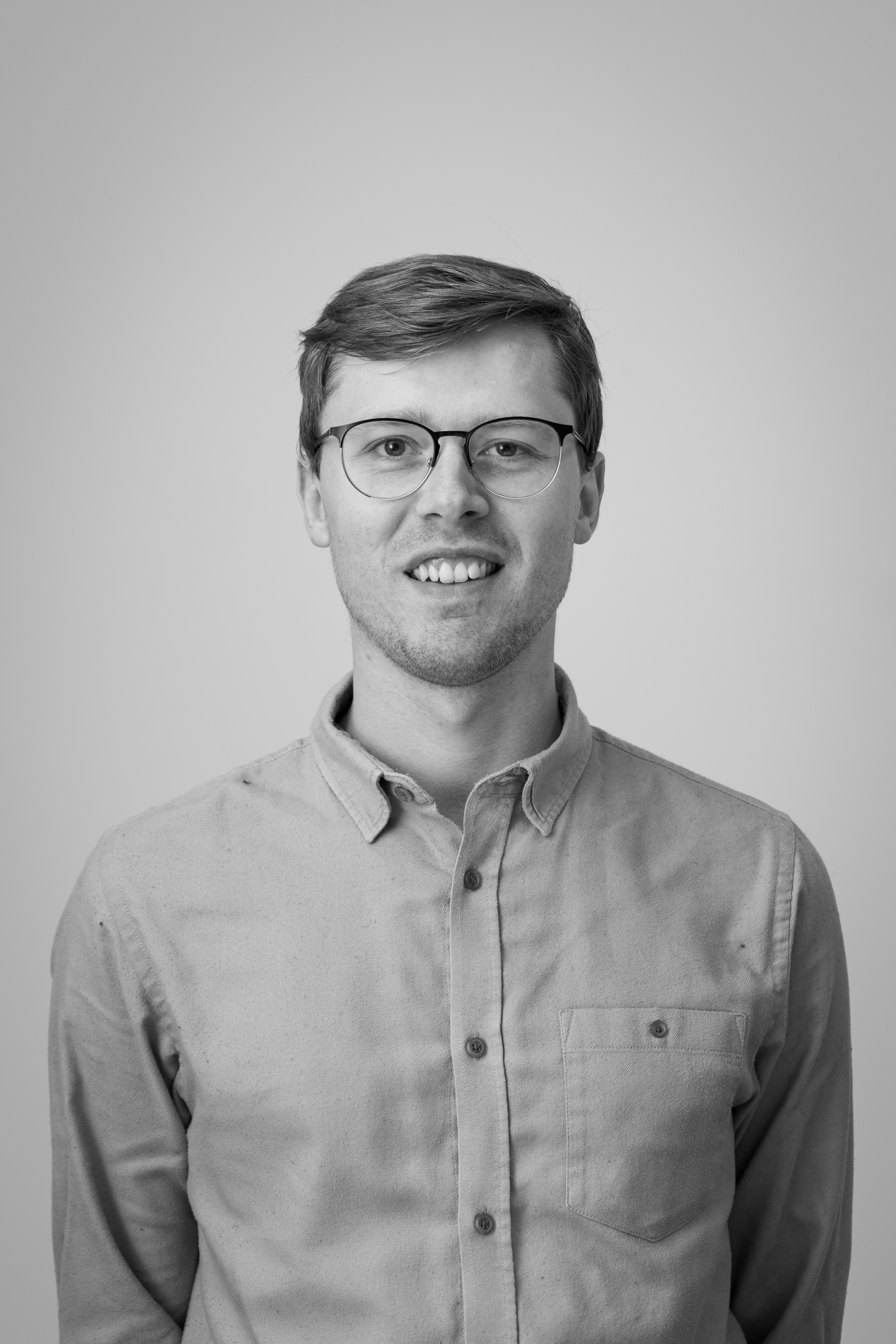}}]{Tijl Schepens} graduated in 2018 at KU Leuven campus Ghent, Belgium earning his M.Sc. degree. Afterwards he joined onsemi as an application engineer. For four years he was responsible for evaluating ICs and developing evaluation kits. In 2022, he joined the startup BelGaN which was founded to build Gallium-Nitride transistors in Belgium. After one year of test development on wafer probing systems, he started a PhD at the DRAMCO research group. His PhD focuses on developing the next generation of battery-less \gls{iot} devices through low-power wireless communication techniques.
  \end{IEEEbiography}

  \begin{IEEEbiography}
    [{\includegraphics[width=1in,height=1.25in,clip,keepaspectratio]{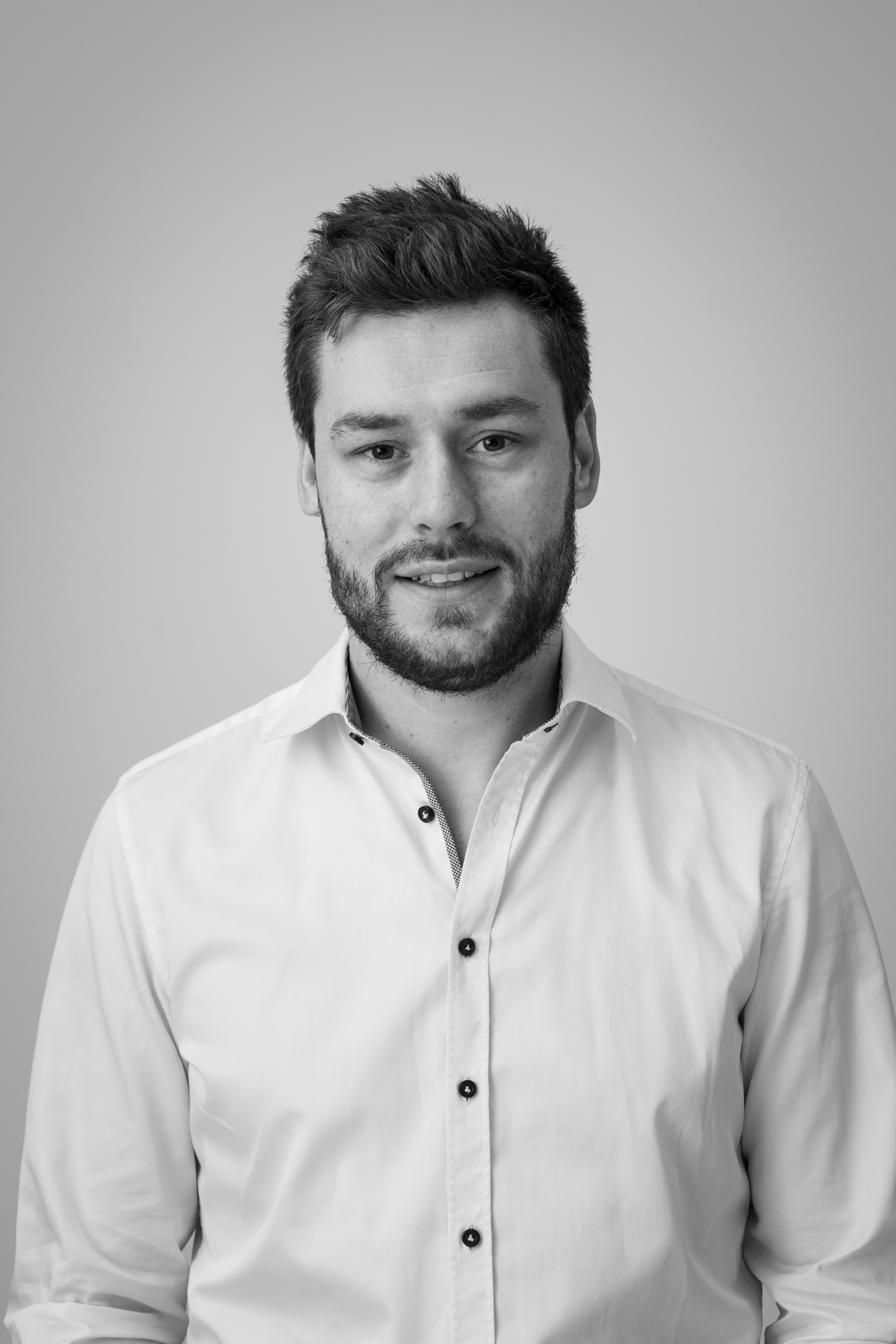}}]{Gilles Callebaut} earned his M.Sc. degree in Engineering Technology from KU Leuven, Belgium, in 2016, graduating summa cum laude. He completed his Ph.D. in Engineering Technology at the same institution in 2021, focusing on optimizing energy efficiency in \gls{iot} devices through single and multi-antenna technologies. Following a brief postdoc period under an FWO fellowship he became an assistant professor in 2025. He is a passionate advocate for bridging theoretical frameworks to practical applications. Gilles focuses on creating scalable, sustainable, and energy-efficient wireless technologies for \gls{iot}, 6G, and beyond, while actively promoting transparency and collaboration through open science. His main focus areas are: sustainability, \gls{iot}, 6G, WPT over \gls{rf} with multi-antenna systems and energy-efficient design.
  \end{IEEEbiography}

  \begin{IEEEbiography}
    [{\includegraphics[width=1in,height=1.25in,clip,keepaspectratio]{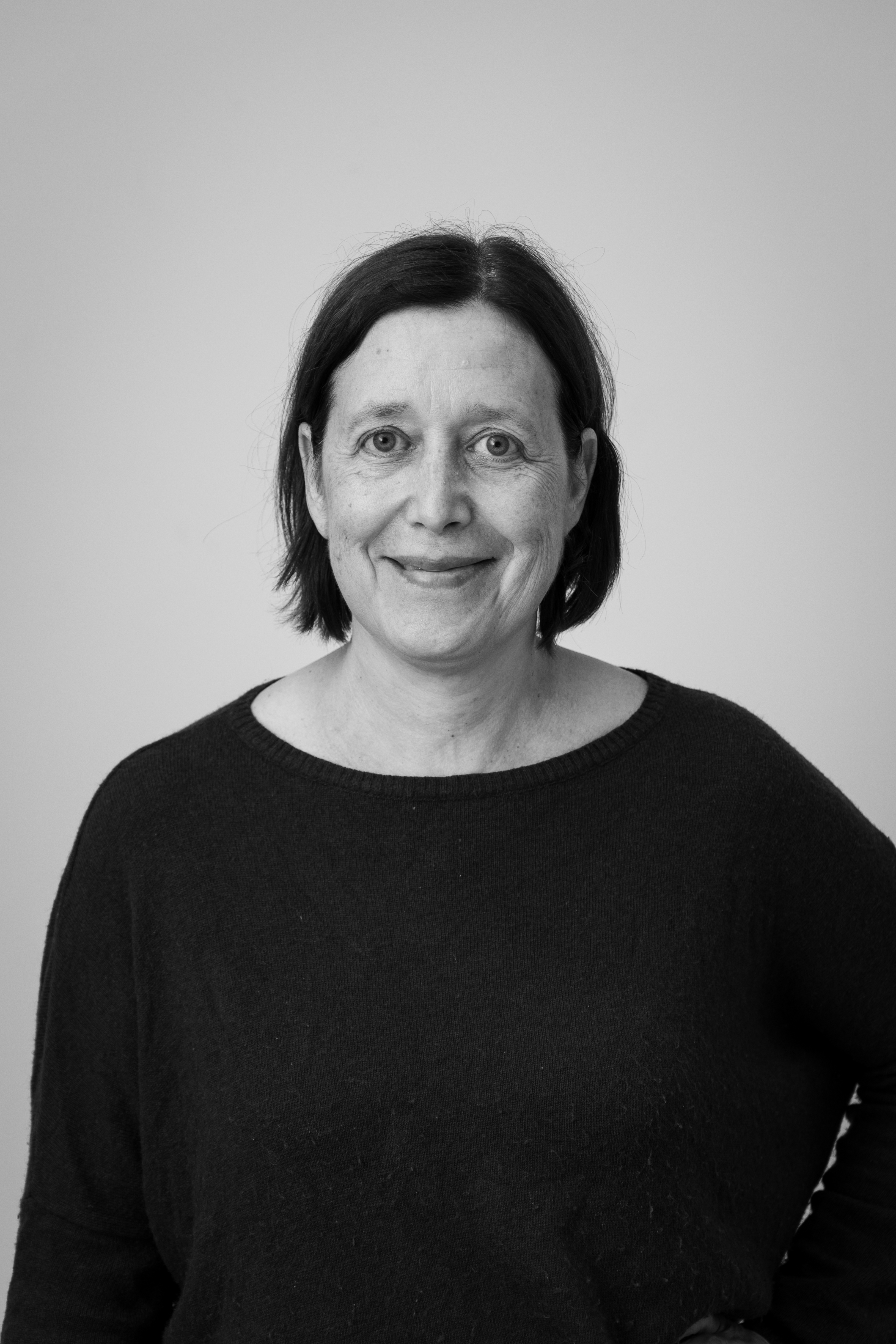}}]{Liesbet Van der Perre} received the M.Sc. and PhD degree in Electrical Engineering from the KU Leuven, Belgium, in 1992 and 1997 respectively. Dr. Van der Perre joined imec’s wireless group in 1997 and took up responsibilities as senior researcher, system architect, project leader and program director, until 2015. She was appointed full Professor in the DRAMCO lab of the Electrical Engineering Department of the KU Leuven and guest professor at the University of Lund in 1996. Her main research interest is in energy efficient wireless connectivity and embedded systems, with applications in multiple antenna and large array systems, \gls{iot} and broadband networks.
  \end{IEEEbiography}

\end{document}